\documentclass[12pt,oneside]{book} 

\usepackage{graphicx}
\usepackage[square,comma,sort&compress]{natbib} 
\usepackage{amsmath} 
\usepackage{amssymb} 
\usepackage{latexsym} 
\usepackage{bm}
\usepackage{wasysym}

\usepackage{lscape}

\usepackage{fancyhdr}

\usepackage{makeidx}

\usepackage{setspace} 
\usepackage{pseudocode}
\makeindex

\begin{document}
\pagenumbering{roman}

\chapter*{Modeling Protein Contact Networks}
\thispagestyle{empty}

\vspace{1cm}
PhD Thesis
\vspace{1cm}

\begin{flushleft} \textbf{Ganesh Bagler} \end{flushleft}
Centre for Cellular and Molecular Biology\\
(Jawaharlal Nehru University, New Delhi.)\\
Hyderabad, India.\\
December 2006\\

\chapter*{To my Gurus\ldots}
\thispagestyle{empty}

\begin{flushright}
who taught me to fly,\\
who gave me the dreams;\\
to whom I belong,\\
not by genes, but by \emph{memes}.\\
\end{flushright}

\chapter*{Acknowledgements}
\parindent0.0cm
\parskip0.2cm

\onehalfspacing

\textheight 23cm
\fancyhf{} 

\pagestyle{fancy}                       

\fancyhead[LE,RO]{\bfseries\thepage} 




\renewcommand{\headrulewidth}{0.3pt}   



I came to CCMB as a complete novice, without any notions about its
greatness or any understanding about its contribution to India's
molecular biology prowess. I was a Physicist after all! It took
some time for the realisation to sink that I am not only in a good,
but one of the best research laboratories I could ever have
been. Today, after close to five years, I can say it has been an
exhilarating journey of biology for me. What is more important is that
the journey seems to have just begun!     

I express my deepest thanks to Dr.\ Somdatta Sinha for her support and
guidance throughout the course of my PhD\@. She initiated me into new
areas of network science, systems biology. I am extremely grateful to
her for giving me a start into these happening ideas of the current
century.

I thank all the present and past members of the Dr.\ Somdatta Sinha's
Lab. It has been enriching to interact and learn from many of them. In
particular I thank Dr.\ Suguna, R.\ Maithreye, Dr.\ Amitava
Mukhopadhyay, Hemant Dixit, Nilanjan Maity, Dr.\ Ramrup Sarkar, Uttam
Maity, Mousumi Bhattacharya, Jeevan Karlos, and Karthik Venkatesh.

Various individuals have encouraged me from time to time; have inspired me
and guided me through my journey of scientific adventure. I thank
Mohan Rao, R.\ Sankaranarayanan, Deepak Dhar, Cosma Shalizi, Marc
Barthelemy, and Wilson Poon. I must also thank Santa Fe Institute and
the organisers of the Complex Systems Summer School, Beijing, 2005,
for providing with excellent intellectual environment. 

Had it not been for the scientific milieu available at CCMB, I would 
not have been able to add the quality to my work that I could. 
Triangle Group is an active group of protein researchers in 
CCMB. Being a part of Triangle Group, presenting my work to them and
getting their critical feedback has been part of learning process for
me through which I have improved upon my work. I thank every member of
Triangle Group for their contribution to my work.

CCMB is one of the best organisational setups that I have come
across. Its scientific team, technical support team, administration
division, maintenance team, all work together in an impeccable
manner. The value of such a setup is perhaps realised only when it is
not working to its full potential. I congratulate and thank the
Director and his entire team for maintaining such high standards.

I acknowledge the financial support of CSIR for the Junior and Senior
Research Fellowships. I also thank Dr.\ Sankaranarayanan and Dr.\
Ramesh Sonti for financial support. 

On personal front, there were times when I needed support. And
every time there was a need, I have had friends and well-wishers
supporting me--sometimes when I asked for it and many times without
even asking  for it. 

I thank P.\ Ramesh, Suguna, Ramesh Sonti, Somdatta Sinha, Jyotsna
Dhawan, Shashidhara, Mohan Rao, and Lalji Singh for taking time out for
me and helping me with their precious experience. 

I thank Praveen, Pallavi, Sumit, Gopal, Ram Parikshan, Pavan, Sathish,
Usha, Tirumal, Shoeb, Rupa, Prashant, Gurudatta, Bony, Raghvendra,
Vineet, Raghu, Shomi, Subhash, and Nishanth for being there for me. 

After working for PhD for close to five years and being associated
with such a large number of people, it is a difficult task to
remember and acknowledge every individual that mattered. But one thing
is for sure. I am not the same person that I was when I came here. As
I go I am taking each one of you with me as my part. That is the
ultimate acknowledgement that I can express.

\tableofcontents

\addcontentsline{toc}{chapter}{List of Figures} 
\listoffigures

\listoftables
\addcontentsline{toc}{chapter}{List of Tables} 

\chapter*{Synopsis}
\addcontentsline{toc}{chapter}{Synopsis} 

\parindent0.0cm
\parskip0.2cm

\onehalfspacing

\textheight 23cm
\fancyhf{} 

\pagestyle{fancy}                       

\fancyhead[LE,RO]{\bfseries\thepage} 




\renewcommand{\headrulewidth}{0.3pt}   



\section*{Introduction}
Proteins are an important class of biomolecules that serve as
essential building blocks of the cells. 
They are structurally complex and functionally one of the
most sophisticated molecules known. 
They perform diverse biochemical functions and also provide structural
basis in living cells. 
These all-pervasive, versatile molecules constitute (barring water) 
the largest fraction of the total mass of the cell.  
Proteins are macromolecules comprised of thousands of atoms.
They are characterised by a specific structure which specifies their
function.\\
\\
In the cell, they are synthesised in a complex multi-step process
starting from DNA to RNA to Protein, thereby giving genetic basis to
the protein sequence. 
Chemically, proteins are linear chains composed of ($20$ types of)
monomeric molecules called `amino acids'. 
These amino acids are linked together with a backbone made of peptide
bonds.    
This polypeptide chain folds into its unique three-dimensional (3-D)
structure, known as the `native state'. How, starting from a linear
chain of molecules, a protein attains its specific 3-D structure is an
unsolved problem in computational biology and is known as the `protein
folding problem'.  It's a system in which there is
an inherent 1-D structure in terms of the polypeptide backbone held
together by covalent peptide bonds. The polypeptide chain folds
onto itself by virtue of the chemical forces acting among the
constituent residues, thereby creating noncovalent `contacts' on
various length-scales as specified by the separation distance between the
contacting residues. These distance scales could be loosely defined as
short-range and long-range.\\
\\ 
Proteins perform an array of functions in the cell. 
They perform these specific functions by virtue of their precise
structure and chemistry. Structures are a critical determinant of their functions. 
Hence the study of structure--function relationship, prediction of
structure given the sequence etc., are important areas of research.\\
\\
Various approaches have been taken for this purpose. Experimentalists have
been performing biophysical experiments supported with genetics to get
answers to questions pertaining to protein structure--function
relationship. 
Theoreticians have believed that with the help of computational
power they will be able to obtain the answers that have been eluding
the experimentalists. Within theory, two distinct approaches have been
used: Forward and Reverse Engineering. Forward is the traditional way
in which one works from sequence to structure in a hope to obtain some
general results to the protein folding problem and other related
questions. 
Reverse Engineering relies on a large pool of structural data (such as
at Protein Data Bank (PDB)) that is made available. It approaches the
problem in reverse, as the name suggests, and tries to uncover the
laws with which the structures were put in place.\\
\\
A complex system could be modelled from various perspectives. Complex 
network analyses one way to study a system such as a protein 
structure. 

\section*{Objectives}
In this thesis we use coarse-grained, reverse engineering as our tool
and investigate experimentally known protein structures in an attempt
to gain better understanding of the processes by which they were
constructed. 
Specifically we focus on following three points.
\begin{itemize}
\item Describe protein structures as `contact networks' at two
  different length scales--Protein Contact
  Networks (PCN) and Long-range Interaction Networks (LIN)
\item Study the general complex network properties of protein structures.
\item Investigate how different secondary and tertiary
  structural features of proteins reflect in their network properties.
\item Investigate relation of network properties with biophysical
  properties, such as rate of folding, of proteins. 
\end{itemize}

\section*{Work Plan}
The work plan was as follows:
\begin{itemize}
\item To develop programs for extracting relevant data from the PDB file,
      to develop the code to build and visualise the complex network
      model of protein structures from the extracted data.  
\item To develop algorithms and programs for complex network
  analyses of PCNs and LINs.
\item To develop appropriate controls of the networks.
\item To calculate and study the relationship of the general network
  parameters for PCNs and LINs.
\item To analyse relationship between the network parameters of PCNs,
  LINs, and their controls to identify topological properties of PCNs
  and their relevance in structure-function relationship of proteins
  of diverse class.
\item To correlate two general network properties (assortativity and
  clustering) with the rate folding of single-domain, two-state
  folding proteins. 
\end{itemize}

\section*{Results}
In \textbf{Chapter~\ref{chap:protnet01}} we lead the reader to the content of the
dissertation by providing the methods and materials. The data
on proteins structures enables us to model them with atomic level
resolution. But we opt to coarse-grain the proteins on two different
scales. First, we model protein structures as Protein Contact Networks
(PCNs) in which the atomic-level details are jettisoned and amino
acids are represented as a point situated at  their respective
$C_{\alpha}$ atoms' coordinates. Noncovalent interactions, responsible
for the folding and stability of proteins, are depicted as spatial
contacts and any two residues are said to be in contact if they are at
a distance of less than or equal to $8$\AA\/.  On a coarser level, we
model Long-range Interaction Networks (LINs) wherein, apart from the
backbone, we consider only those contacts in the PCN that exist between
residues that are distant from each other along the backbone. 
We present computational procedures for creating PCNs, LINs as well
their random controls.
We also present various ways of visualising PCN data while
highlighting its various features.
We define various network parameters and illustrate them.
Finally we present the data that would be used for analyses in the
rest of the dissertation.\\
\\
In \textbf{Chapter~\ref{chap:protnet02}} we investigate the ``small-world''
nature of PCNs from proteins of various structural and functional
classes.   
All PCNs, irrespective of their classes, showed high clustering ($C$)
and low characteristic path length ($L$) compared to their random and
regular control networks. We also show that $L$ increases with the
logarithm of the size of the PCNs.
We emphasise the fact that the ``small-world'' result is a general
result and not restricted to globular proteins alone as shown
earlier. 
The question, then, follows is that whether non-globular
proteins such as fibrous proteins too would have small-world nature.
We investigate this question in this chapter. 
Other than $L$--$C$ properties, we also investigate degree
distributions of PCNs and LINs,  
hierarchical nature of the PCNs and other relevant network features.\\
\\
Amongst all the complex network systems studied proteins structures
are special because of their biological importance. 
Hence some unique property is anticipated in network properties of 
proteins. We do indeed find such a property.
In \textbf{Chapter~\ref{chap:protnet03}} we discuss this property,
assortative mixing in the contact networks of proteins at both short
and longer length scales (PCN \& LIN). 
We show that proteins are assortative in nature, i.e. rich nodes tend
to make contact with other rich nodes and poor nodes tend to make
contact with each other. Assortative degree correlations of proteins is an
exceptional property in the field of complex networks
as other networks (except for social networks) are known to be of
disassortative nature. 
Since it is known that assortative mixing plays a role in information
transfer across network, it implies that proteins are structurally (and
hence functionally) are different from other networks. 
We further explore topological origins of assortativity by
constructing appropriate controls. 
Random controls in which the degree distribution of the nodes is
conserved regain the assortative mixing which otherwise is not there
in the null model. This indicates that degree distribution is a crucial
feature that specifies assortativity in proteins.   
We also discuss other possible properties that might be conferred onto
proteins by virtue of their assortative mixing. \\
\\
The fact that proteins have special network property leaves us with
more questions. Natural selection not necessarily is a causal factor
for assortative mixing in complex systems.
There are network systems of biological origin, (e.g.\ as yeast gene
regulatory network, protein interaction network)  
that have been subjected to natural selection, but are known to be
disassortative. 
In \textbf{Chapter~\ref{chap:protnet04}} we ask: Do biophysical properties have any
bearing on network properties of proteins? \\

For this we chose $30$ single-domain two-state folding proteins whose
rate of folding is available.
We notice that as opposed to the clustering coefficients of
PCNs~($C_{PCN}$), which are indistinguishably clustered, those of
LINs~($C_{LIN}$) are sparse and unique. We show that clustering
coefficients of LINs, $C_{LIN}$s, are negatively correlated with the
rate of folding~($ln(k_F)$).  Each protein's departure from mean
compaction in its LIN is associated to rate of folding: the more the
departure the faster is the folding. Also, we find that coefficient of
assortativity of LINs ($r_{LIN}$) is positively correlated with the
$ln(k_F)$. Thus we identify two general network property (clustering
coefficient and coefficient of assortativity) that have negative and
positive association with the rate of folding of proteins.

\section*{Conclusions}
In this thesis we have investigated the protein structures using a
network theoretical approach. While doing so we used a coarse-grained 
method, viz., complex network analysis. We found that proteins by virtue
of being characterised by high amount of clustering, are small-world
networks. We also found that regardless of structural classification
all proteins, even fibrous proteins have signature of small-world
nature. Apart from the small-world nature, we found that proteins have
another general property, viz., assortativity. This is an interesting
and exceptional finding as all other complex networks (except for social
networks) are known to be disassortative. Importantly, we could
identify one of the major topological determinant of assortativity by
building appropriate controls. In our controls the assortativity is
partially recovered. Small-world nature and assortativity together
could be useful in dissipating mechanical disturbances across sparsely
distributed amino acids.\\
\\
The interesting question is if these general network parameters can 
offer any meaningful insight into the specific system properties
--the biophysical properties of proteins, in this case--which is a
naturally evolved intra-cellular network.
In this thesis we have shown that such correlations can be observed
even at a coarse grained model of protein structures at different
length scales.
Our results indicate that clustering coefficient ($C_{LRI}$) of the
LINs of the single-domain two-state folding proteins is negatively
correlated, and the coefficient of assortativity ($r_{LIN}$) are
positively correlated with the rate of folding of these proteins
($ln(k_F)$).
At PCN level, $C_{PCN}$ show no significant correlation, but $r_{PCN}$
has low but significant association with the rate of folding.
This indicates that our reverse engineering approach can offer
significant understanding of the differential role of contact
formations (``folding'') at different length scales in proteins.
We discuss our results in the light of some open questions in
modularity in protein structure, folding process and evolutionary
conservation of residues.

\parindent0.0cm
\parskip0.2cm

\onehalfspacing

\textheight 23cm


\fancyhf{} 


\pagestyle{fancy}                       

\fancyhead[LE,RO]{\bfseries\thepage} 

\renewcommand{\chaptermark}[1]{\markboth{\chaptername\ \thechapter:\ #1}{}}

\renewcommand{\sectionmark}[1]{\markright{\thesection \ #1}}

\fancyhead[RE]{\leftmark}      
\fancyhead[LO]{\rightmark}

\renewcommand{\headrulewidth}{0.3pt}   



\chapter*{List of Publications}
\addcontentsline{toc}{chapter}{List of Publications} 

\begin{itemize}
\item \textbf{Ganesh Bagler} and Somdata Sinha, ``Assortative mixing in protein Contact Networks and protein folding kinetics'', 
Bioinformatics, vol.\ $23$, no.\ $14$, $1760$--$1767$ ($2007$).

\item \textbf{Ganesh Bagler} and Somdata Sinha, ``Network properties of
  protein structures'', Physica A, $346$, $27$--$33$ ($2005$).

\item \textbf{Ganesh Bagler} and Somdatta Sinha, ``The Role of Host
  Migration on Host-Parasite Population Dynamics'', in \emph{Proceedings of
  the Second Conference on Nonlinear Systems and Dynamics}, $37$--$40$
  ($2005$).

\item Hemant Dixit, \textbf{Ganesh Bagler}, and Somdatta Sinha,
  ``Modelling the Host-Parsite Interaction'', in \emph{Proceedings of the
  First Conference on Nonlinear Systems and Dynamics}, $321$--$324$
  ($2003$).
\end{itemize}

\pagenumbering{arabic}

\chapter{\label{chap:protnet00} Introduction}

\section{Protein: An Important Biomolecule}
\label{chap00:sec:protein}

Proteins are an important class of biomolecules and serve as essential
building blocks of the cell. 
They are structurally the most complex and functionally one of the
most sophisticated molecules known. 
They perform diverse biochemical functions and also provide structural
basis in living cells. 
Barring water, these all-pervasive, versatile molecules constitute
the largest fraction of the total mass of the cell.  

Chemically, proteins are linear chains composed of monomeric molecules
called `amino acids'. 
These amino acids are linked together with a backbone made of peptide
bonds.    
This polypeptide chain folds into its unique three-dimensional
structure, known as the `native state'.   
Proteins are practically involved in every function performed by a cell,
such as gene regulation, signal transduction, metabolism etc. 
These functional abilities (listed below) of the proteins
are specified by their detailed three-dimensional (3-D) structures.  

Following is a partial list of the roles/functions that proteins, by
virtue of their structure, are known to be part of. 

\begin{itemize}
\item Enzymes (eg.: biological catalysts)
\item Antibodies (eg.: immune system molecules)
\item Regulation (egs.: transcription, translation)
\item Messengers (egs.: transmission of nervous impulses, harmones)
\item Transport (eg.: transportation of molecules ranging from
  electrons to macromolecules)
\item Storage (egs.: hemoglobin stores oxygen, iron stores ferritin)
\item Mechanical Support (eg.: structural proteins used in skeletons
  such as collagen )
\end{itemize}

Studying protein structures is not only of fundamental scientific
interest in terms of understanding biochemical processes, but also
produces practical benefits. Understanding proteins' structural
properties, their relation to function (as well as loss of
function), folding kinetics, relevance of specific (sometimes also
referred to as `hot') residues, and collection of residues (such as folding
nucleus, active sites), are of considerable importance in
biotechnology industry, agriculture, medicine, to name a few. The
knowledge gained from such an understanding can be put to use for
`protein engineering'. The properties of proteins could be modified,
enhanced, and in fact proteins of novel and desired properties could
be designed \emph{de novo} with better understanding of the areas
mentioned above.  

It is important to understand how proteins consistently fold into
their native-state structures and the relevance of structure to their
functions.
The folding mechanism, kinetics, structure, and function of proteins
are intimately related to each other.  
Misfolding of proteins into non-native structures can lead to several
disorders~\cite{taubes_Science1996}.  
Understanding of the folding process will provide clues to misfolding
and resulting disorders. 
Correlating sequence with structure, as well as understanding of
folding kinetics has been an area of intense activity for  
experimentalists and
theoreticians~\cite{fersht_book,branden_and_tooze_book}.  
Among the different theoretical approaches used for studying protein
structure, function, and folding kinetics, a graph theoretical
approach based on perspectives from complex networks has been used
recently to study protein
structures~\cite{cabios,kannan,protnet:PRE,protnet:JMB,protnet:Biophys,Bagler2005,Brinda2005,Amitai2004}. 
 
\section{Protein: A Complex System}
Proteins could be regarded as complex dynamical systems which is
reflected in their surprisingly fast folding process by which they attain
their native-state structure. 
Despite large degrees of freedom, surprisingly, proteins fold into
their native state in a very short time which is known as the  
Levinthal's Paradox~\cite{levinthal}.
All the information needed to specify a protein's three-dimensional
structure is contained within its amino acid sequence. 
Given suitable conditions, most small proteins fold to their native
states~\cite{anfinsen_science}.  
The spontaneous folding of proteins into their elaborate
three-dimensional structures, starting from linear chains, is  
one of the remarkable examples of biological self-organisation.

Not only individual proteins, but multi-protein units,
together, are proposed to be working as `computational elements in
living cells'~\cite{Bray1995}. Many proteins that appear to have as their
primary function the transfer and processing of information, are
functionally linked through allosteric or other mechanisms into
biochemical `circuits' that perform a variety of simple computational
tasks including amplification, integration, and information
storage~\cite{Bray1995}. 

\section[Complex Network Models: A Brief Historical
Perspective]{Complex Network Models: \\A Brief Historical Perspective} 
Complex systems, that are characterised by discrete constituents and
their inter-relationships, have been traditionally studied in the
field of graph theory~\cite{Bollobas2001,reka:thesis}. Erd\"{o}s and
R\'{e}nye proposed that connectivity in  the large-scale, real-world
networks are random.  
For decades this proposition remained unchallenged~\cite{Bollobas2001}.  
Systems of high complexity and that coming from diverse origins are
known to be driven by networks of elements. 
Living cells, eventually, are the outcome of dynamic interactions
among various networks such as protein-protein interaction, 
gene regulatory network, signal transduction pathways, metabolite
networks and such.  
Many other non-biological systems are also amenable to complex
network analysis. Few examples of such networks are:
Internet~\cite{Internet:yook,satorras:knn,Internet:faloutsos},
World-Wide Web~\cite{WWW:reka,WWW:adamic,WWW:physicaa}, Software
Network, Power-Grid Network, Transportation
(Railways~\cite{transport:manna},
Airlines~\cite{transport:barrat,transport:ANI}) Networks, Social
Networks~\cite{newman:society01,socialNW:newman2001,r:newman} to
name a few.  
Each of these networks is, shaped by physical, spatial (geographical), 
technological, even political influences, that are specific
to that system. 
The question then is, can the networks driving such systems be
inherently random?  It seems logical that the processes and dynamics
responsible for these networks wire them in a non-random
fashion. Lately, it has been  shown that complex networks are indeed
non-random~\cite{reka:thesis,dorogovtsev:book}. 

In recent years, there has been considerable
interest~\cite{reka:thesis,dorogovtsev:book} in structure and dynamics  
of networks, with application to systems of diverse origins such as 
society (actors' network, collaboration networks, etc.), technology  
(world-wide web, Internet, transportation infrastructure), biology
(metabolic networks, gene regulatory networks, protein--protein  
interaction networks, food webs) etc. 
These are characterised by some universal properties, 
such as small-world nature~\cite{watts:nature,watts:book} and a
scale-free degree distribution~\cite{WWW:reka,reka:thesis}. 

Below we briefly summarise a few of the important network features. 

\subsubsection{Small-World Networks}
\begin{figure}[!tbh]
\begin{center}
\includegraphics[scale=1.0]{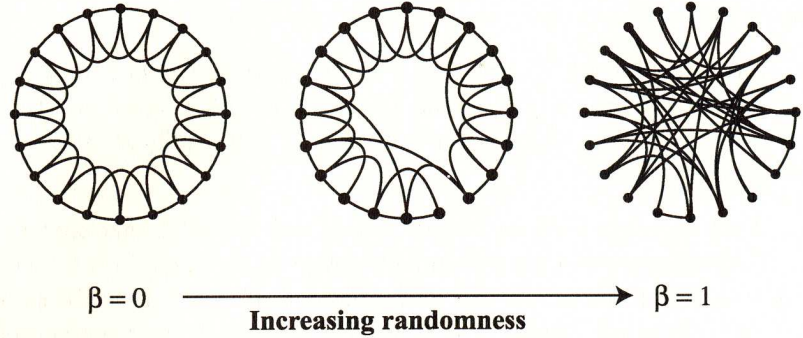}
\end{center}
\caption[The Small World.]
{The Small World Topology.} 
\label{fig:small_world}
\end{figure}

Erd\H{o}s and R\'{e}nye~\cite{Bollobas2001} define a random graph
as $N$ nodes connected by $n$ edges which are chosen randomly from the
$\frac{N(N-1)}{2}$ possible edges. But in reality, the connections in
the networks are not random and are dictated by various forces. One
way it reflects is that real-world networks have unusually high
clustering coefficients. These networks with high amount of clustering
are classified as ``small-world networks''. Watts and
Strogatz~\cite{watts:nature} visualised them as depicted in
Figure~\ref{fig:small_world}. $\beta$ is the edge rewiring
probability. Small-world graphs are the systems
obtained midway between regular ($\beta=0$) and random ($\beta=1$)
graphs, when, starting from regular networks, the edges are rewired
with increasing probability $\beta$. Networks of diverse origins have been
shown to be having a small-world 
nature~\cite{reka:thesis,dorogovtsev:book}.

\subsubsection{Scale-Free Networks}
While modelling systems as random graphs and the small-world models,
the emphasis was modelling the network topology. The scale-free model
put the emphasis on modelling network assembly and evolution. While
the goal of the former models was to construct a graph with correct
topological features, modelling scale-free networks put the emphasis
on capturing the network dynamics~\cite{reka:thesis}. 
The Scale-Free model is composed of two constituents:
\begin{enumerate}
\item \emph{Growth}: Starting with a small number ($m_0$) of nodes, at
  every time-step one adds a new node with $m$($ \le m_0$) edges that link
  the new node to $m$ different nodes already present in the system.
\item \emph{Preferential Attachment}: When choosing the nodes to which
  the new node connects, one assumes that the probability $\prod$ that
  a new node will be connected to node $i$ depends on the degree $k_i$
  of node $i$, such that $$\prod(k_i)=\frac{k_i}{\sum_j k_j}.$$
\end{enumerate}
Scale-free networks are characterised with power-law degree
distribution. 

\subsubsection{Modularity}
The concept of modularity assumes that the system's functionality
could be seamlessly partitioned into collection of modules~\cite{ravsaz:science}.
Various networks that have been investigated so far have been found to
be modular in nature, where a module is a discrete entity with several
elementary components and performs an identifiable task, separable
from the functions of the other modules~\cite{ravsaz:science,Shen-Orr2002}.   

\subsubsection{Degree Correlations}
A measure that expresses degree-degree correlation feature of
the network is assortativity. It exhibits whether, in a network, nodes with poor
degrees tend to connect to those with poor degrees, or, those with
rich degrees. A network can be assortative, dominated by rich-to-rich
node connections, or it could be disassortative with more rich-to-poor
connections. A random network has no preferred degree correlation
tendencies. It is known that most real-world networks (except for
social networks) are disassortative~\cite{r:newman} and the origin of
disassortativity in real-world networks is  listed as ``one of the
ten most leading questions for network
research''~\cite{EPJB:round_table}. 
This property has been shown to be having a bearing over
the percolation threshold of the network~\cite{r:newman}. 

\subsubsection{Some of the leading questions in complex network research}
Apart from origin of disassortativity, as mentioned above, many other
questions are unsolved and are considered to play a potentially 
important role in the field of network
research~\cite{EPJB:round_table}. Many networks in nature are found to
be modular as well as hierarchical. The emergence of modularity and
how it could be reconciled with other properties of networks are basic
questions in network research. Networks are characterised by topology
as well as the dynamics that is taking place over it. Are there universal
features to the network dynamics similar to the topology? Compared to
the technological networks, the evolution of biological networks is
much more complex. What could be the evolutionary mechanisms that
shape the topology of the biological networks? These and many other
questions remain at the forefront of the network research.

\subsubsection{Biological Networks}
Among the various networks studied, biological networks are of special
interest as they are the product of long evolutionary history. 
The mode of creation, evolution and functionality of these
networks are distinct from those of technological networks.


\begin{itemize}
\item Biological networks are the products of natural selection as
  opposed to the rest of the networks. 
\item The time-scale at which these have evolved is orders of
  magnitudes larger than that of non-biological networks.
\item Since each of these evolutionary machines are the outcome of
  `survival of the stable'~\cite[page 12]{book_the_selfish_gene} rule, these
  systems are the most stable and robust (against all natural
  detrimental sources) systems known to us. 
\end{itemize}

These are of academic interest for their complex, versatile,
dynamic, and evolvable nature. On the practical side, understanding
the nature of biological systems have direct or indirect implications
to drug design, disease diagnosis and cure, epidemic control, and
biotechnological applications.  

Biological networks~\footnote{We classify `social networks' as
  non-biological keeping in view our criterion, that a system that
  is shaped by natural selection, for biological networks.} 
could be characterised by the length-scale as follows. 

\begin{itemize}
\item Ecological (Food Webs)
  Networks~\cite{Hastings1991,Berlow2004,Sole2000,Camacho2002,Williams2002,Dunne2002}; (in metres)  
\item Inter-cellular Networks (between micrometres to millimetres)
\item Protein-Protein Interaction Networks~\cite{protein:yook,Mering2002,Uetz2000}, Metabolic Pathways
  Networks~\cite{metabolic:jeong,ravsaz:science,Almaas2004,Almaas2005}, 
 Gene Regulation Networks~\cite{gene:reka,Farkas2003,Shen-Orr2002} ( in micrometres) 
\item Macromolecular Networks--Complex Chemicals, Polymers~\cite{Scala2000} (in nanometres)
\end{itemize}

Our focus is on `protein structures', an interesting class of
macromolecular networks. 
Proteins are unique among all other networks. They are constituted
from a linear polymer chain of amino acids as opposed to sparsely 
distributed unconnected nodes as in most other networks. They evolve
by changing their conformation and not by addition or removal of
nodes. Their polypeptide backbone attains a stable shape through
well-defined secondary structures and tertiary folds.

It is important to understand how proteins consistently fold
into their native-state structures and the relevance of structure to
their biological function. Network analysis of protein structures is
an attempt to study the networks as complex dynamical systems composed
of a web of interacting elements, and, thereof, to understand
possible relevance of various network parameters. 

\section{Protein Structures as Complex Networks}

Many efforts have been done to model biological systems from complex
systems
viewpoint~\cite{Atlan2003,Oltvai2002,Koch1999,Smaglik2000,Knight2002,Csete2002}. 
Specifically they have been increasingly studied as complex
networks~\cite{Alm2003,Proulx2005,Barabasi2004,Strogatz2001}. 
Amongst all the biological systems, a system of special interest is
that of a `protein' for its structure, function, kinetics and
stability.   
With its omnipresence in the cell and diverse functionality, it is a
biomolecular system with immense implications to the cellular
dynamics.  
It's function is specified by the structure. 
The structure is also associated with it's kinetic properties and 
stability. 
All these make a protein a very interesting system to study as a `complex
dynamical system'. 
Here, we are specifically interested in studying protein structures as
networks of noncovalent contacts, and the covalent backbone contacts.  

\subsection{Fine-grained vs. Coarse-grained Models}
Various approaches~\cite{fersht_book,baker:nature} have been used for
studying protein structures as well as the protein folding
dynamics. Apart from other differences, these methods vary in the
extent of detail with which the structure is modelled. Some consider
atomic-level details~\cite{Orengo1999,Murzin1999}, whereas some reduce
the structure to a chain of beads spatially constrained to a
rectangular lattice with a limited number of attainable
conformations~\cite{mirny_shakh_PNAS_1998}.   
The relevance and applicability of each of these models, of
course, rests on the kind of questions that are asked. While
\textbf{fine-grained models} are heavy on resources (computer memory,
time needed, complexity of coding etc.), they are better suited for
questions that involve aspects of protein that, from experimental
studies, are known to be dependent on fine structural details. On the
other hand, \textbf{coarse-grained models} are of special value as
they make it feasible to work with a large and complex system and
offer a systems-level insight.


By virtue of a large number of constituent atoms and complexity of
chemical interactions amongst them, a protein structure
is a system with large degrees of freedom, rendering it immensely
difficult for detailed modelling and analyses.  Given the diversity in
functional roles of protein and fairly large number of structural
units (Number of Unique Folds, as defined by SCOP~\cite{SCOP},
$\approx~1000$ as of Oct.\ 2006~\cite{PDB}) that proteins are composed
of, it makes sense to consider coarse-grained models as a viable
option for modelling protein structures.  
Today, thanks to the untiring efforts of crystallographers and
structural biologists across the world, and because of the advances
in techniques and instruments, one has open access to a huge
repository of interesting protein structures such as PDB, MSD-EBI,
PDBJ, BMRB. 
As on $15^{th}$ August $2006$, there were a total of $38198$
structures deposited in  the Protein Data Bank~\cite{PDB}. Also, from earlier
studies~\cite{Alm1999,Riddle1997,Kim1998,Perl1998,Chiti1999}, it is 
known that protein structures are amenable to coarse-graining while
being of practical use. These facts strengthen the case of
`coarse-grained models' vis-\`{a}-vis detailed, fine-grained ones.

\subsection{Range of Interactions in Proteins}
\label{subsec:range}
Proteins are characterised by interactions happening at various
ranges. Here, in the context of linear chain nature of proteins, range
is defined as the distance between two interacting residues along the
polypeptide. Interactions, then, can be divided into long- and
short-range interactions. Apart from interactions that take place in
the process of folding, many NMR experiments have shown that even
after reaching the native state, proteins undergo conformational
fluctuations with time scales from several nanoseconds to
milliseconds. It has been suggested that such functionally important
fluctuations are triggered by long-range interactions among a network
of residues~\cite{Dima2006}. Communication happening via such
long-range interactions is central to protein function and proteins
have evolved specific mechanisms to address this constraint. It has
been shown that~\cite{Suel2002} information about these mechanisms are
embedded in the evolutionary record of a protein family. In our work,
we delineate a range of interactions to study their individual importance
and contribution.

\subsection{Earlier studies on Protein Contact Networks}


So far many studies have been undertaken to investigate protein
structures as complex networks of interacting residues. 

In an early study~\cite{Crippen1978}, Crippen analysed protein
structure in which effort was done to offer an objective definition of
the domain of a protein. The author studied the structural
organisation through a binary tree clustering algorithm for the
residues of a single polypeptide chain. It was found that the 
protein structure is constituted of a hierarchy of segments that group
together, then these clusters merge together eventually to form the
complete chain.  

In another similar study, Rose~\cite{Rose1979} developed an automated
procedure for the identification of domains in globular
proteins.Through a slightly different approach Rose reached the
conclusion that hierarchic organisation of structural domains is an
evidence in favour of an underlying protein folding process that
proceeds by hierarchic condensation.    

Asz\'{o}di and Taylor~\cite{cabios} modelled linear polypeptides as
well as 3-D proteins as non-directed graphs. They defined two
topological indices, one (\emph{connectedness number}) for
residue-distance measure and another (\emph{effective chain length})
as a foldedness measure, to compare folding topologies. They could
reveal the hierarchical structure in the non-backbone connections of
proteins. 

Kannan and Vishveshwara~\cite{kannan} have used the graph spectral
method to detect side-chain clusters in three-dimensional structures
of proteins. The approach they described is used to detect a variety of
side-chain clusters and to identify the residue which makes the
largest number of interactions among the residues forming the cluster. 
Vishveshwara and
others~\cite{Brinda2005,Brinda2005a,Brinda2006,Krishnadev2005,Sistla2005} 
have consducted many studies with amino-acid networks.

Vendruscolo et al.\/~\cite{protnet:PRE} showed that protein
structures have small-world~\cite{watts:book} topology.  
They studied transition state ensemble (TSE) structures to
identify the key residues that play an important role of ``hubs'' in
the network of interactions to stabilise the structure of the
transition state. They also showed that, though homopolymers have high
clustering comparable to those of the proteins, their betweenness
profile is uniform unlike that of the proteins.

Greene and Higman~\cite{protnet:JMB} studied the short-range and
long-range interaction networks in protein structures and showed 
that long-range interaction network is \emph{not} small world and its
degree distribution, while having an underlying scale-free 
behaviour, is dominated by an exponential term indicative of a
single-scale system.  

Atilgan et al.\/~\cite{protnet:Biophys} studied the network properties
of the core and surface of globular protein structures,  
and established that, regardless of size, the cores have the same
local packing arrangements. They showed that connectivity distribution
of residues is independent of their spatial location.
They also explained, with an example of binding of two proteins, how
the small-world topology could be useful in efficient and effective
dissipation of energy, generated upon binding.

Aftabuddin and Kundu~\cite{Aftabuddin2006,Kundu2005} have studied
protein structures as made of three classes of amino acids:
hydrophobic, hydrophilic and charged. They found that average degree
of the hydrophobic networks has significantly larger value than that
of hydrophilic and charged networks. They also found that all amino
acids' networks and hydrophobic networks bear the signature of
hierarchy; whereas the hydrophilic and charged networks do not have
any hierarchical signature.   

Shakhnovich and others~\cite{dokholyan_pnas2002} have studied protein
conformation network to study features that make a protein
conformation on the folding pathway to become committed to rapidly
desceding to the native state. They used a macroscopic measure of the
protein contact network topology, the \emph{average graph
  connectivity}, by constructing graphs that are based on the geometry of
protein conformations. They found that average connectivity is higher
for conformations with a high folding probability than for those with
a high probability to unfold.   

Jung et. al.~\cite{jung_lee_moon} studied the protein structures in
search of identification of topological determinants of protein
unfolding. They find that a newly introduced quantity, \emph{the impact
edgre removal per residue}, has a good overall correlation with
protein unfolding rates.

Amitai \emph{et. al.}~\cite{Amitai2004} found that active site,
ligand-binding and evolutionarily conserved residues, typically have
high closeness, a network property, value. What separates this method
from others is that this method solely depends on single protein
structure's information while making such a conclusion and does not rely
on sequence conservation, comparison to other similar structures, or
any prior knowledge.  

In a recent paper Sol \emph{et. al.}~\cite{Sol2006} study proteins as
systems that have a permanent flow of information between amino
acids. By doing removal experiments in seven protein families they
find that many of the centrally conserved residues are also important
for allosteric communication. They put these results in perspective
in view of network dynamics, topology, constraints on the evolution of
protein structure and function.

\section{In this thesis}
\label{sec:this_thesis}
The aim of this work has been to describe and study the
three-dimensional native-state structures of proteins of different
structural and functional classes as complex networks, enumerate the
general network parameters, and study the relation of these parameters
to their structural, functional, and kinetic properties at different
length scales.

In our studies, we model protein structure as networks of interacting
residues. We start from detailed fine-grained protein structure with
atomic level details and obtain Protein Contact Network (PCN) by
coarse-graining. In this process we keep positional information of
$C_{\alpha}$ atoms which are representatives of the amino acids and
disregard all the other information. The `contacts' between any two
residues represents a possible noncovalent interaction happening
between them. The cut-off threshold ($R_c$) for deciding a contact is
chosen accordingly at $8$\AA\/. We describe the construction of PCN
model in Chapter~\ref{chap:protnet01}. 
We consider interactions happening at various length-scales as
described in Subsection~\ref{subsec:range}. 
Long-range Interaction Network (LIN) is a subset of PCN and comprises
only of the backbone and the long-range interactions. In this
chapter, we also describe the construction of LINs and different
control networks. Further, we
explain various visualisation schemes that we have used in our
studies. Then we define and illustrate various network parameters
and properties. Finally, we present the data of the proteins that will
be used in our studies.

In Chapter~\ref{chap:protnet02} we describe our results related to
small-world nature of the PCNs. We find that protein structures of
diverse structural and functional classification display small-world
nature. We observe that all the $80$ proteins of different classes
have very high clustering coefficient. Despite being structurally
different from globular proteins, even fibrous proteins are found to
be having small-world signature. We find that PCNs have a clear
signature of hierarchical nature on the clustering versus size profile.  

PCNs are an unique class of macromolecular complex networks
characterised by biological origin and evolutionary pressure. Hence
one expects PCNs to show their unique nature through network
properties. In Chapter~\ref{chap:protnet03} we find that PCNs are
`assortative', i.e.\ rich nodes tend to connect to rich nodes and poor
nodes tend to make contact with each other. This is an exceptional
observation as it is known that (except for social networks) all other
complex networks are `disassortative'. We find that LINs, despite
their very different degree distribution, are also assortative. This is an
interesting observation as it indicates that the short-range
interactions possibly don't contribute towards the observed
assortativity. In our study to investigate the role of
various network features in bringing in assortativity, we show that
degree distribution has a major contribution towards conferring
assortativity in PCNs as well as in LINs.

In Chapter~\ref{chap:protnet04} we investigate biophysical correlates of
the topological parameters of PCNs and LINs. 
For this study we use $30$ single-domain two-state folding proteins
whose rate of folding is known.
We find that the exceptional topological property, assortativity, has
a positive correlation with the rate of folding ($ln(k_F)$) for both
PCN and LIN of the proteins. Also, we find that clustering
coefficients of LINs has a very good negative correlation with the
$ln(k_F)$. 

Thus with the help of our coarse-grained, complex network models we
analyse protein structures and study questions relating to structure,
function and kinetics.

\chapter{\label{chap:protnet01}Materials and Methods}

In our aim of analysing the protein structures, we developed various
network models of protein structures, as well as their controls, 
and defined various network properties of these models. 
In this chapter, in Section~\ref{chap01:sec:nw_construction}, we
describe the models that we have used and the procedure of
constructing the models. Wherever required we also mention the
algorithms that were used for this purpose. 
In Section~\ref{chap01:sec:nw_visualisation}, we describe and
illustrate a few ways of visualising the network.
Throughout our study we use various network parameters and properties
to characterise the network system under study. 
In Section~\ref{chap01:sec:network_parameters} we define and
describe these parameters. In
Section~\ref{cha01:sec:data_analyses} we present the data, along with
other relevant information, of the proteins that we have used in our
studies. We mention the details of programming languages and the
software used in Section~\ref{cha01:sec:softwares}. 
The pseudocodes of all the programmes (written in FORTRAN90 \& MATLAB)
are given in the Appendix~\ref{app:pseudocode}.     

\section[Construction of Protein Networks and Controls]
{Construction of Protein Contact Networks and Controls}
\label{chap01:sec:nw_construction}

In our studies, we have used graph theory to model protein structures.
Graphs, in general, could be used to model various kinds of systems in
which nodes (vertices) represent discrete network elements and links
(edges) represent a well-defined relationship between any two
nodes. Below we explain two coarse-grained models of protein
structure controls that were used in our studies.

\begin{figure}[!tbh]
\begin{center}
\includegraphics[width=14cm]{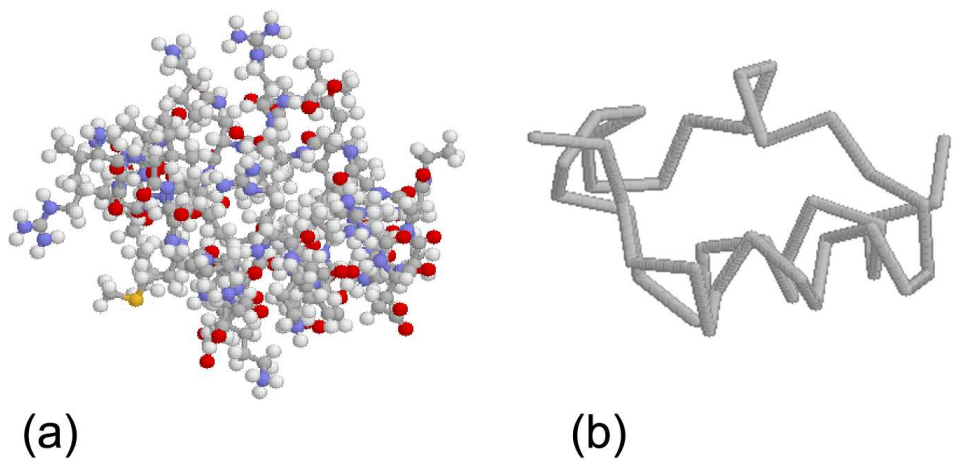} 
\end{center}
\caption[Two representations of Acyltransferase (2PDD)]
{Two representations of Acyltransferase (2PDD) 
(a) Ball and stick representation. The colours of the atoms are
attributed as specified by RasMol's `CPK colour scheme':
hydrogen~(white), carbon~(gray), oxygen~(red), nitrogen~(light blue),
and  sulphur~(yellow). 
(b)~the backbone. } 
\label{fig:pcn}
\end{figure}

\subsection{Protein Contact Network (PCN)}
\label{chap01:subsec:PCN}
We modelled the native-state protein structure as a network made of
its constituent amino-acids and their noncovalent
interactions. Protein Contact Network (PCN) is a graph-theoretical
representation of the protein structure, where
each amino acid is a `node' and spatial proximity of any two amino
acids is a `link' between them. Any two amino acids were considered to be in
`spatial contact' if the distance ($R_c$) between their $C_{\alpha}$
atoms was less than or equal to $8$\AA. The choice of $R_c$ was 
based on the range at which non-covalent interactions, which are
responsible for the polypeptide chain to fold into its native-state,
are effective. 

A point to note is that, apart from the noncovalent
interactions, we considered the covalent peptide bonds between
consecutive amino acids as links, thus representing the backbone of
the protein. This chain of backbone-links was left unaltered while
creating the controls, thus reflecting an important aspect of protein
folding dynamics: throughout the folding process, the peptide backbone
is unbroken and the protein goes through structural changes
by making and breaking the noncovalent contacts.  

\subsubsection*{Contact Map}
Contact Map
(CM)~\cite{Vendruscolo1999,Domany2000,Banavar2001,DEMIREL1998,Gupta2005,Hu2002}
is a 2-D, binary, symmetric representation of the 
protein structure in terms of pair-wise, inter-residue contacts. 
Any two residues are defined to be in `contact' with each other if the
$C_{\alpha}$ atoms of these two residues are within a cut-off 
distance~($R_c$).    
Thus the contact map is a coarse-grained representation of the 3-D
structure of a protein.  
A contact map~($\mathbf{M}$) for a protein with $n_r$ residues is a
matrix $\mathbf{M}$ of the order $n_r \times n_r$,  whose elements are  
defined as, 

\begin{equation}
\mathbf{M}_{ij} = 
   \begin{cases}
     1 & \textit{if residues $i$ and $j$ are in contact}\\
     0 & \textit{otherwise}
   \end{cases}
\label{contact_map_defn}
\end{equation}

\subsubsection*{The Choice of Cut-off Distance  ($R_c$)}
The choice of cut-off distance was done based on the chemical
interactions that are responsible for folding, unfolding, stability, 
function, etc. The chemistry of these processes is
primarily dictated by chemistry of noncovalent interactions, viz.,
Van der Waals interactions, hydrogen bonds, ionic bonds, hydrophobic
interactions. The cut-off threshold could be varied from a very
high, fine-grained resolution (say, $R_c \approx 4$) to a very low,
coarse-grained resolution. There is lower as well as upper limit to
the cut-off. 
A value of $R_c$ that is less than the resolution of the protein model
doesn't make sense. And a threshold larger than the size of the
protein, again, is meaningless. For our purpose, to retain the meaningful
information specified by the noncovalent interactions, while at the
same time not be bogged down by the atomic level details, a threshold
of $R_c=8$\AA\ is an ideal choice.

In our studies we have used $7$\AA\ or $8$\AA\ as a cut-off threshold
depending on the data-set, though the results are valid for a range of
thresholds between (at least) $7$--$9$\AA\/.  
For practical purposes the threshold should be considered $R_c=8$\AA\
throughout our studies. Various cut-offs ranging from
$5$\AA~\cite{protnet:JMB}, to $7$\AA~\cite{protnet:Biophys}, to
$8.5$\AA~\cite{protnet:PRE} have been used in earlier studies.

\begin{figure}[!tbh]
\begin{center}
\includegraphics[scale=0.5]{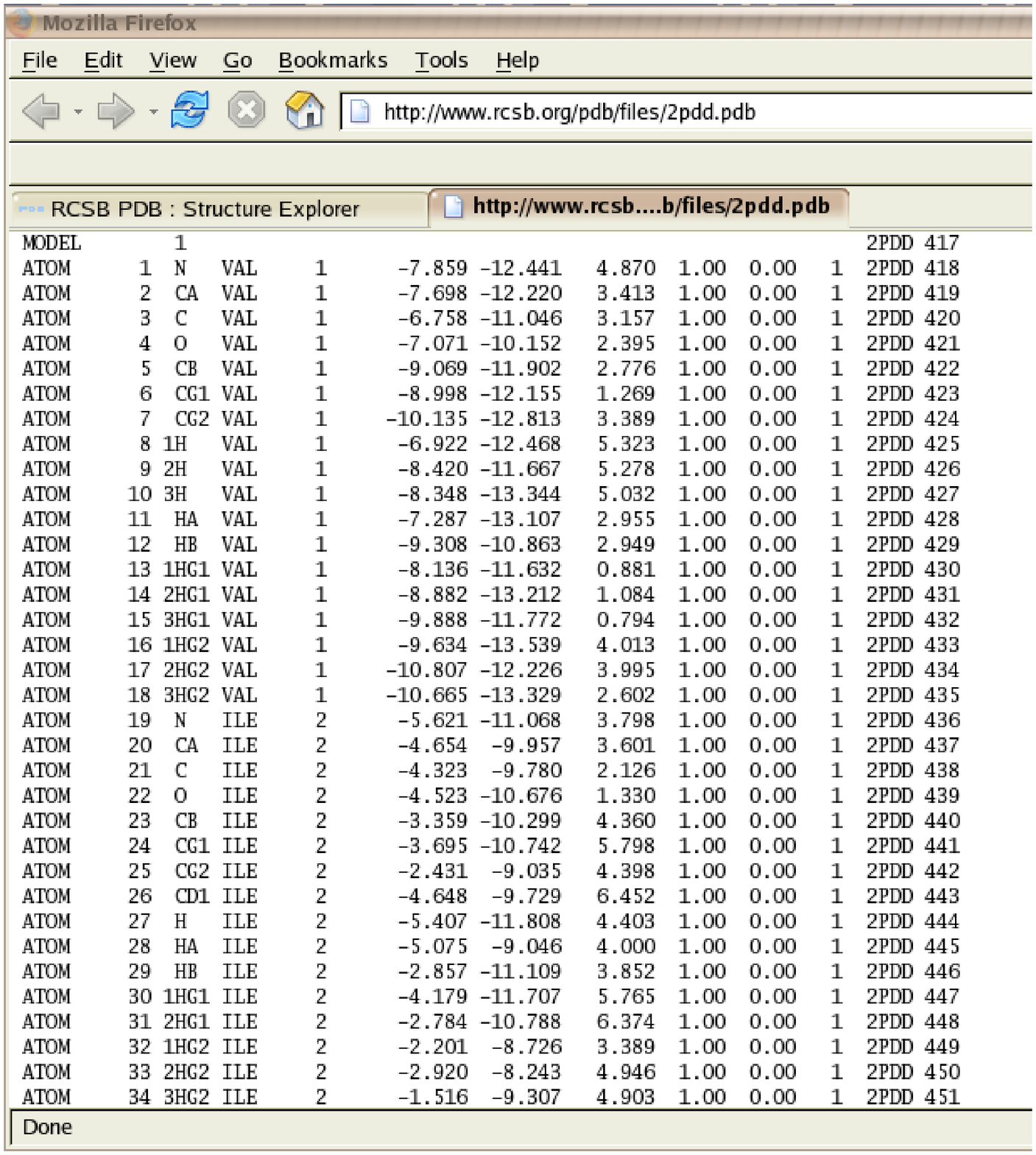}
\end{center}
\caption[PDB file: Screenshot of a PDB file.]
{The Protein Data Bank (PDB) file containing the atomic coordinates of
  2PDD (Acyltransferase).} 
\label{fig:PDB_file}
\end{figure}

\subsubsection*{Computational Procedure}
\begin{figure}[!tbh]
\begin{center}
\includegraphics[width=12cm]{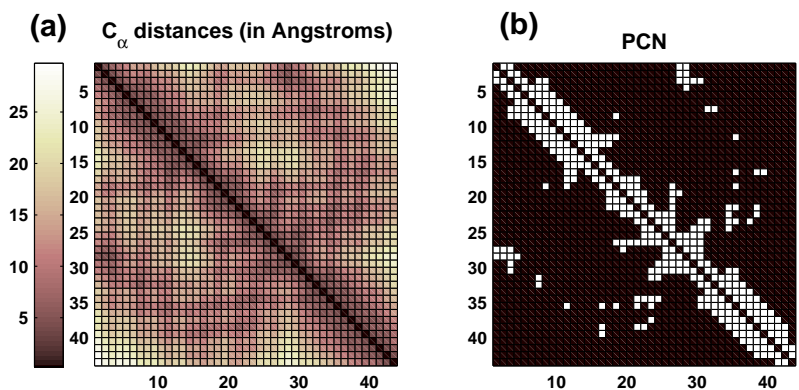}
\end{center}
\caption[From `Pair-wise distance matrix' to the Contact Map]
{`Pair-wise distance matrix' of all $C_{\alpha}$ atoms from
  Fig.~\ref{fig:PDB_file} and Contact Map after thresholding with a
  cut-off of $8$\AA\/.} 
\label{fig:coarse-graining01}
\end{figure}

The information required for building models of protein structures was 
extracted from its PDB (Protein Data  Bank; http://www.rcsb.org/pdb/) file. 
The PDB file contains a large amount of structural details obtained
from X-ray diffraction or NMR method. 
We explain the methodology with the example of the protein
Acyltransferase (2PDD) as shown in Fig.~\ref{fig:pcn}.
Figure~\ref{fig:PDB_file} shows the `Model' section of the PDB file in
which, apart from other details,  atom number, atom label, amino acid
type, amino acid number, and coordinates are shown.  
The amino acids are labelled in increasing order
from N-terminal to C-terminal residue, starting from $1$ upto $n_r$
($1$ to $43$ for 2PDD), the total number of residues.
First, we extracted three-dimensional coordinates of the
$C_{\alpha}$ atoms (CA in Fig.~\ref{fig:PDB_file}), the structural
representatives, of amino acids in the network models. 
Next, we calculated the Cartesian distances
between all pairs of $C_{\alpha}$ atoms of the residues. 
Using a threshold $R_c$, (as described earlier) we computed the
`Contact Map~($\mathbf{M}$)'.  
Fig.~\ref{fig:coarse-graining01}~(a) shows the pairwise
distance (in $\AA$) matrix for Acyltransferase (2PDD) and (b) the
corresponding Contact Map with distance threshold $R_c=8\AA$. 

\begin{figure*}[!tbh]
\begin{center}
\includegraphics[width=14cm]{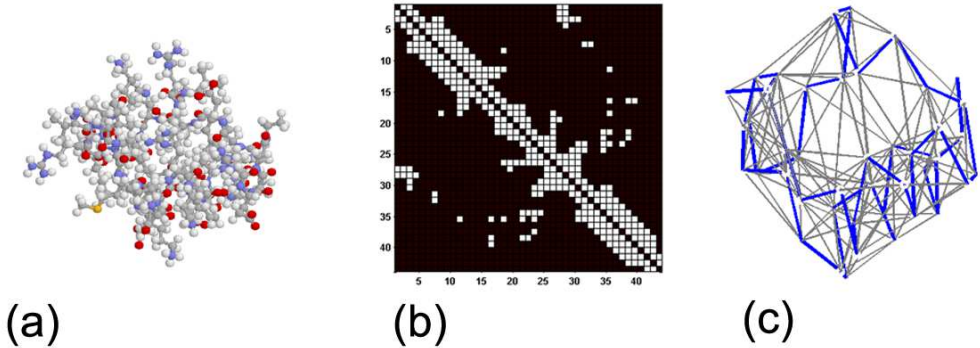} 
\end{center}
\caption[Coarse-graining of the protein structure data. (a)~Ball and
stick model of 2PDD, (b)~its Contact Map, and (c)~the PCN.] 
{Coarse-graining of the protein structure data. (a)~Ball and
  stick model of 2PDD, (b)~its Contact Map, and (c)~the PCN with $R_c=8$\AA\, 
where the backbone contact are shown with a blue line and the rest of
the non-covalent contacts are shown in gray.} 
\label{chap01:coarse-graining02}
\end{figure*}

The Contact Map then serves as the adjacency matrix for drawing the
nodes and links of the contact network.
Coarse-graining is inherent in the process of construction of PCN. 
Figure~\ref{chap01:coarse-graining02} summarises the process of
coarse-graining involved in the making of PCN.
PCN, is created by ignoring a large amount of positional information
of atoms in the X-Ray data. Starting from atomic-level details
 (Fig.~\ref{chap01:coarse-graining02}(a)), we 
jettison a large amount of structural details to go through
residue-level details to finally arrive at the two-dimensional Contact
Map (Fig.~\ref{chap01:coarse-graining02}(b)). 
The protein contact network (PCN) can  be reconstructed given the
coordinates of ($C_{\alpha}$ atoms of) the residues in the structure
to which the Contact Map corresponds (Fig.~\ref{chap01:coarse-graining02}(c)).

\subsection{Long-range Interaction Network (LIN)}
\label{chap01:subsec:LIN}
\begin{figure}[!tbh]
\begin{center}
\begin{tabular}{cc}
\includegraphics[width=7cm]{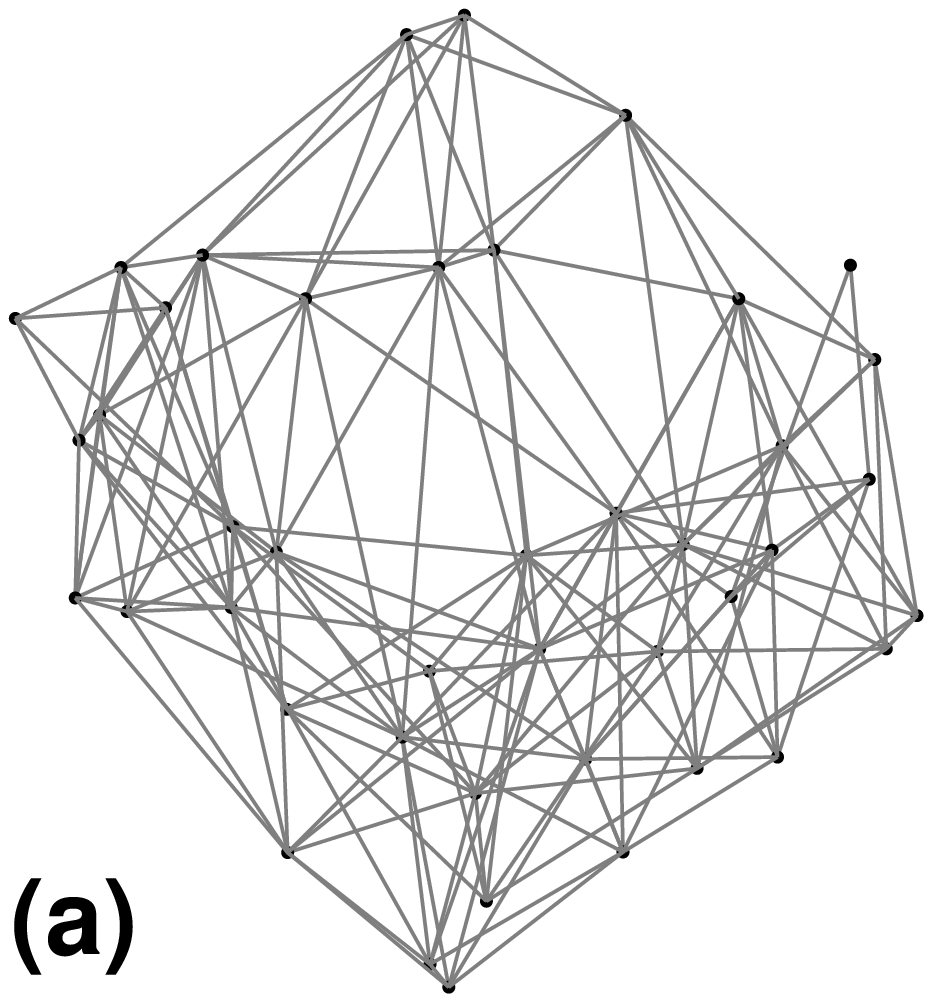} &
\includegraphics[width=7cm]{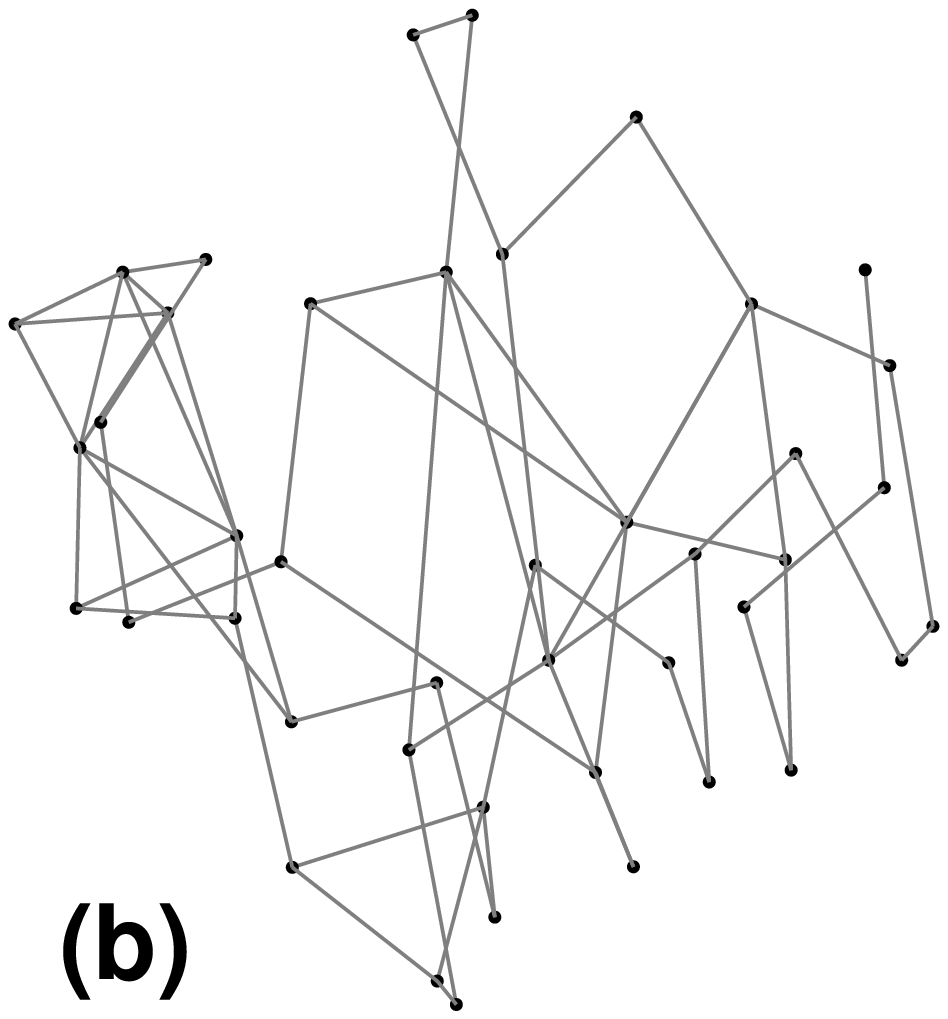} 
\end{tabular}
\end{center}
\caption[(a) PCN and (b) its LIN.]
{(a) PCN and (b) its LIN.} 
\label{fig:lin}
\end{figure}

The Long-range Interaction Network (LIN) of a PCN was obtained by
considering, other than the backbone links, only those `contacts'
which occur between amino acids that are `distant' from each other. 
i.e. residue pairs that are, along the backbone,
separated by a threshold, termed $LRI_{Threshold}$ of $12$ amino
acids~\cite{protnet:JMB} or more amino acids. Here, $LRI_{Threshold}$
stands for the Long-range 
Interaction threshold, measured in terms of the number of residues
along the backbone, that is used to decide the range upto which the
`long-range effects' are taking place. 
Thus formed, a LIN is a subset of its PCN with same number of nodes
($n_r$) but fewer number of links (contacts) due to removal of
short-range contacts. Fig.~\ref{fig:lin} shows the PCN and its LIN of
2PDD. 

This network is of special significance in the context of a linear
chain (1-D network) model that has additional long-range links  
happening between nodes that are separated along the chain. A protein is
one such network system in which there is an inherent 1-D 
structure in terms of the polypeptide backbone held together by
covalent peptide bonds. The polypeptide chain folds onto itself  
by virtue of the chemical forces acting among the constituent
residues, thereby creating `contacts' on various scales as specified  
by the separation distance between the contacting residues.

\subsection{Random Controls of PCN}
\label{chap01:subsec:Random_Controls}
We created two types of random networks as controls for the PCNs. 
The polypeptide backbone connectivity was kept intact in both the
random controls, while randomising the noncovalent contacts.  
For every protein, $100$ instances of each type of random control were
generated. An average of all the instances were used as a representative
of the parameters and properties that were compared with PCNs and
their LINs. 


\subsubsection*{Type I Random Control}
\begin{figure}[!tbh]
\begin{center}
\begin{tabular}{ccc}
\includegraphics[width=4.3cm]{pcn.eps} &
\includegraphics[width=4.3cm]{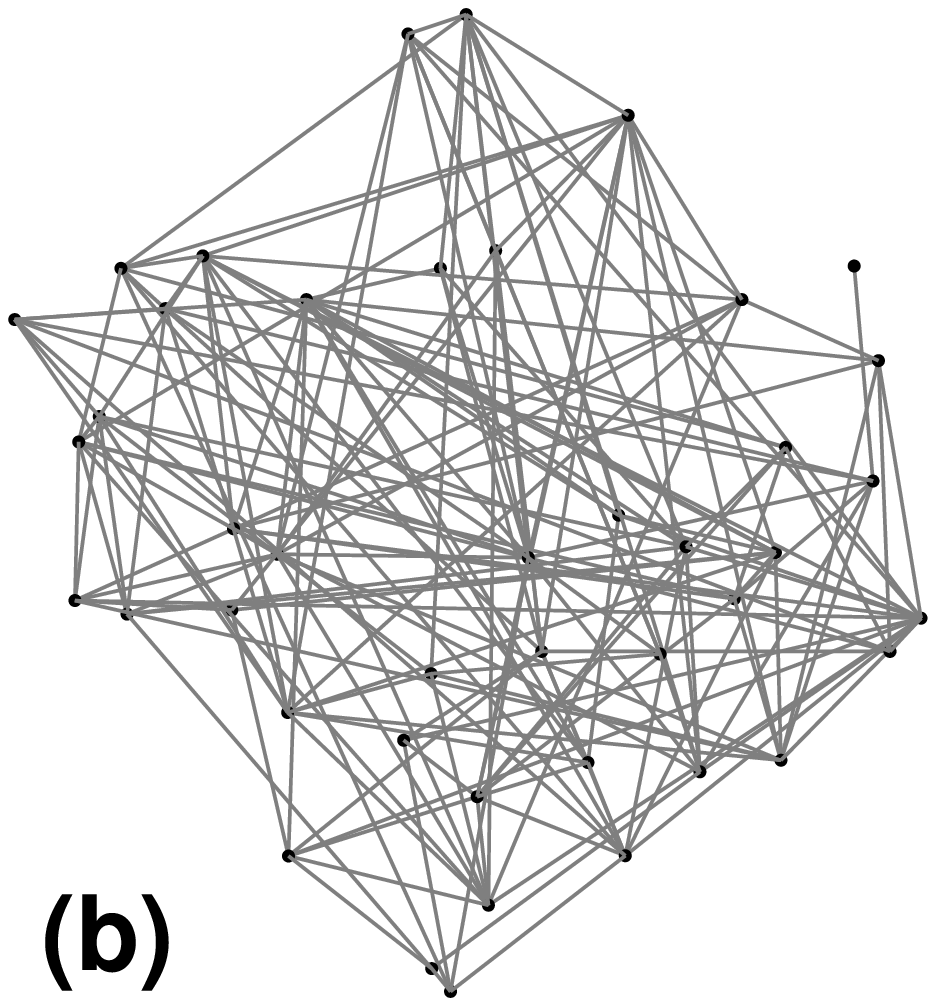} &
\includegraphics[width=4.3cm]{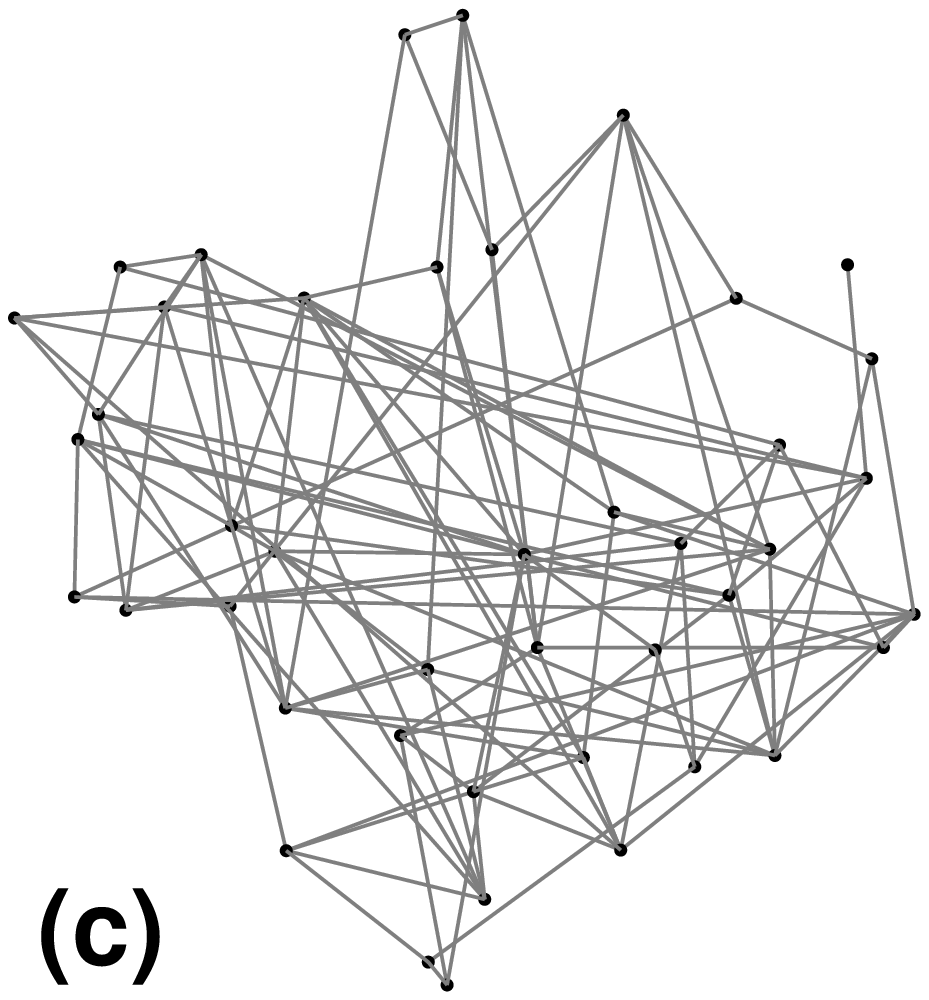} 
\end{tabular}
\end{center}
\caption[(a) PCN, (b) Type-I Random Control and (c) its LIN.]
{(a) PCN, (b) Type-I Random Control and (c) its LIN.} 
\label{fig:pcn_rand01_lin}
\end{figure}

The Type I random control has the same number of residues ($n_r$) as
well as number of contacts ($n_c$) as those of PCN, except that the
contacts are created randomly by avoiding self-contacts or duplicate
contacts. The connectivity distribution of the Type I random controls,
in general, is not the same as that of PCNs. The algorithmic steps
used for creating the Type I random controls were as follows. 
We started the network with $n_r$ number of nodes and $n_r-1$ covalent
contacts representing the backbone. The covalent contacts were put in
place by sequentially connecting residues from $1$ to $2$ to $3$, and
so on till $n_r$. Further, we added all the noncovalent contacts in a  
random manner. First we chose two unique residues using a uniform
pseudo-random number generator. A noncovalent contact was created
between these residues provided they were not part of the
backbone-forming contacts and if they were not already connected. This
process was repeated till the total number of contacts in the random
control is same as those in the PCN. The `LINs of Type I random
controls' were obtained in the same fashion by which the LINs were
obtained from PCNs. Figure~\ref{fig:pcn_rand01_lin} shows a typical
Type-I random control (b) of 2PDD's PCN (a) and its LIN (c). 

\subsubsection*{Type II Random Control}
\begin{figure}[!tbh]
\begin{center}
\begin{tabular}{ccc}
\includegraphics[width=4.3cm]{pcn.eps} &
\includegraphics[width=4.3cm]{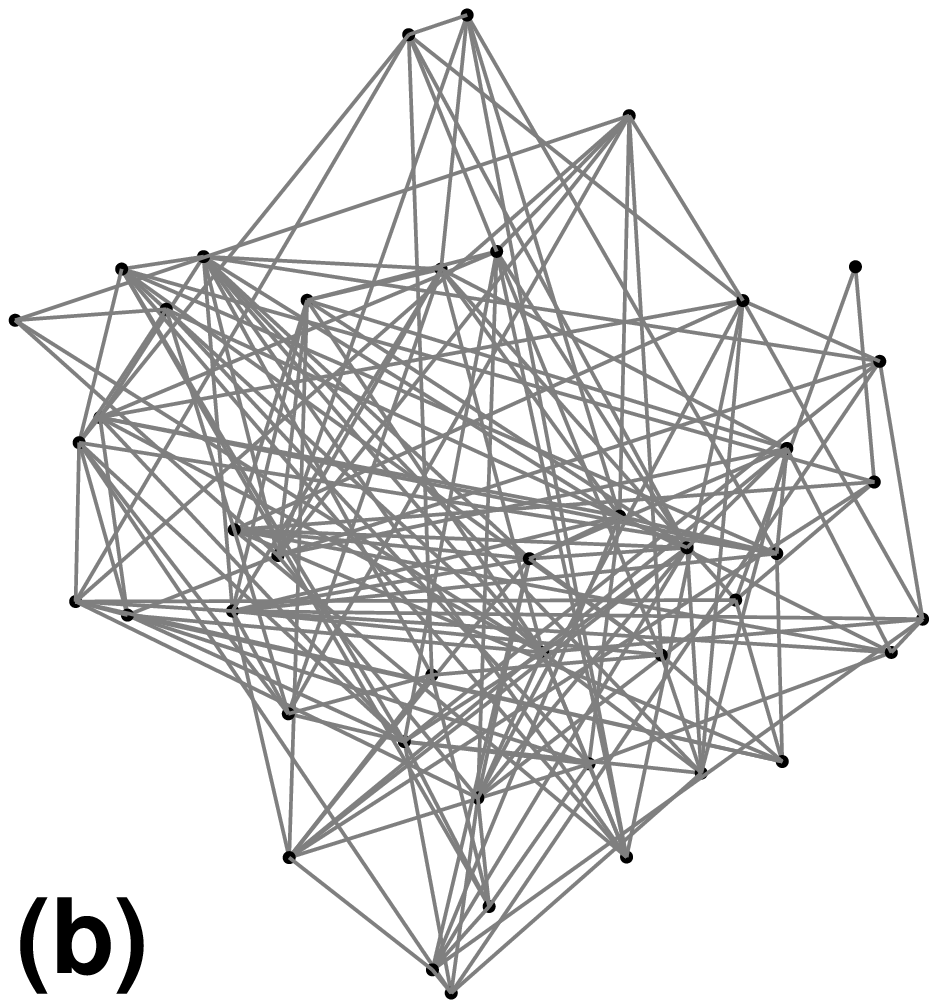} &
\includegraphics[width=4.3cm]{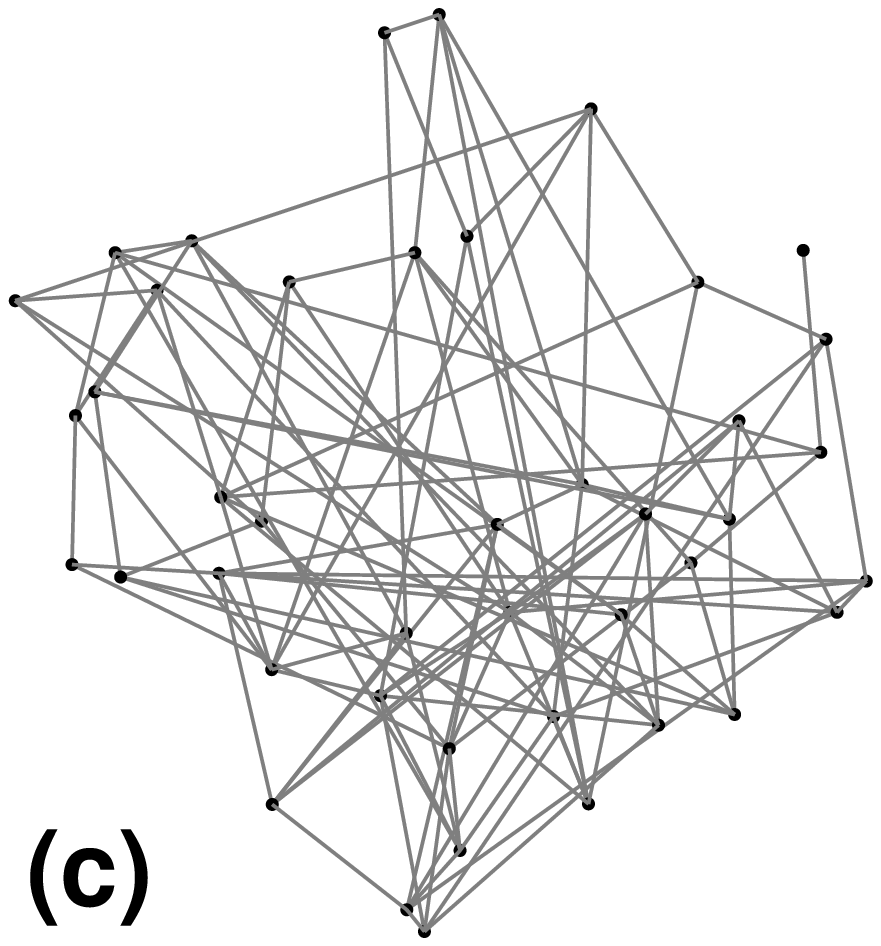} 
\end{tabular}
\end{center}
\caption[(a) PCN, (b) Type-II Random Control and (c) its LIN.]
{(a) PCN, (b) Type-II Random Control and (c) its LIN.} 
\label{fig:pcn_rand02_lin}
\end{figure}

In Type II random controls, apart from maintaining the number of nodes 
($n_r$) and contacts ($n_c$), the connectivity distribution as well as
individual connectivity of PCNs was also conserved. We started with
the original PCN and then the non-covalent contacts were randomised
while maintaining the degree of 
individual nodes. To ensure adequate randomisation of the
connectivity, the pattern of pair-connectivity was randomised $2000$
times. In Type II random controls, degree distributions of only the
PCNs were conserved. For the LINs obtained from these controls of
PCNs, the degree distributions were \emph{not} 
explicitly conserved in the randomisation procedure. 
Figure~\ref{fig:pcn_rand02_lin} shows PCN (a), it's typical Type-II
random control (b), and its LIN (c).

\section{Network Visualisation}
\label{chap01:sec:nw_visualisation}
There are various ways a graph (network) can be
visualised. Depending on the purpose of the visualisation, one may
want to choose appropriate visualisation method. Network systems could
be classified based on the way the nodes are, if at all, positionally 
related to each other.
A network with (a) no positional relationship among its nodes, (b) a
linear relationship--a 1-D chain, (c) a 2-D order--and, finally (d)
each node characterised by  positional coordinates in a 3-D space.   
 
When studying a system in which there is no structural order of the
nodes, a 2-D or 3-D visualisation with arbitrary node positions
optimised for minimal crossings of the links is one of the suitable
choices. For a 
system in which a linear order is specified, a chain-like or ring-like
representation would capture the necessary details. A system with 2-D
(3-D) order could be represented in the 2-D (3-D) space with
appropriate positions of nodes and, if necessary, with optimised
edge-crossings.  

\subsection*{Contact Map Visualisation}
Contact Map has been defined and used earlier. Here we mention the
visualisation aspects and its relationship to the proteins that they
model.
The principle diagonal of the contact map corresponds to the  
self-contacts which by definition are zero: $\mathbf{M}_{ii}=0.$ The
positions parallel and next to the diagonal correspond to  
contact separation of ($|j-i|=$)~$1,$ which is equivalent to the
polypeptide chain that is held together with the covalent  
peptide bonds. The elements diagonally parallel and next to backbone
represent contacts happening between residues which are one  
residue apart ($|j-i|=2$) along the backbone. The procedure continues 
so on and so forth till one reaches the single contact possible with
$|j-i|=n_r-1$ which, when existent in a protein, indicates a contact
between the N-terminal and C-terminal residue of the protein. This
understanding could be used for creating appropriate models with
desired types of chemical connectivities. 

\subsection*{Chain Representation}
\begin{figure*}
\begin{center}
\includegraphics[scale=1.05]{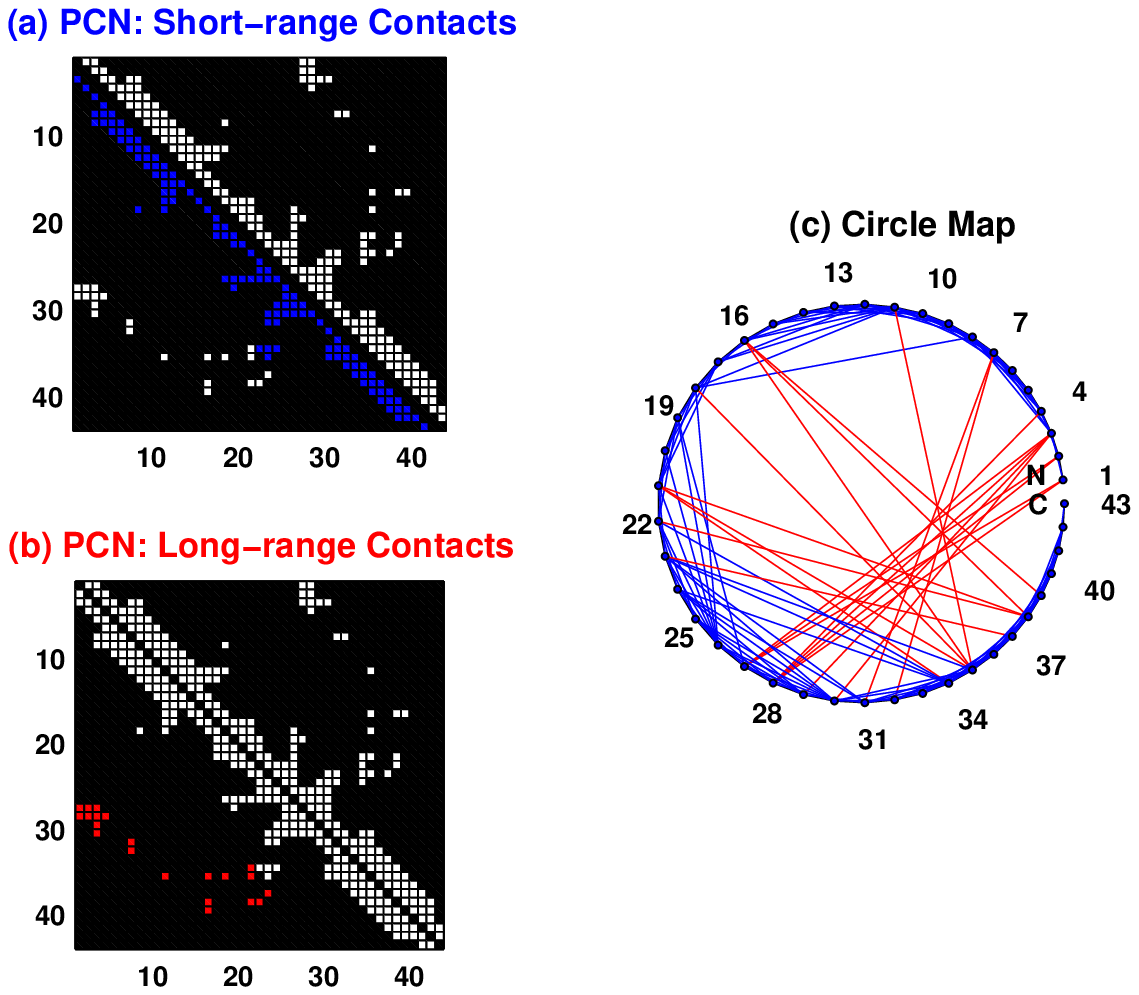} 
\end{center}
\caption[Drawing `Chain Contact Map' and colour code for short- and
long-rage interactions.]  
{Drawing `Chain Contact Map'.
Contact Map of PCN with (a) short-range contacts, and (b) long-range
contacts highlighted.(c) `Circle' or `Chain Contact Map'}
\label{chap01:chain_map}
\end{figure*}

The information content of the protein's contact map can be
transformed into a representation that offers similar insight about
the range at which contacts are taking place in the protein
backbone. It also gives a hint about the locations where the secondary
structures are taking place. This `Chain Representation' of the
protein structure, though similar to that of the Contact Map, is
sometimes more useful as it represents the protein structure in a less
abstract and easily accessible fashion. 
Fig.~\ref{chap01:chain_map} depicts, for Acyltransferase (2PDD), the
parts that Chain Representation is composed of.
 
Figure~\ref{chap01:chain_map} shows contact map of the network with
(a) short-range contacts ($|j-i|<12$) in `blue'. 
Fig.~\ref{chap01:chain_map}(b) shows long-range contacts which   
($|j-i| \ge 12$) are shown in `yellow'. 
Finally, Fig.~\ref{chap01:chain_map}(c) shows the `Chain Contact Map'
that is a combination of the above two.
The polypeptide backbone of Acyltransferase~(2PDD) is aligned as a
chain of residues along a circular curve. The residues are labelled in
an increasing order (anti-clockwise) from N-terminal to C-terminal.  
The backbone contacts, which trace the circle, are shown in black. 
The short-range contacts ($|j-i|<LRI_{threshold}$) are shown in blue,
and those with long-range are shown in red.

\subsection*{3-D Representation}
As mentioned earlier, the positional information of the residues in
the protein's 3-D structure is lost in the contact map as well as in
chain representation. Owing to the relevance of the positional
information, PCNs can be better visualised  in 3-D space. This is
achieved by superimposing `positional information' with that in the
`contact map', as shown in  Fig.~\ref{chap01:coarse-graining02}~(c).

\section{Network Parameters and Properties}
\label{chap01:sec:network_parameters}
Various features of network's topology and dynamics could be measured
by defining parameters that capture appropriate aspects of it. 
Below, we describe properties that are typically used to characterise
a network. Since a network could be a directed/undirected and  
weighted/unweighted, the parameters need to be appropriately
defined. The following definitions are valid for any undirected and
unweighted network.  

Here, we explain the implication of each of these parameters in the
context of the network model that we build for the protein structures.   

\subsection{Distance Measures}
\label{chap01:subsec:distance}
Here, Distance is measured in terms of the number of edges that are
needed to be traversed, to reach to one node from the other  
node. Many distance measures could then be defined which measure
different aspects of the protein structure.  

\subsubsection{Characteristic Path Length}
Shortest path length ($L_{ij}$), between any two pairs of nodes $i$ \&
$j$, is defined as the number of links that must be  
traversed, by the shortest route, from one node to another. 
The average of shortest path lengths, known as `characteristic path
length' ($L$), is an indicator of compactness of the network,  
and is defined as~\cite{watts:nature},
\begin{equation}
L = \frac{2 ~ \sum_{i=1}^{n_r-1}\sum_{j=i+1}^{n_r}L_{ij}}{n_r(n_r-1)},
\end{equation}
where $n_r$ is the number of residues in the network. 

\begin{figure*}[!tbh]
\begin{center}
\begin{tabular}{c}
\textbf{(a)}\includegraphics[width=12cm]{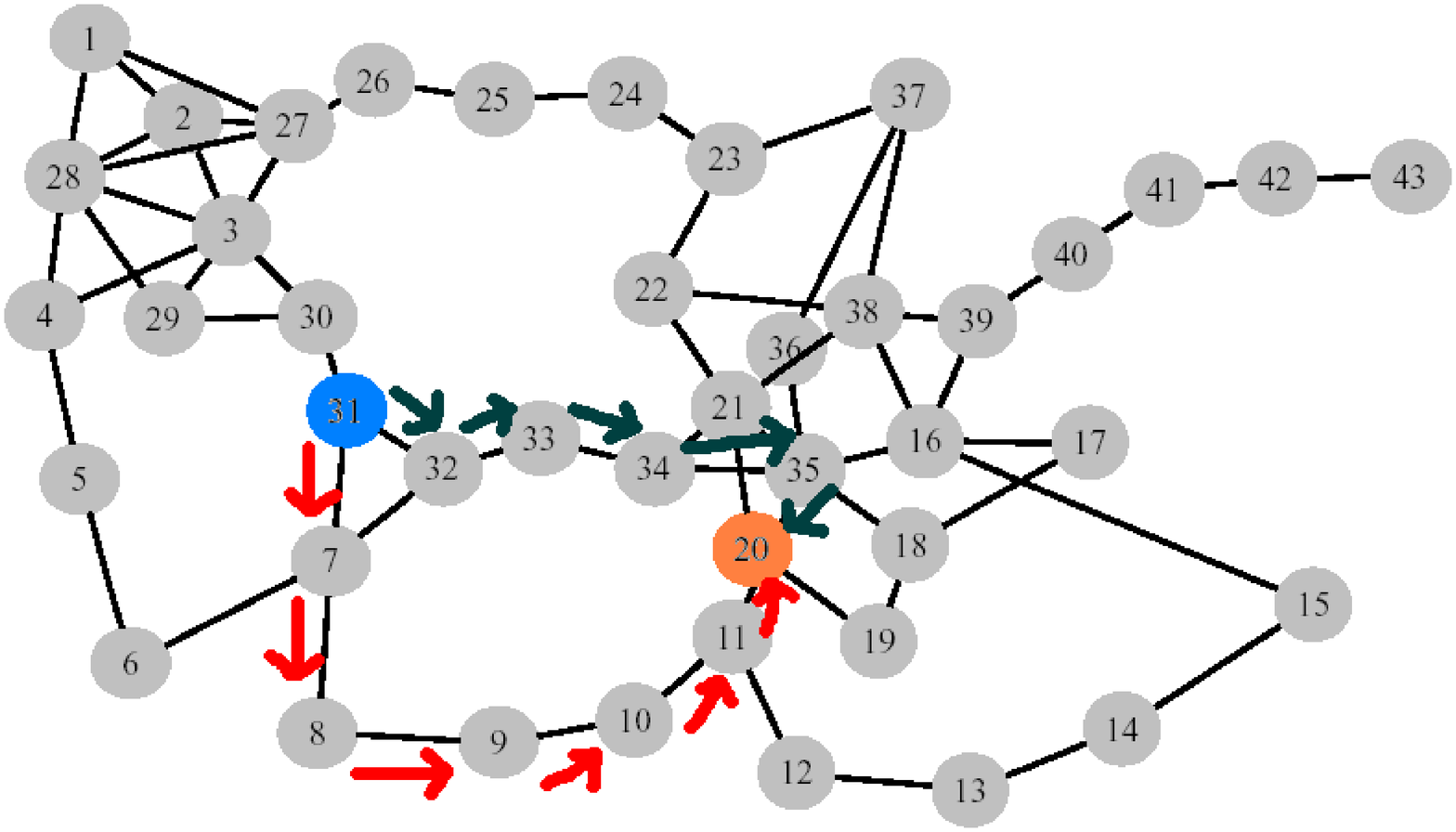} \\
\textbf{(b)}\includegraphics[width=10cm]{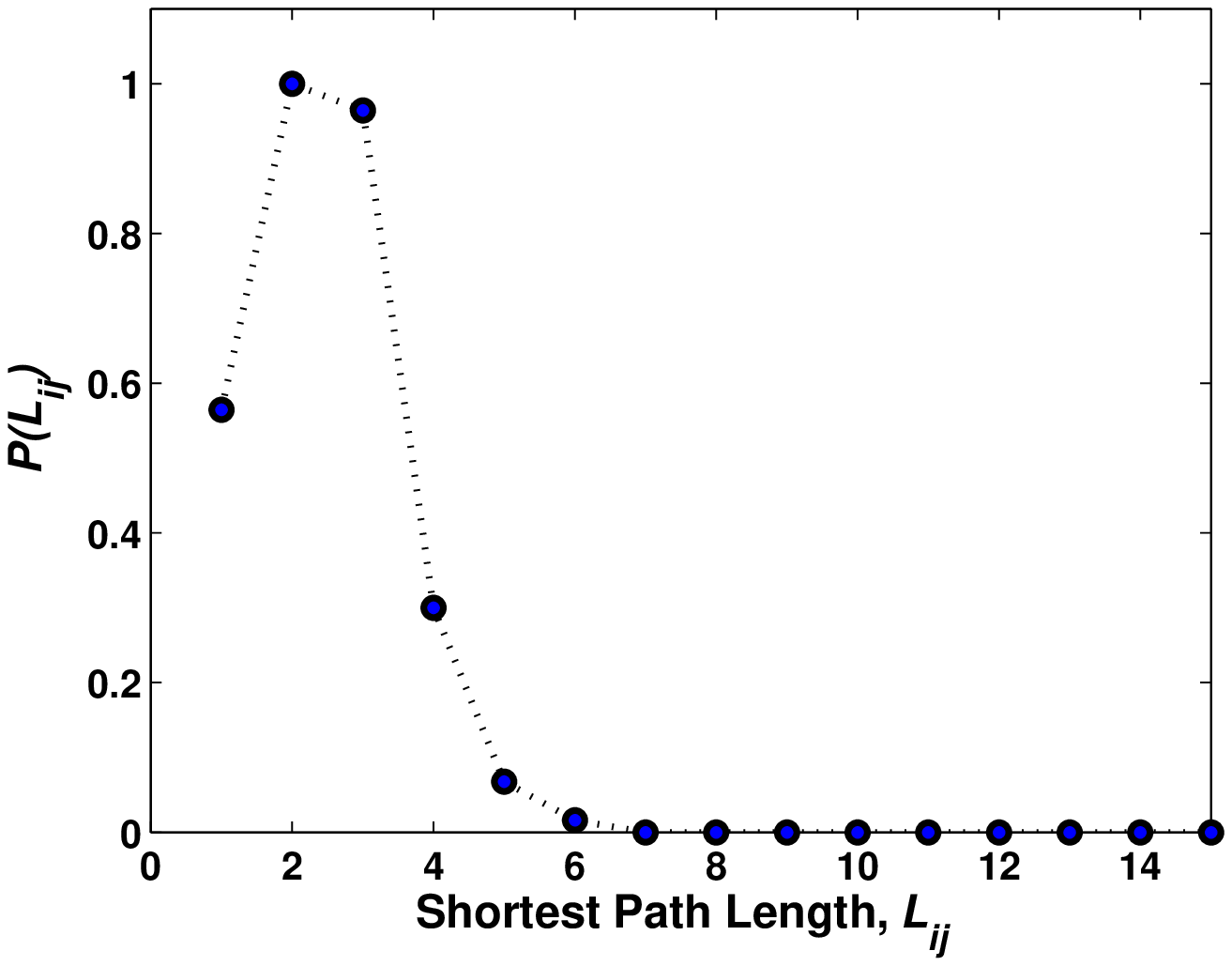}
\end{tabular}
\end{center}
\caption[Topological Properties: Definition of Characteristic Path Length ($L_{ij}$).]
{(a) Illustration of the Characteristic Path Length~($L_{ij}$), (b)
  Shortest Paths Distribution for 2PDD.}
\label{fig:chap01:L}
\end{figure*}

This definition is illustrated in Fig.~\ref{fig:chap01:L}~(a). 
The figure shows two of all the possible `paths' between nodes $31$
and  $20$ which are two of the nearest to `the shortest path'. Shown
with blue-coloured arrows is the path
$31$~$\rightarrow$ $32$~$\rightarrow$ $33$~$\rightarrow$
$34$~$\rightarrow$ $35$~$\rightarrow$~$20$,  
with path-length of $5$. Whereas the path with red-coloured arrows, 
$31$~$\rightarrow$ $7$~$\rightarrow$ $8$~$\rightarrow$
$9$~$\rightarrow$ $10$~$\rightarrow$ $11$~$\rightarrow$~$20$, 
has path-length of $6$. Hence the shortest path length between node
$31$ and $20$ is, $L_{31,20}=5$. 

Fig.~\ref{fig:chap01:L}~(b) shows the shortest paths
distribution for the example protein network, 2PDD. Analytically, the
$L$ is defined for a network with number of nodes $n_r$ and average
degree $\langle k \rangle$ as, 
$$ L=\frac{n_r(n_r+ \langle k \rangle  -2)}{2\langle k \rangle(n_r-1)} $$

\begin{figure*}[!tbh]
\begin{center}
\begin{tabular}{c}
\textbf{(a)}\includegraphics[width=12cm]{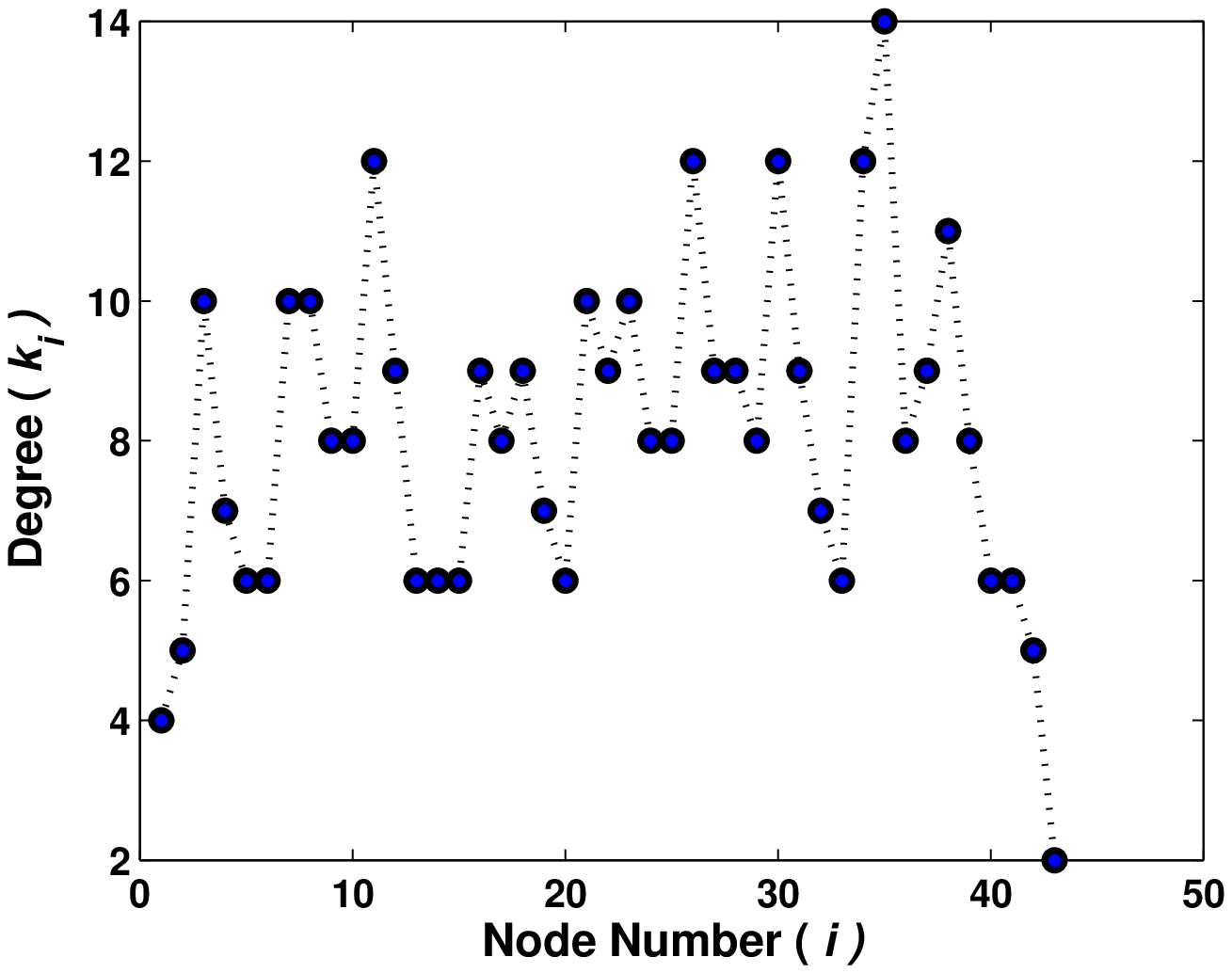} \\
\textbf{(b)}\includegraphics[width=12cm]{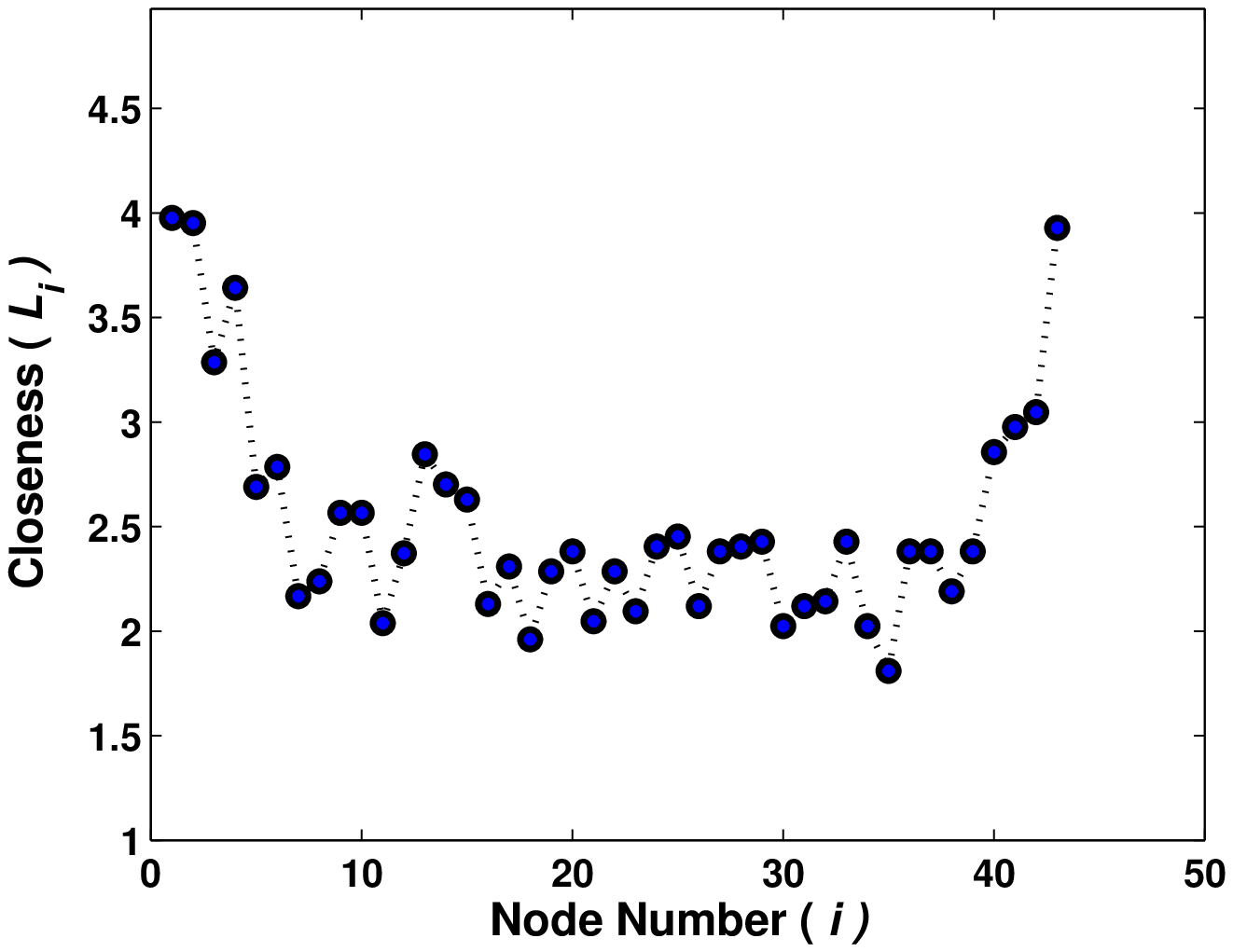}
\end{tabular}
\end{center}
\caption[Topological Properties: Closeness~($l_i$) and Degree~($k{i})$)] 
{(a) Degree~($k{i}$) and  (b) Closeness~($l_i$).}
\label{fig:chap01:closeness_ki}
\end{figure*}

\subsubsection{Diameter}
Another measure for compactness of the network is Diameter ($D$), 
which is defined as the largest of all the shortest paths in the network.
\begin{equation}
D = \max{L_{ij}}, \forall~ \text{\emph{i--j} pairs of shortest paths.}
\end{equation}

\subsection{Centrality Measures}
\label{chap01:subsec:centrality}
Networks representation of a complex system, by definition, embeds the
complex interactions happening among the various elements of  
the system. Despite the distributed nature of the elements, some of
them could potentially hold a `central' position in the network, 
thereby being crucial to the topology and/or the dynamics. Centrality
refers to the structural attribute of nodes in the network and  
not to the attribute of the node itself.

\subsubsection{Degree and Average Degree}
Degree~($k_i$) of a node $i$ is the total number of neighbours (linked
nodes) it has. 

\begin{equation}
k_i = \sum_{j=1}^{n_r} A_{ij}.
\end{equation}

Thus defined, degree captures the centrality of the node in terms its
connectivity. The more is the degree, the better connected it is. 
In PCN, degree ($k_i$) measures the number of other amino acids that
amino acid $i$ is spatially proximal to (with given $R_c$) in the
native state  protein
structure. Figure~\ref{fig:chap01:closeness_ki}(a) shows the degrees
of individual nodes of 2PDD.

It may be noted that as compared to
other biological as well as technological networks, the process of
formation and the constraints which shape the network structure are
very different for the PCNs. Owing to the covalent backbone
connections and steric and space constraints, the typical degree in
PCN is much lower than that found in other networks. 
  
Average degree, $\langle k \rangle$, of a network with $n_r$ nodes is
defined as  
\begin{equation}
\langle k \rangle=\frac{1}{n_r}{\sum_{i=1}^{n_r} k_{i}}.
\end{equation}

\subsubsection{Closeness}
Closeness is defined based on the measurements of shortest path length
between pairs of amino acids.  
It is a measure that computes the average connectivity of \emph{a
  residue} with the rest of the network.  
It integrates the effect of the entire protein, measured in terms of
its shortest distance from every other node, on a single residue.  
It is defined as,

\begin{equation}
L_i = \frac{\sum_{j=1}^{n_r}L_{ij}}{(n_r-1)}.
\end{equation}

Figure~\ref{fig:chap01:closeness_ki}(b) shows the closeness values of
individual residues of 2PDD.
Any property of an amino acid that is dependent on the average
connectivity of amino acids could potentially be related to
closeness. 
It is known that the kinetics and stability of a protein is often
dependent on the chemical properties of one or a few amino acids.  
Given this fact the property of closeness acquires a special meaning and
could possibly be used to explore functional relevance of  
individual amino acids. 

\subsection{`Pattern of Connectivity' Measures}
\label{chap01:subsec:connectivity_pattern}
The parameters defined so far characterise individual nodes or pairs
of nodes, measuring their distances or centrality in the network. 
On a level above this, the network is put in place by pattern of
connectivities among nodes. This pattern could be characterised in  
following ways. 

\subsubsection{Degree Distribution}
\begin{figure*}[!tbh]
\begin{center}
\begin{tabular}{cc}
\textbf{(a)}\includegraphics[width=6.5cm]{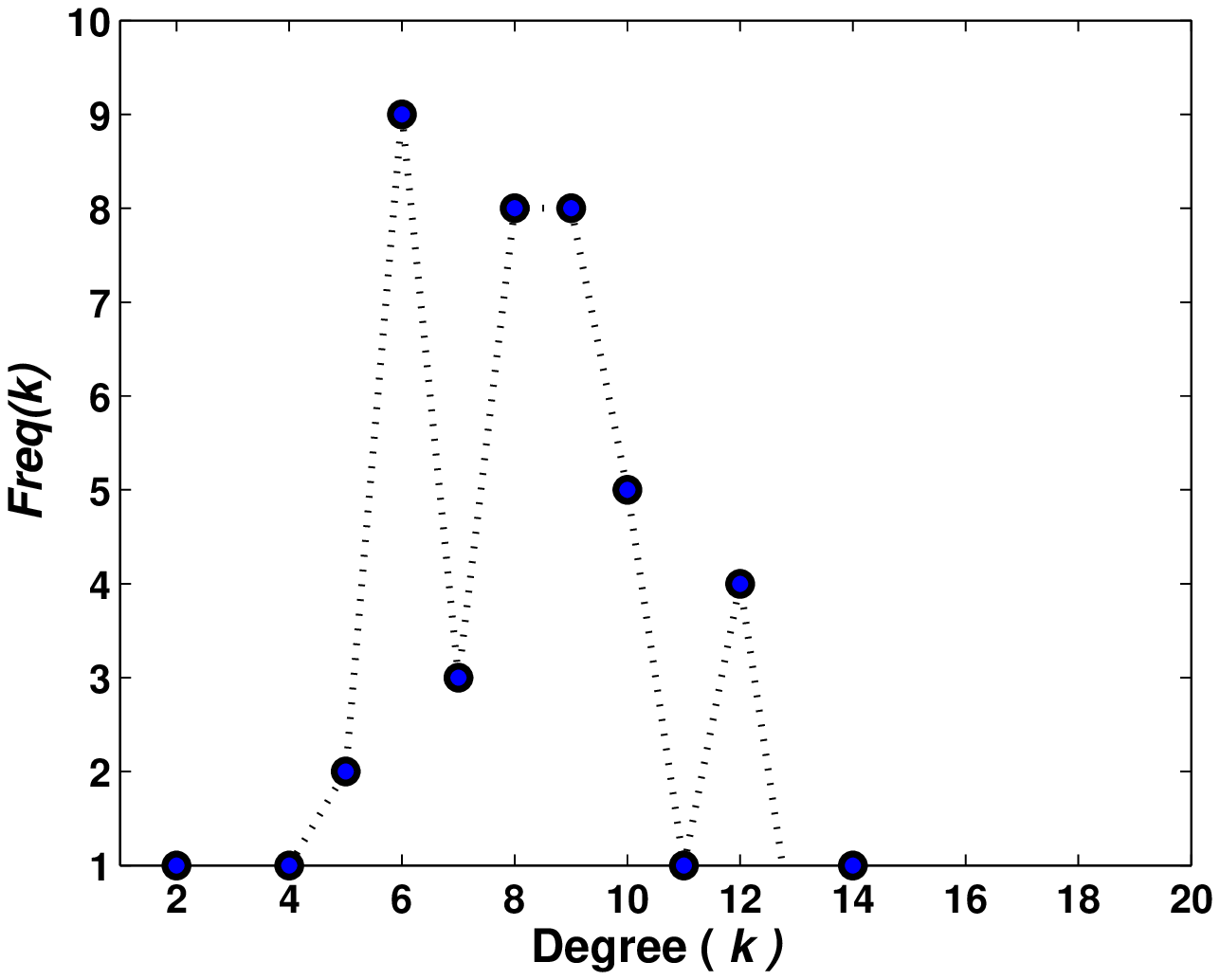} &
\textbf{(b)}\includegraphics[width=6cm]{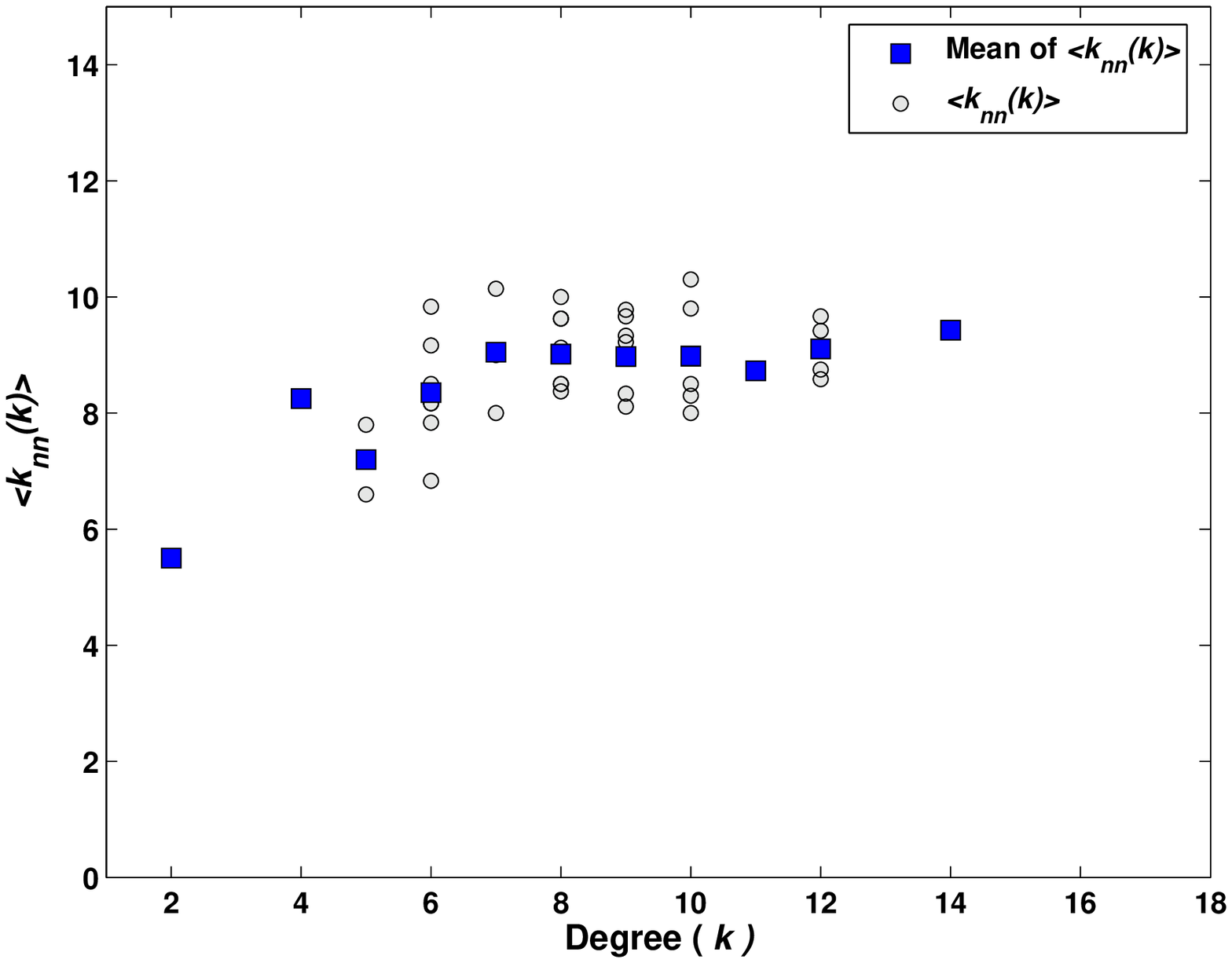} 
\end{tabular}
\end{center}
\caption[Topological Properties: Degree distribution~($P(k)$) and
Degree Correlations~($\langle k_{nn}(k) \rangle$)]
{Topological Properties: (a) Degree Distribution ($P(k)$) and (b) Degree
  Correlations~($\langle k_{nn}(k) \rangle$)} 
\label{chap01:PCNprop02}
\end{figure*}

Degree symbolises the importance of a node from the perspective of
mere connectivity---the larger the degree, the more important it is. 
The distribution of degrees in a network is an important feature which
characterises the topology of the network.  
It could possibly reflect on the processes by which the network has
evolved to attain the present topology.  
The networks in which the links between any two nodes are assigned
randomly have a Poisson degree distribution~\cite{bollobas1981}  
with most of the nodes having similar degree.
Fig.~\ref{chap01:PCNprop02}(a) shows the degree 
distribution pattern of 2PDD.

\emph{Normalised Degree Distribution} is the degree distribution
normalised with the $Freq(max)$, the maximum frequency of the
distribution. Henceforth $P(k)$ would denote the normalised degree
distribution. $P(k)$ allows one to compare networks with disparate
degree distribution profiles. 

\emph{Remaining degree} is simply one less than the total degree
of a node~\cite{r:newman}. 
If $p_k$ is the distribution of the degrees, then the normalised
distribution, $q_k$, of the \emph{remaining degree} is 
$$q_k=\frac{(k+1)p_{k+1}}{\sum_{j}jp_j}.$$

\subsubsection{Coefficient of Assortativity}
A network is said to show assortative mixing, or simply `assortative',
if the high-degree nodes in the network  
tend to be connected with other high-degree nodes. On the other hand,
the network is said to be `disassortative' if the high-degree  
nodes tend to be connected with other low-degree nodes. 
The Coefficient of Assortativity ($r$) measures the tendency of degree
correlation.  
It is the Pearson correlation coefficient of the degrees at either end
of a link and is defined~\cite{r:newman} as, 
\begin{equation}
 r=\frac{1}{\sigma_{q}^{2}} \sum_{jk}jk(e_{jk}-q_j q_k),
\end{equation}
where $r$ is the coefficient of assortativity, $j$ and $k$ are the
degrees of nodes, $q_j$ and $q_k$ are the \emph{remaining degree}  
distributions, $e_{jk}$ is the joint probability distribution of the
remaining degrees of the two nodes at either end of a randomly  
chosen link, and $\sigma_{q}^2$ is the variance of the distribution
$q_k$. 
$r$ is a normalised degree correlation function, a global quantitative
measure of degree correlations in a network, and takes values as $-1
\leq r \leq 1$. The value of $r$ is zero for no specific trend in
degree correlations, positive or negative for assortative or  
disassortative mixing, respectively.

\subsubsection{Degree Correlations}
Another way to assess the degree correlation pattern in a network is
to visualise it by measuring the average degree of nearest
neighbours, $k_{nn}(k)$, for nodes of degree $k$. In presence of
correlation, $k_{nn}(k)$ increases with increasing $k$ for  
an `assortative network' whereas it decreases with $k$ for a
`disassortative network'. Fig.~\ref{chap01:PCNprop02}(b) shows the degree 
correlations pattern of 2PDD.

\subsection{Compactness Measures}
\label{chap01:subsec:compactness_measures}

\subsubsection{Clustering Coefficient}
Clustering coefficient of a node $i$, $C_i$,  
is defined~\cite{watts:nature} as $C_i~=~2*n/k_i(k_i-1)$, where $n$
denotes the number of contacts amongst the $k_i$ neighbours  
of node $i$. 
$C_i$ of a node is equal to $1$ for a node whose neighbours are fully
interlinked, and zero if none of the neighbouring nodes do not  
share any contacts.  Average clustering coefficient of the network
($C$) is defined as the average of $C_i$s of all the nodes in the  
network and will be referred to as `clustering coefficient' unless
specified otherwise. Clustering coefficient is the measure of
\emph{cliquishness} of the network.

Numerically the clustering coefficient is computed as follows using
the contact map. 
\begin{equation} 
C_i = \frac{\frac{1}{2}\sum_{j=1}^{n_r} \sum_{k=1}^{n_r}
  \mathbf{M}_{ij}\mathbf{M}_{ik}\mathbf{M}_{kj} }{^{k_i}C_{2}}, 
\label{eq:fig01:}
\end{equation}
where, $\mathbf{M}$ is the symmetric, binary, adjacency matrix
representation of the network. 

Analytically, the $C$ for a network of average degree $\langle
k \rangle$ is given by,  
$$ C=\frac{3}{4}\frac{(\langle k \rangle  -2)}{(\langle k \rangle -1)} $$

\begin{figure*}[!tbh]
\begin{center}
\includegraphics[width=11cm]{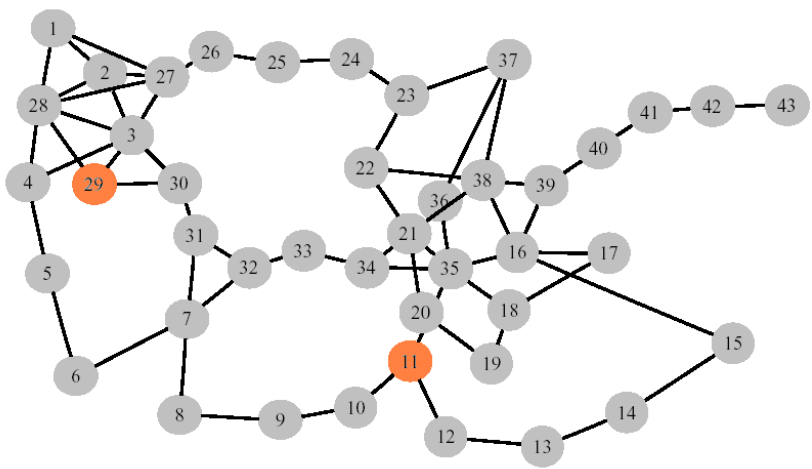} 
\end{center}
\caption[Topological Properties: Definition of Clustering Coefficient
of a node~($C_{i}$).]
{Illustration of the Clustering Coefficient~($C_i$).}
\label{fig:chap01:CC}
\end{figure*}

Fig.~\ref{fig:chap01:CC} illustrates the definition of $C$. 
The figure shows a network with $43$ nodes of which node number
$29$ and $11$ are highlighted.
With the given definition of $C_i$, we find that
$C_{29}=2/C^3_2=0.66,$ and that for node $11$ is $C_{11}=0/C^3_2=0.$ 
Obviously, $C$ is `not defined' for isolated nodes~($k=0$), and is $0$
for nodes with degree $1$. 

\begin{figure*}[!tbh]
\begin{center}
\includegraphics[width=11cm]{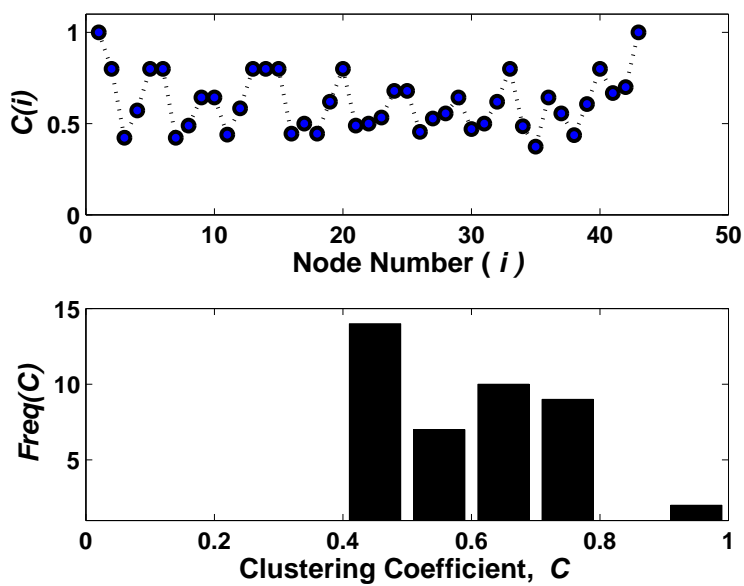} 
\end{center}
\caption[Topological Properties: Clustering Coefficient~($C_{i}$) and
its distribution~($P(C)$)]
{Clustering Coefficient~($C_{i}$) and its distribution~($P(C)$)}
\label{chap01:PCNprop03}
\end{figure*}

Figure~\ref{chap01:PCNprop03} shows $C_i$'s of individual residues
of 2PDD and the their histogram.

\newpage
\section{Data}
\label{cha01:sec:data_analyses}
For most part of the studies in (Chapter \ref{chap:protnet02} and \ref{chap:protnet03})
we analysed a total of $80$ proteins belonging to different functional
categories. Of these 80 proteins, we had 20 each from $\alpha$~(Table
No.~\ref{tab:data_80ptns_A}), $\beta$~(Table No.~\ref{tab:data_80ptns_B}), 
$\alpha/\beta$~(Table No.~\ref{tab:data_80ptns_ApB}), and
$\alpha+\beta$~(Table No.~\ref{tab:data_80ptns_AsB}) structural
class. Here, we followed SCOP~\cite{SCOP} classification of proteins.
$\alpha$ and $\beta$ class of proteins consist of proteins that are
made of $\alpha$ helices and $\beta$ sheets
respectively. 
$\alpha/\beta$ class consists of mainly parallel beta sheets
($\beta$-$\alpha$-$\beta$ units). 
$\alpha+\beta$ class consists of mainly anti-parallel beta sheets
(segregated $\alpha$ and $\beta$ regions).

For Chapter~\ref{chap:protnet04} we considered only small globular proteins. These
were $30$ single-domain two-state folding proteins (Table
No.~\ref{tab:data_30ptns_A} and \ref{tab:data_30ptns_B}). 
Following tables categorise each protein in terms of name and other
classification details. 

\begin{landscape}
\begin{table}[htp]
\begin{center}
\begin{tabular}{|l|l|l|c|c|l|} \hline \hline
PDB ID	&Name				&Functional Class  & Size ($n_r$) & No. of & Resoln.\\ 
        &           &           &        & Chains & (if X-ray) \\ \hline
1A6M	& Oxy-Myoglobin				& Oxygen Transport		&151		&1	& 1.00\AA\\ \hline 
1ALL	& Allophycocynin			& Light Harvesting Protein    	&321		&2	& 2.30\AA\\ \hline 
1B33	& Allophycocynin $\alpha$ and $\beta$ chains  & Photosynthesis		&2058		&14	& 2.30\AA\\ \hline 
1C75	& Cytochrome				& Electron Transport		&73		&1	& 0.96\AA\\ \hline
1DLW	& Truncated Hemoglobin			& Oxygen Storage/transport    	&116		&1	& 1.54\AA\\ \hline 
1DO1	& Carbonmonoxy-Myoglobin Mutant 	& Oxygen Storage/transport    	&153		&1	& 1.50\AA\\ \hline 
1DWT	& Myoglobin Complex			& Oxygen Transport		&152		&1	& 1.40\AA\\ \hline 
1FPO	& J-Type co-chperone			& Chaperone			&499		&3	& 1.80\AA\\ \hline
1FXK	& Archael Prefoldin (GIMC)		& Chaperone			&349		&3	& 2.30\AA\\ \hline 
1G08	& Bovine Hemoglobin			& Oxygen Storage/transport 	&572		&4	& 1.90\AA\\ \hline 
1H97	& Trematode Hemoglobin			& Non-Vertebrate Hemoglobin 	&294		&2	& 1.17\AA\\ \hline 
1HBR	& Oxygen Storage/Transport		& Chicken Hemoglobin D		&570		&4	& 2.30\AA\\ \hline 
1IDR	& Oxygen Storage/Transport		& Truncated Hemoglobin  	&253		&2	& 1.90\AA\\ \hline 
1IRD	& Oxygen Storage/Transport		& Human Carbonmonoxy && &\\
&&Haemoglobin&287		&2	& 1.25\AA\\ \hline 
1KR7	& Oxygen Storage/Transport		& Nerve Tissue Mini-Hemoglobin 	&110		&1	& 1.50\AA\\ \hline 
1KTP	& Photosynthesis			& C-Phycocyanin			&334		&2	& 1.60\AA\\ \hline 
1LIA	& Light Harvesting Protein		& R-Phycoerythrin		&664		&4	& 2.73\AA\\ \hline 
1MWC	& Oxygen Storage/Transport		& Wild Type Myoglobin		&310		&2	& 1.70\AA\\ \hline 
1NEK	& Oxidoreductase			& Succinate Dehydrogenase 	&1070		&4	& 2.60\AA\\ \hline 
1PHN	& Electron Transport			& Phcocynin			&334		&2	& 1.65\AA\\ \hline 
\end{tabular}
\end{center}
\caption[Data table for 20 proteins of $\alpha$ structural class]
{Data table for 20 proteins of $\alpha$ structural class.} 
\label{tab:data_80ptns_A}
\end{table}
\end{landscape}

\begin{landscape}
\begin{table}[htp]
\begin{center}
\begin{tabular}{|l|l|l|c|c|l|} \hline \hline
PDB ID	&Name				&Functional Class  & Size ($n_r$) & No. of & Resoln.\\ 
        &           &           &        & Chains & (if X-ray) \\ \hline

1AUN&	 Pathogenesis-related Protein 5D &	                Antifungal Protein   &	                208&	1&	1.80\AA\\ \hline
1BEH&	 Human Phosphatidylethanemine-&	                        Lipid Binding    &	                367&	2&	1.75\AA \\
    &    Binding Protein              &  & & &  \\ \hline 
1BHU&	 Streptomyces Metalloproteinase & Metalloproteinase Inhibitor   &	        102&	1&	NMR\\ 
    &    Inhibitor                    &  & & &  \\ \hline   
1DMH&	 Catechol 1&	                                        2-Dioxyoxygenase  Oxidoreductase    &	628&	2&	1.70\AA\\ \hline
1DO6&	 Superoxide Reductase &	                                Oxidoreductase&	                        248&	2&	2.00\AA\\ \hline
1F35&	 Murine Olfactory Marker &	                        Signaling Protein    &	                162&	1&	2.30\AA\\ \hline
1F86&	 Transthyretin&	                                        Transport Protein    &	                231&	2&	1.10\AA\\ \hline
1G13&	 Human GM2 Activator &	                                Ligand Binding Protein &	        486&	3&	2.00\AA\\ \hline
1HOE&	 $\alpha$-Amylase Inhibitor &	                        Glycosidase Inhibitor &	                74&	1&	2.00\AA\\ \hline
1I9R&	 CD40 Ligand &	                                        Cytokine/Immune System &	        1731&	9&	3.10\AA\\ \hline
1IAZ&	 Equinatoxin II &	                                Toxin&	                                350&	2&	1.90\AA\\ \hline
1IFR&	 Lamin--Globular Domain &	                        Immune System    &	                113&	1&	1.40\AA\\ \hline
1JK6&	 Bovine NeuroPhysin   &	                                Neuropeptide    &	                160&	2&	2.40\AA\\ \hline
1KCL&	 Cyclodextrin glycosyl transferase &	                Transferase&	                        686&	1&	1.94\AA\\ \hline
1KNB&	 Adenovirus Type 5 Fiber Protein &	                Cell Receptor Recognition    &	        186&	1&	1.70\AA\\ \hline
1SFP&	 Acidic Seminal Fluid Protein &	                        Spermadhesin&	                        111&	1&	1.90\AA\\ \hline
1SHS&	 Small Heat Shock Protein  &	                        Heat Shock Protein    &	                920&	6&	2.90\AA\\ \hline
1SLU&	 Rat Trypsin   &	                                Complex (serine Protease)  &	        345&	2&	1.80\AA\\ \hline
2HFT&	 Human Tissue Factor &	                                Coagulation Factor &	                207&	1&	1.69\AA\\ \hline
2MCM&	Macromomycin&	                                        Apoprotein&	                        113&	1&      1.50\AA\\ \hline
\end{tabular}
\end{center}
\caption[Data table for 20 proteins of $\beta$ structural class]
{Data table for 20 proteins of $\beta$ structural class.} 
\label{tab:data_80ptns_B}
\end{table}
\end{landscape}

\begin{landscape}
\begin{table}[htp]
\begin{center}
\begin{tabular}{|l|l|l|c|c|l|} \hline \hline
PDB ID	&Name				&Functional Class  & Size ($n_r$) & No. of & Resoln.\\ 
        &           &           &        & Chains & (if X-ray) \\ \hline

1AL8&	 Glycolate Oxidase  &	 Flavoprotein    &	344&	1&2.20\AA\\ \hline
1BQC&	$\beta$-Mannanase&	 Hydrolase    &	302&	1&	1.50\AA\\ \hline
1BWK&	 Old Yellow Wnzyme Mutant    &	Oxidoreductase&	399&	1&	2.30\AA\\ \hline
1C9W&	 CHO Reductase &	Oxidoreductase&	315&	1&	2.40\AA\\ \hline
1D8C&	 Malate Synthase G &	Lyase&	709&	1&	2.00\AA\\ \hline
1E0W&	 Xylanase 10A &	 Glycoside Hydrolase Family 10 &	302&	1&	1.20\AA\\ \hline
1EDG&	 Catalytic Domain of Celcca  &	 Cellulose Degradation  &	380&	1&	1.60\AA\\ \hline
1F8F&	 Benzyl Alcohol Dehydrogenase &	Oxidoreductase&	362&	1&	2.20\AA\\ \hline
1FIY&	 Phospoenolpyruvate Carboxylase  &	 Complex (Lysase/Inhibitor) &	874&	1&	2.80\AA\\ \hline
1FRB&	 FR-1 Protein/NADPH/ & Oxidoreductase (NADP) &	315&	1&	1.70\AA\\ 
    &    Zopolrestat Complex & & & & \\ \hline
1GAD&	 Dehydrpgenase  &	Oxidoreductase&	656&	2&	1.80\AA\\ \hline
1HET&	 Liver Alcohol Dehydrogenase  &	Oxidoreductase&	748&	2&	1.10\AA\\ \hline
1HTI&	 Triosephosphate Isomerase (TIM)  &	Isomerase&	496&	2&	2.80\AA\\ \hline
1N8F&	 Mutant of Phosphate Synthase  &	 Metal Binding Protein  &	1372&	4&	1.75\AA\\ \hline
1OY0&	 Ketopantoate  &	Transferase& 1240&	5&	2.80\AA\\ 
    & Hydroxymethyltransferase & & & & \\\hline
1QO2&	 Ribonucleotid Isomerase  &	Isomerase&	482&	2&	1.85\AA\\ \hline
1QTW&	 DNA Repair Enzyme  &	Hydrolase&	285& 1&	1.02\AA\\ 
    &    Endonuclease IV & & & & \\ \hline
1YLV&	COMPLEX&	Lyase&	341&	1&	2.15\AA\\ \hline
2TPS&	 Tiamin Phosphate Synthase &	 Thiamin Biosynthesis &	452&	2&	1.25\AA\\ \hline
8RUC&	 Spinach Rubisco Complex  &	 Lyase (Carbon-Carbon) &	2359&	8&	1.50\AA\\ \hline

\end{tabular}
\end{center}
\caption[Data table for 20 proteins of $\alpha/\beta$ structural class]
{Data table for 20 proteins of $\alpha/\beta$ structural class.} 
\label{tab:data_80ptns_AsB}
\end{table}
\end{landscape}

\begin{landscape}
\begin{table}[htp]
\begin{center}
\begin{tabular}{|l|l|l|c|c|l|} \hline \hline

PDB ID	&Name				&Functional Class  & Size ($n_r$) & No. of & Resoln.\\ 
        &           &           &        & Chains & (if X-ray) \\ \hline
1ALC&	 Baboon Alpha-Lactalbumin  &	 Calcium Binding Protein & 122&	1&	1.70\AA\\ \hline
1AVP&	 Human Adenovirus 2  &	Hydrolase&	215&	2&	2.60\AA\\ 
    &	Proteinase&	&	&	&	\\ \hline
1BRN&	Barnase&	Endonuclease&	216&	2&	1.76\AA\\ \hline
1CNS&	Chitinase&	 Anti-Fungal Protein &	486&	2&	1.91\AA\\ \hline
1CQD&	 Cysteine Protease &	Hydrolase&	864&	4&	2.10\AA\\ \hline
1EUV&	 ULP1 Protease Domain &	Hydrolase&	300&	2&	1.60\AA\\ \hline
1F13&	 Human Cellular  &	 Coagulation Factor &	1441&	2&	2.10\AA\\ 
    &	 Coagulation Factor XIII &	&	&	& \\ \hline
1GCB&	 DNA-Binding Protease &	 DNA-Binding Protein &	452&	1&	2.20\AA\\ \hline
1GOU&	 Ribonuclease Binase &	Hydrolase&	218&	2&	1.65\AA\\ \hline
1IWD&	 A Plant Cysteine  &	Hydrolase&	215&	1&	1.63\AA\\ 
    &	 Protease Ervatamin &	&	&     &	\\ \hline
1K3B&	 Human Dipeptidyl  &	Hydrolase&	352&	3&	2.15\AA\\ 
    &	 Peptidase I &	&	&	& \\ \hline
1LNI&	 A Ribonuclease &	Hydrolase&	219&	2&	1.00\AA\\ \hline
1LSD&	 Lysozyme  &	Hydrolase&	129&	1&	1.70\AA\\ \hline
1ME4&	COMPLEX&	Hydrolase&	204&	1&	1.10\AA\\ \hline
1MZ8&	COMPLEX&	 Toxin	 Hydrolase &	435&	4&	2.00\AA\\ \hline
1PPN&	 Papain Cys-25 &	 Hydrolase &	212&	1&	1.60\AA\\ \hline
1QMY&	 FMDV Leader Protease  &	 Hydrolase  &	468&	3&	1.90\AA\\ \hline
1QSA&	 Lytic Transglycosylase &	Transferase&	618&	1&	1.65\AA\\ \hline
1UCH&	 Deubiquitinating  &	 Cysteine Protease &	206&	1&	1.80\AA\\ 
&	 Enzyme UCH-L3 &	&	&	& \\ \hline
2ACT&	Actinidin&	 Hydrolase (Protease) &	218&	1&	1.70\AA\\ \hline

\end{tabular}
\end{center}
\caption[Data table for 20 proteins of $\alpha+\beta$ structural class]
{Data table for 20 proteins of $\alpha+\beta$ structural class.} 
\label{tab:data_80ptns_ApB}
\end{table}
\end{landscape}

\begin{landscape}
\begin{table}[htp]
\begin{center}
\begin{tabular}{|l|r|l|} \hline \hline 
  PDB ID&	$n_r$&   Name  \\ \hline
1HRC&	104&	 Horse heart cytochrome C 				 \\ \hline
1IMQ&	86&	 Colicin e9 immunity protein IM9  \\ \hline
1YCC&	108&	 Yeast ISO-1-cytochrome C  \\ \hline
2ABD&	86&	 Acyl-coenzyme a binding protein from bovine liver  \\ \hline
2PDD&	43&	 Acetyltransferase \\ \hline
1APS&	98&	 Acylphosphatase \\ \hline
1CIS&	66&	 Chymotrypsin inhibitor 2 and Helix E  \\ \hline
1COA&	64&	 The hydrophobic core of chymotrypsin inhibitor 2  \\ \hline
1FKB&	107&	 Rapamycin human immunophilin FKBP-12 complex  \\ \hline
1HDN&	85&	 Phosphocarrier protein HPR from e.~\emph{coli}  \\ \hline
1PBA&	81&	 Activation domain of porcine procarboxypeptidase B  \\ \hline
1UBQ&	76&	 Ubiquitin  \\ \hline
1URN&	96&	 U1A mutant/RNA complex + glycerol  \\ \hline
1VIK&	99&	 HIV-1 protease  \\ \hline
2HQI&	72&	 Oxidized form of MERP  \\ \hline
2PTL&	78&	 Immunoglobulin light chain-binding domain of protein L \\ \hline
2VIK&	126&	 Actin-severing domain villin 14T  \\ \hline
\end{tabular}
\end{center}
\caption[Data table for single-domain, two-state folding proteins, belonging to $\alpha$ and $\beta$ class.]
{Data table for single-domain, two-state folding proteins, belonging to $\alpha$ and $\beta$ class.} 
\label{tab:data_30ptns_A}
\end{table}
\end{landscape}

\begin{landscape}
\begin{table}[htp]
\begin{center}
\begin{tabular}{|l|r|l|} \hline \hline 
PDB ID&	$n_r$&   Name \\ \hline
1AEY&	58&	 Alpha-spectrin SRC homology 3 domain  \\ \hline
1CSP&	67&	 Bacillus subtilis major cold shock protein  \\ \hline
1MJC&	69&	 The major cold shock protein of e.~\emph{coli}  \\ \hline
1NYF&	58&	 SH3 domain from fyn proto-oncogene tyrosine kinase  \\ \hline
1PKS&	76&	 The PI3K SH3 domain  \\ \hline
1SHF&	59&	 The SH3 domain in Human FYN  \\ \hline
1SHG&	57&	 SRC-homology 3 (SH3) domain  \\ \hline
1SRL&	56&	 The SRC SH3 domain  \\ \hline
1TEN&	89&	 Fibronectin Type III domain from tenascin  \\ \hline
1TIT&	89&	 Titin, IG repeat 27  \\ \hline
1WIT&	93&	 Twitchin immunoglobulin superfamily domain  \\ \hline
2AIT&	74&	 Alpha-amylase inhibitor tendamistat  \\ \hline
3MEF&	69&	 Major cold-shock protein from escherichia coli  \\ \hline
\end{tabular}
\end{center}
\caption[Data table for single-domain, two-state folding proteins, belonging to $\alpha\beta$ class.]
{Data table for single-domain, two-state folding proteins, belonging to $\alpha\beta$ class.} 
\label{tab:data_30ptns_B}
\end{table}
\end{landscape}

\section{Software used}
\label{cha01:sec:softwares}
Following is the list of the software (programming languages and
software utilities) used in various parts of the study.

\begin{itemize}
\item FORTRAN90\\
Fortran90 was used to program most of the algorithms needed for the
network analyses. 
\item MATLAB\\
MatLab was primarily used for the visualisation purpose. Though
extensive programming had to be done for creating intricate and
detailed graphics to complement the  analyses. 
URL: www.mathworks.com
\item Gnuplot\\
Since most of the work was done on Linux platform, mainly
Gnuplot was used for plotting purpose. Extensive coding was done to
automate the graph generation process on mass scale.
URL: www.gnuplot.info 
\item Graphviz\\
The Fortran90 code was programmed to generate standard Graphviz input
file. Files so generated were fine-tuned while laying out the graphs
in Graphviz.
URL: www.graphviz.org 
\item PERL\\
It was primarily used for extraction of the required data from the PDB
files. URL: www.perl.com
\item Octave\\
Octave was used as a replacement for MatLab whenever required on the
Linux platform. URL: www.octave.org
\item Pajek\\
This useful graph layout package was used many times, though
Graphviz was preferred over Pajek. URL: vlado.fmf.uni-lj.si/pub/networks/pajek
\end{itemize}

\chapter{\label{chap:protnet02}Small-World Nature of Protein Contact Networks}

\section{Introduction}

There have been several efforts to study protein structures as (graphs) 
networks. In these studies the effort has been to analyse globular
proteins as systems composed of interacting parts. 
In recent years, with the elaboration of network properties in a
variety of real networks,  Vendruscolo et al.\/~\cite{protnet:PRE}
showed that protein structures have small-world topology.
Greene and Higman~\cite{protnet:JMB} studied the short-range and
long-range interaction networks in protein structures of $65$
proteins and showed that long-range interaction network is \emph{not}
small world and its degree distribution, while having an underlying
scale-free behaviour, is dominated by an exponential term indicative
of a single-scale system. 
Atilgan et al.\/~\cite{protnet:Biophys} studied globular protein
structures and analysed the network properties of the core and surface
of the proteins. They established that, regardless of size, the cores
have the same local 
packing arrangements. They also explained, with an example of binding
of two proteins, how the small-world topology could be useful in
efficient and effective dissipation of energy, generated upon
binding.

\section{Small-World topology of PCNs}
\label{chap02:sec:small-world}
The small-world nature of protein networks is a basic finding.
The small-world nature of a network is reflected in two properties: high
clustering compared to their random controls, and a logarithmic increase
in the characteristic path length with increase in the size of the
network.    

\begin{figure*}[!tbh]
\begin{center}
\begin{tabular}{c}
\includegraphics[width=10cm]{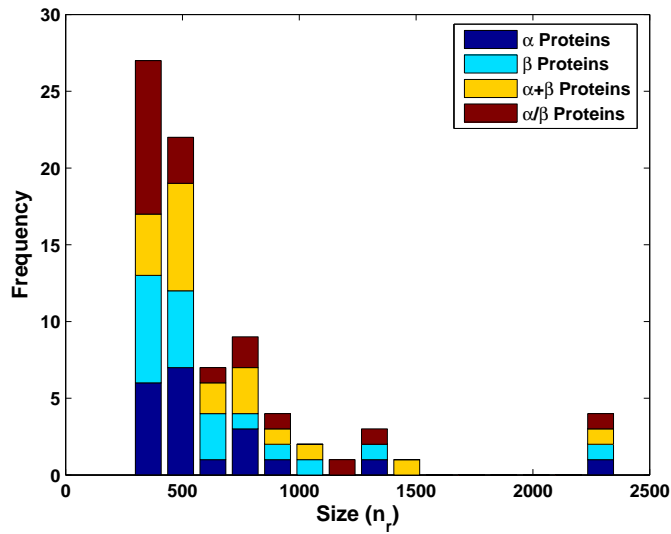}
\end{tabular}
\end{center}
\caption[Distribution of sizes of proteins analysed]
{Distribution of sizes of proteins analysed.}
\label{chap02:nr_hist}
\end{figure*}

The function that a protein serves in the cell is decided by the structure
of the protein. Proteins, owing to oft-repeated structural constructs,
could be classified~\cite{SCOP} (Structural Classification of Proteins,
http://scop.mrc-lmb.cam.ac.uk/scop/) based on their structural composition. 
We analysed $80$ proteins~(listed in Table Nos.\/~\ref{tab:data_80ptns_A},
\ref{tab:data_80ptns_B}, \ref{tab:data_80ptns_AsB},
\ref{tab:data_80ptns_ApB}), $20$ each from four major
categories ($\alpha$. $\beta$, $\alpha/\beta$, $\alpha+\beta$) of the
SCOP structural classification. These are from diverse functional
groups: hydrolase, transferase, protease, calcium binding,
oxydoreductase, antifungal, signalling, transport, toxin, coagulation
factor etc.\ to name a few. The size of these 
proteins  varied from $73$ to $2359$ amino acids. 
Fig.~\ref{chap02:nr_hist} shows the histogram of size of these proteins
and their break-up across the structural classes.

\begin{figure*}[!tbh]
\begin{center}
\begin{tabular}{c}
\textbf{(a)}\includegraphics[width=11cm]{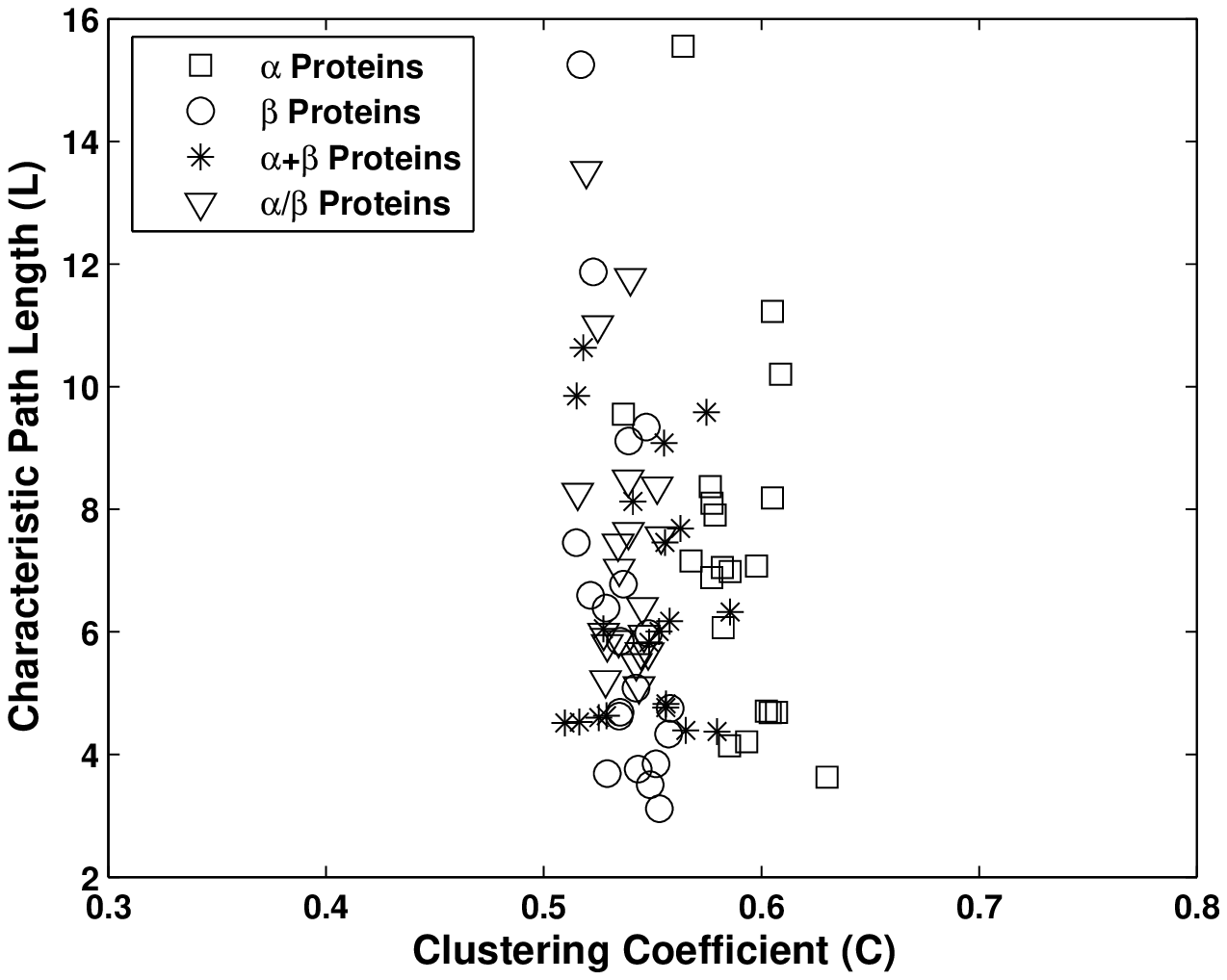} \\
\textbf{(b)}\includegraphics[width=11cm]{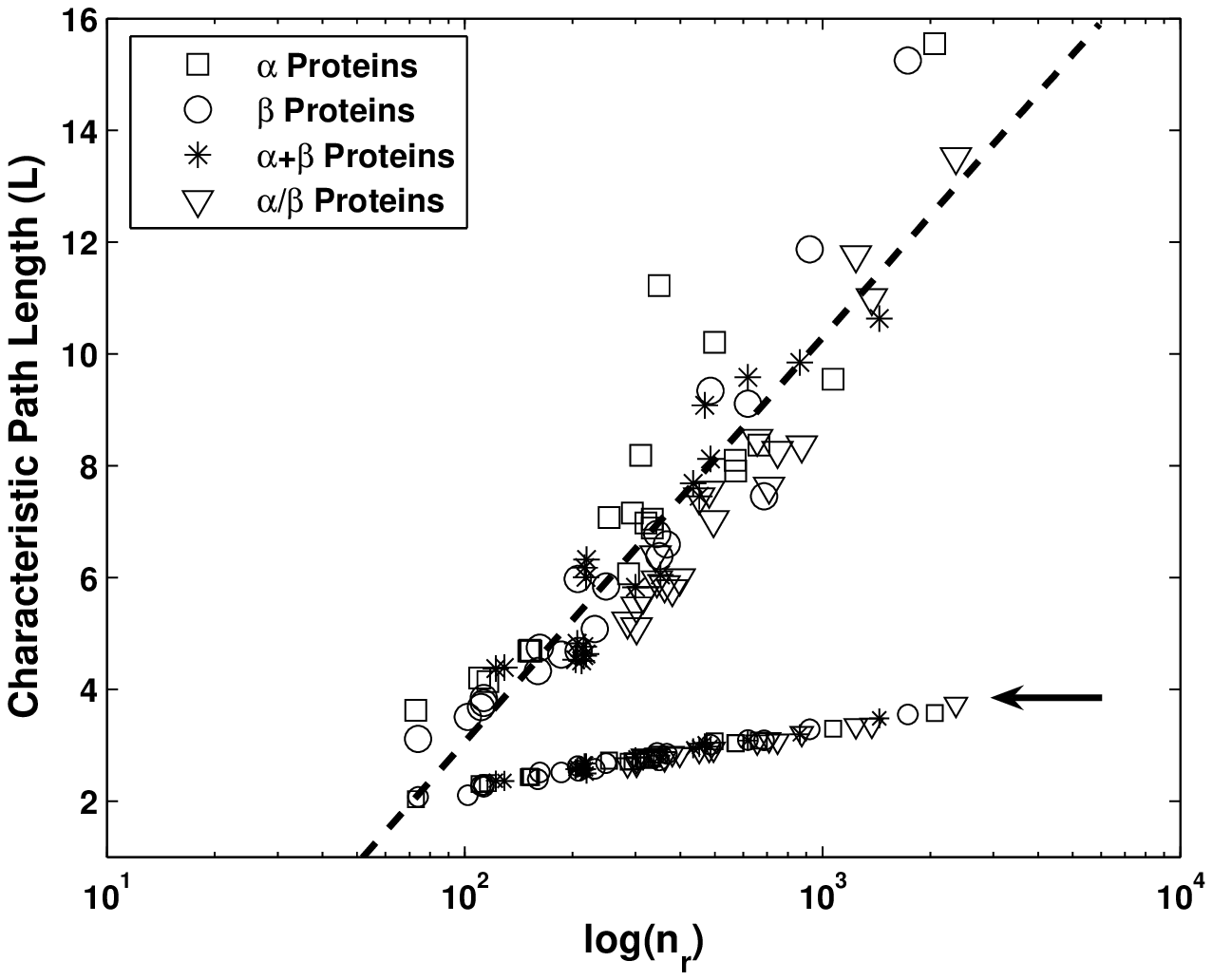}
\end{tabular}
\end{center}
\caption[{(a) $L$--$C$ plot and, (b) $L$--$ln(n_r)$ plot of proteins of
  four structural classes.}] 
{(a) $L$--$C$ plot of proteins from four structural classes. (b)
  Increase in the $L$ of proteins with logarithmic increase in  
size ($n_r$). The dotted line is a log-linear fit to the PCN data.} 
\label{chap02:lcln}
\end{figure*}

\begin{figure*}[!tbh]
\begin{center}
\begin{tabular}{c}
\includegraphics[width=10cm]{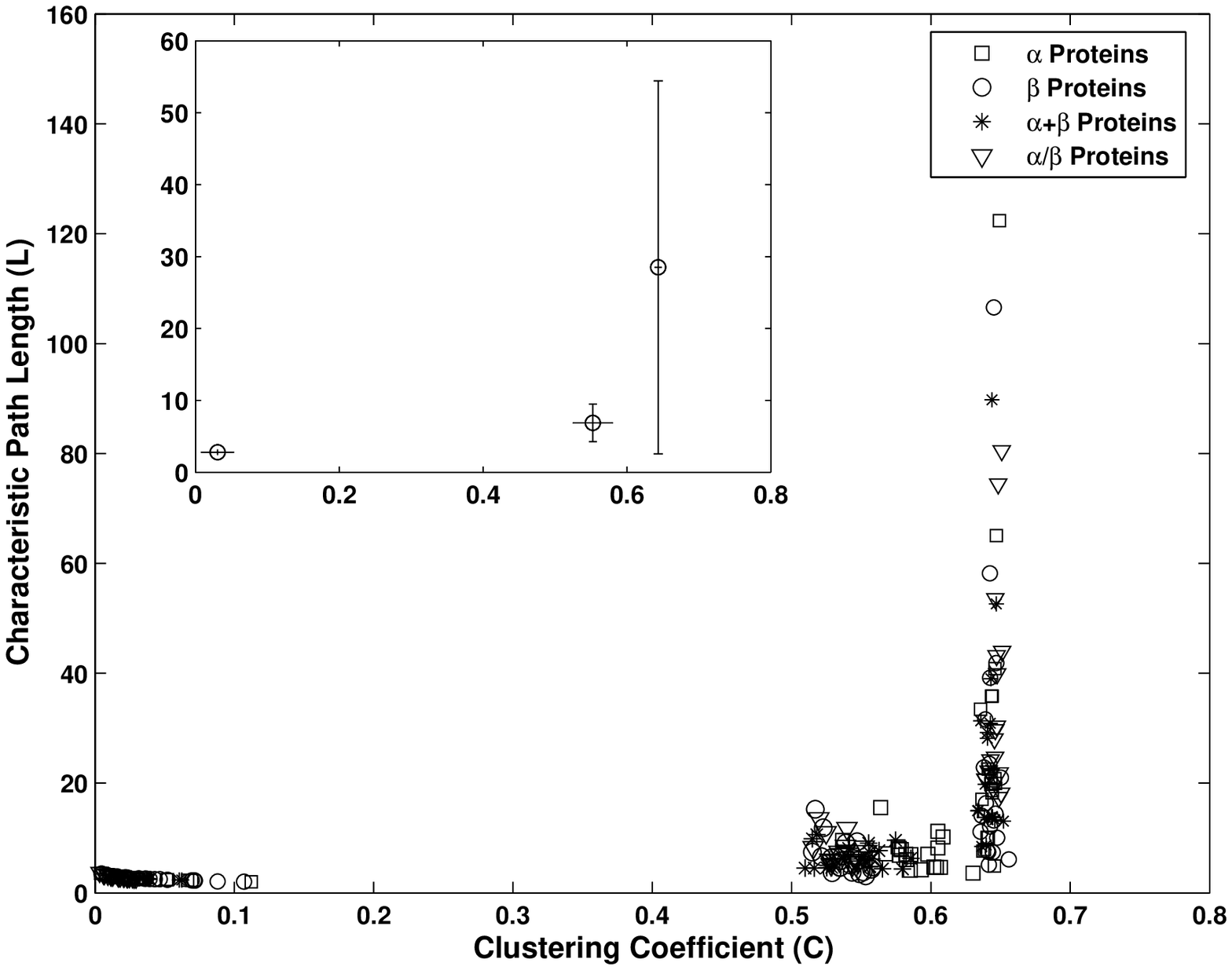} 
\end{tabular}
\end{center}
\caption[The small world nature of the proteins.]
{The small world nature of the proteins. Inset: Standard deviations of
  the corresponding data.} 
\label{chap02:lc_PCN_RAN_REG}
\end{figure*}

We calculated the average clustering coefficient~($C$) and the
characteristic path length ($L$) of the proteins. 
Fig.~\ref{chap02:lcln}~(a) shows the $L$ versus $C$ plot.
As seen in the figure, on the scale of $0$ to $1$, the proteins have
a very high value of clustering coefficients. 
Apart from very high $C$, what is interesting
is that these $80$ proteins are almost indistinguishable with this
parameter. Thus while presenting a generic property (that of high
clustering), of proteins similar to that of a large number of other
complex networks, the small-world network result provides a grim
picture in terms of our ability to correlate this specific network
(geometric) parameter to the proteins' structure and function.  

For a network to be classified as a small-world network, apart from high
clustering, its $L$ should increase only as a $log(n_r)$. Such a
logarithmic scaling of $L$ with $n_r$, makes it a small-world network,
i.e. any node on the network could be reached from any other node in an
exceptionally few number of steps. 
Fig.~\ref{chap02:lcln}~(b) shows that the $L$ of these $80$ PCNs scale
logarithmically with the size of the network. 
These two properties thus ascertain the small-world nature of the PCNs
across structural classes. Fig.~\ref{chap02:lcln}~(b) also shows $L$
of random controls of PCNs (marked with an arrow). 

Fig.~\ref{chap02:lc_PCN_RAN_REG} shows the summary plot of $L$--$C$
for all $80$ proteins, with their Type-I random controls in the
bottom-left, regular controls in the extreme-right, and PCNs in the
middle. The inset of the figure shows the means and standard
deviations of  $L$ and $C$ of the corresponding data.  
As seen in the figure the $L$ of PCNs are of the same order of
magnitude as those of their Type I random controls. 
PCNs of these proteins have very high clustering coefficients
compared to their random controls (statistically significant,
$p<0.001$; Two-Sample Kolmogorov-Smirnov Test). 
The $L$ and $C$ computed here and in the rest of this chapter, for
random and regular controls, were computed based on analytical
formulae mentioned in Subsection~\ref{chap01:subsec:distance} and
Subsection~\ref{chap01:subsec:compactness_measures} respectively.

\section[Globular and Fibrous Proteins]{Globular and  Fibrous Proteins} 
\begin{figure*}[!tbh]
\begin{center}
\begin{tabular}{cc}
\includegraphics[width=8cm]{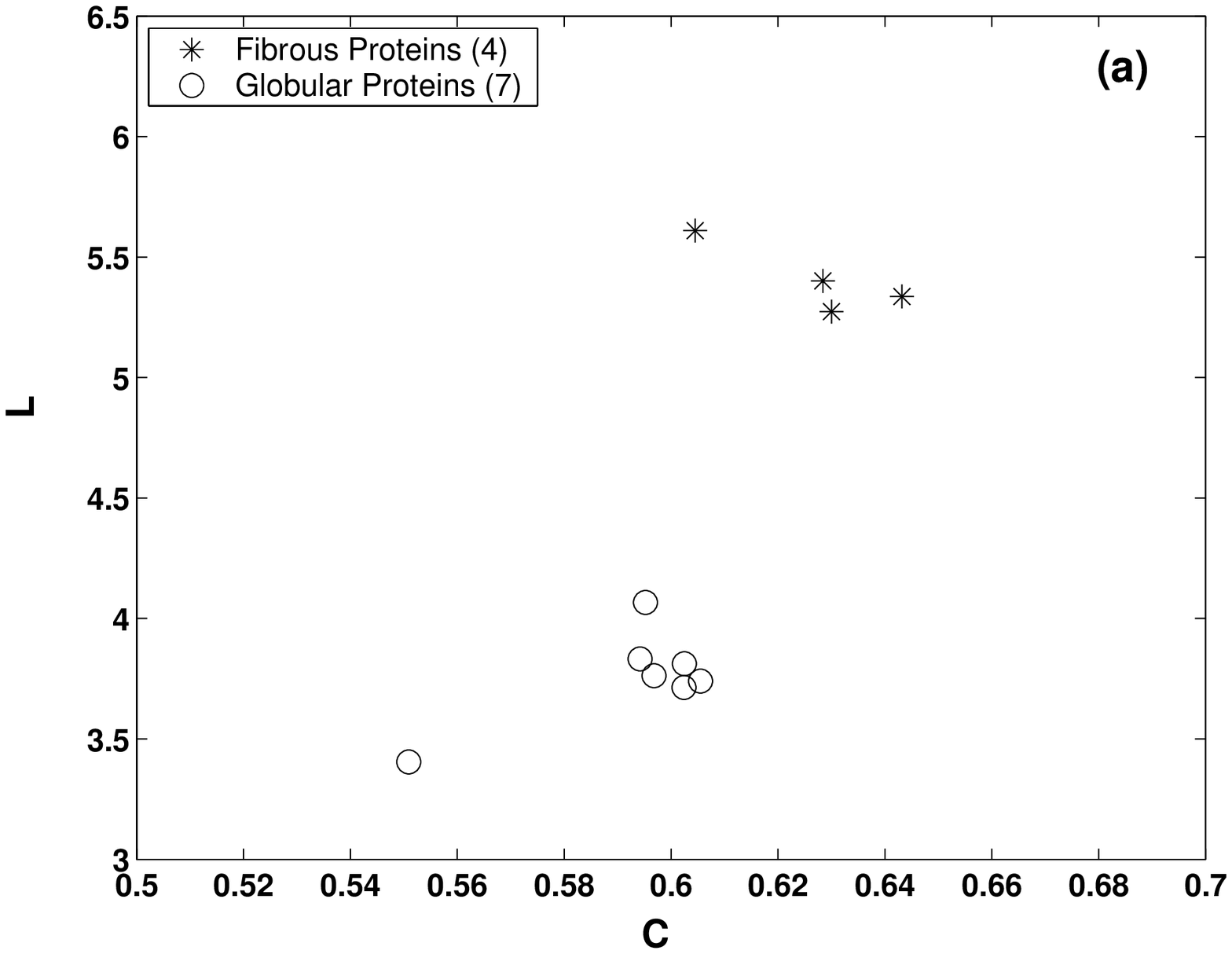} &
\includegraphics[height=6cm]{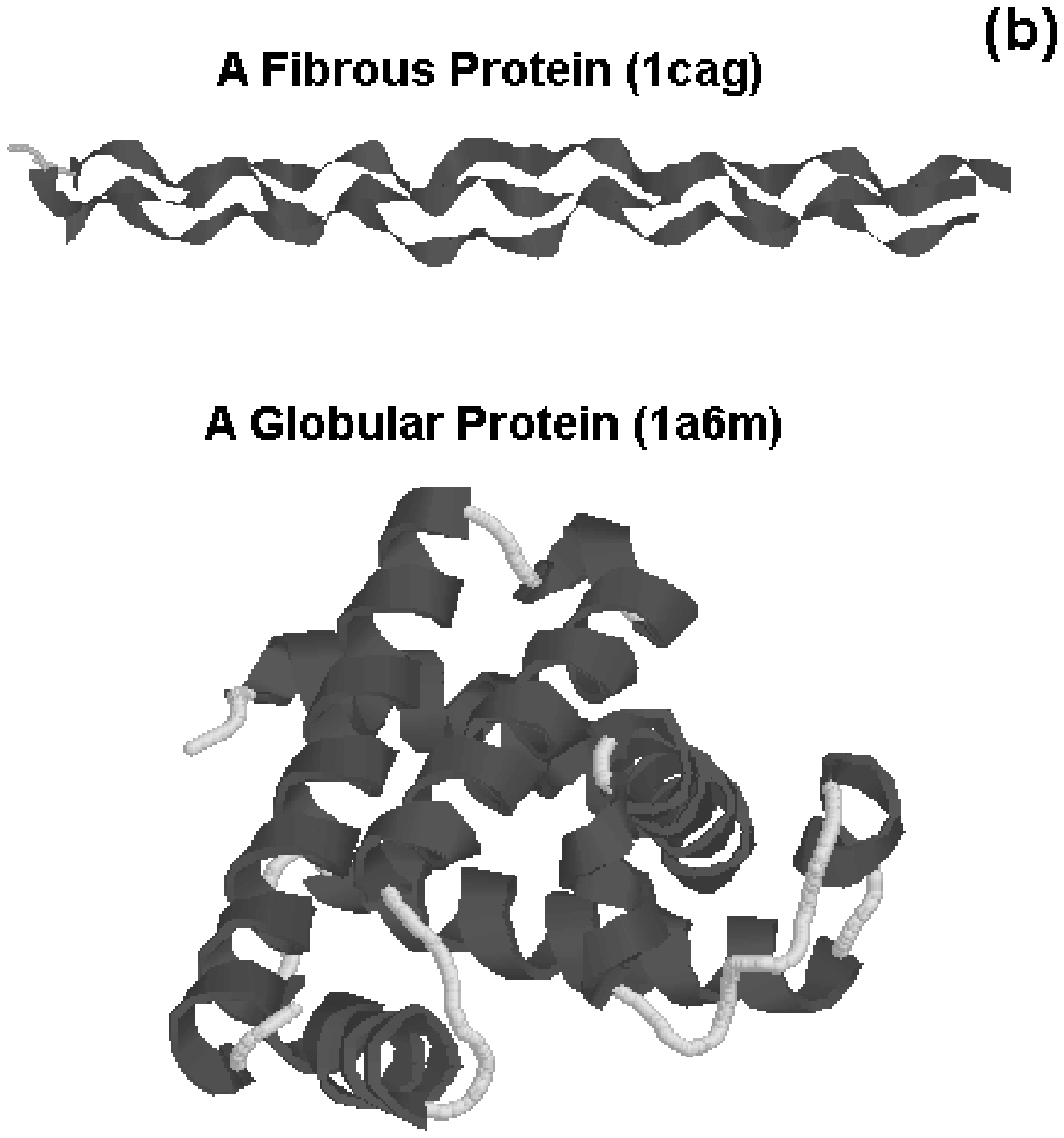}
\end{tabular}
\end{center}
\caption[L--C plot comparing the Fibrous and Globular proteins]
{(a) L--C plot of Fibrous and Globular proteins of comparable
  sizes. (b) Examples of  three-dimensional structures of a fibrous
  and globular protein (not to the scale) with their PDB codes.}
\label{chap02:lc_fib_glob01}
\end{figure*}

Most proteins are ``globular'' in their three-dimensional structure,
into which the polypeptide chain folds into a compact shape. 
In contrast, ``fibrous'' proteins have relatively simple, elongated
three-dimensional structure suitable for their biological  
function (see Fig.~\ref{chap02:lc_fib_glob01}~(b)). 
The ``small-world'' nature of globular proteins was
argued~\cite{protnet:Biophys} to be required for enhancing the ease of
dissipation of disturbances.  
If that were true, the fibrous proteins should depart from the
small-world nature. 
We studied fibrous proteins and compared their network properties with
globular proteins of comparable sizes.  
Table~\ref{tab:fiber_globular_list} shows the details of these proteins.
As shown in the L--C plot in Fig.\/~\ref{chap02:lc_fib_glob01}(a),
fibrous proteins have larger $L$, although the $C$ are  similar to
those of globular proteins.  
Thus, in this respect, the fibrous proteins also show ``small-world''
properties. The average diameter for the fibrous proteins ($D=15$) was
found to be larger than that of the globular proteins ($D=8.57$). This
is expected because of the elongated  structure of fibrous proteins. 
Despite this major difference in structure, the network properties
of fibrous proteins and globular proteins are not very
different. This indicates that the ``small-world'' property of
proteins is generic and persists irrespective of structural
differences. 

\begin{table}[!htb]
\begin{center}
\begin{tabular}{|c|c|r|r|r|}
\hline
Sr.No.& PDB ID & $n_r$ & $L$ & $C$ \\
\hline
F1 &1CGD & 90 & 5.401 & 0.7463 \\
F2 &1CAG & 88 & 5.274 & 0.6933 \\
F3 &1EI8 & 172& 5.610 & 0.6045 \\
F4 &1QSU & 89 & 5.337 & 0.6432 \\ \hline\hline
G5 &1ABA & 87 & 3.382 & 0.5942 \\
G6 &1AE2 & 86 & 4.066 & 0.5952 \\
G6 &1AYI & 86 & 3.812 & 0.6025 \\
G7 &1C6R & 88 & 3.740 & 0.6055 \\
G8 &1CEI & 85 & 3.713 & 0.6024 \\
G9 &1CTJ & 89 & 3.763 & 0.5968 \\
G10&1DSL & 88 & 3.404 & 0.5509 \\ \hline
\end{tabular}
\end{center}
\label{tab:fiber_globular_list}
\caption[List of fibrous proteins and the control globular
proteins]{List of four fibrous(F1--F4) and seven globular proteins(G5--G10) analysed.}
\end{table}

\section{ $\alpha$ and $\beta$ Proteins}

\begin{figure*}[!tbh]
\begin{center}
\begin{tabular}{c}
\includegraphics[width=10.0cm]{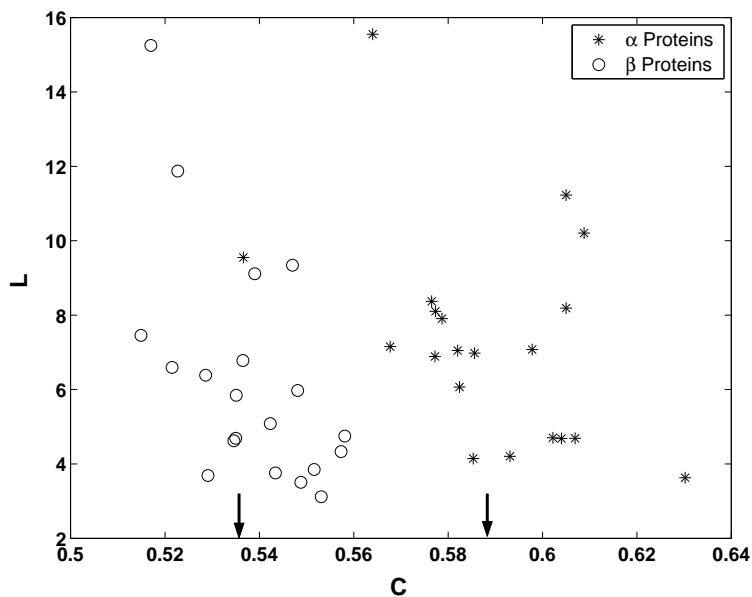} 
\end{tabular}
\end{center}
\caption[L--C plot comparing the $\alpha$ and $\beta$ class of proteins]
{L--C plot for $\alpha$ and $\beta$ proteins. Arrows indicate the means of $C$ for
$\alpha$ and $\beta$ proteins.}
\label{chap02:lc_alpha_beta}
\end{figure*}

As seen earlier~(Figure~\ref{chap02:lcln}(a)), both $\alpha$ and $\beta$
class of proteins show small-world properties. Given that these are
two distinct structural units one would want to know how that reflects on
the global network parameters of $\alpha$ and $\beta$ proteins. 
On finer analysis of $L$--$C$ properties, we find that, while they are
indistinguishable on $L$-scale, there is a marginal, yet consistent
difference in the clustering coefficients of $\alpha$ and $\beta$
proteins as shown in Fig.~\ref{chap02:lc_alpha_beta}. The mean of $C$
for $\alpha$ and  $\beta$ proteins studied are $0.588$ and $0.538$,
respectively. According to Kolmogorov-Smirnov test, this difference is
statistically significant ($p<0.001$). 

\section{Degree Distributions of PCNs and LINs}
The distribution of the degrees is an important property which
characterises the network topology. The degree distribution of a 
random network is characterised by a Poisson distribution. 
The degree distribution of many real-world networks has been shown to
be that of the scale-free type~\cite{alberts}. Many models have been
proposed to explain the evolution of network and the degree
distribution with which they are characterised at present. 

\begin{figure*}
\begin{center}
\begin{tabular}{c}
\includegraphics[width=12cm]{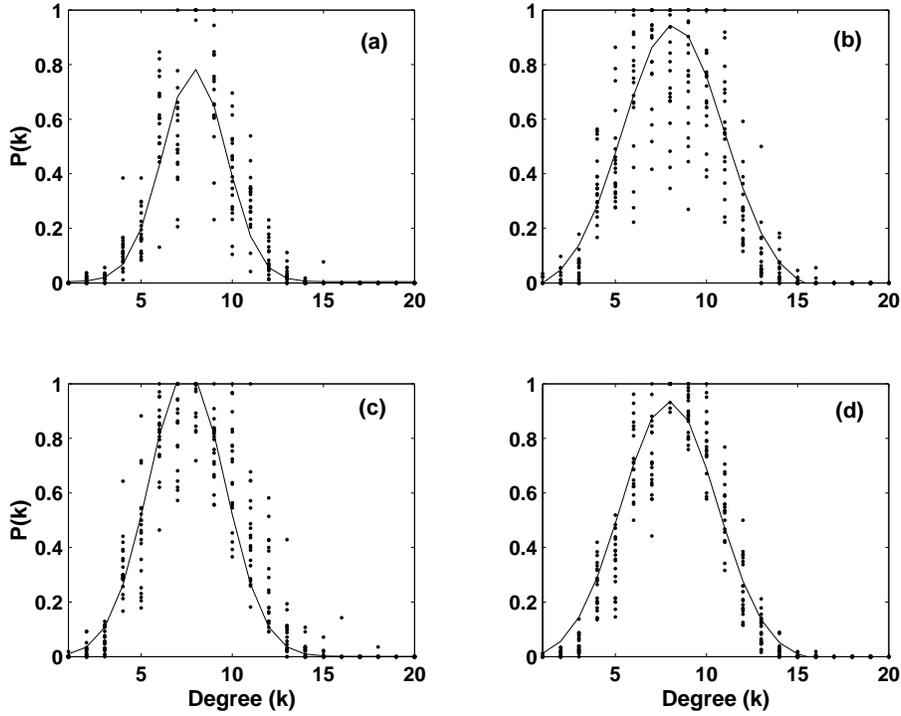}
\end{tabular}
\end{center}
\caption[Scatter plot of degree distributions for (a) $\alpha$, (b) $\beta$, (c)
$\alpha+\beta$, and (d) $\alpha/\beta$ proteins] 
{Scatter plot of degree distributions for (a) $\alpha$, (b) $\beta$, (c)
  $\alpha+\beta$, and (d) $\alpha/\beta$ proteins, $20$ of each
  class. The solid line is the Gaussian fit through the means.}
\label{chap02:degreedist01}
\end{figure*}

We analysed the degree distribution of the $80$ proteins mentioned
above.  Figure~\ref{chap02:degreedist01} shows the normalised degree
distributions of $\alpha$, $\beta$, $\alpha+\beta$, and $\alpha/\beta$
protein networks.  
Figure~\ref{chap02:degreedist01} shows the scatter plot of normalised
degree distributions ($P(k)$) of all $80$ proteins of four different classes.
Data points in each plot indicate $P(k)$ values for all the residues
of $20$ proteins of the respective class. 
Solid line is a Gaussian fit to the mean of $P(k)$ for each  value of
$k$. 

As seen, shapes of these distributions are single
humped, Gaussian-like~\cite{protnet:JMB}. Importantly, unlike in
scale-free degree distributions the number of nodes with very high
degree falls off rapidly. This is interesting as in scale-free
networks high-degree nodes (hubs) are known to be the facilitators of
communication across the network by providing shorter routes through
them.  Hence hubs would partially explain small-world nature. 
But, clearly, the distribution of contacts in proteins is dominated by
an exponential term. 

\begin{figure*}[!htb]
\begin{center}
\begin{tabular}{c}
\includegraphics[width=12cm]{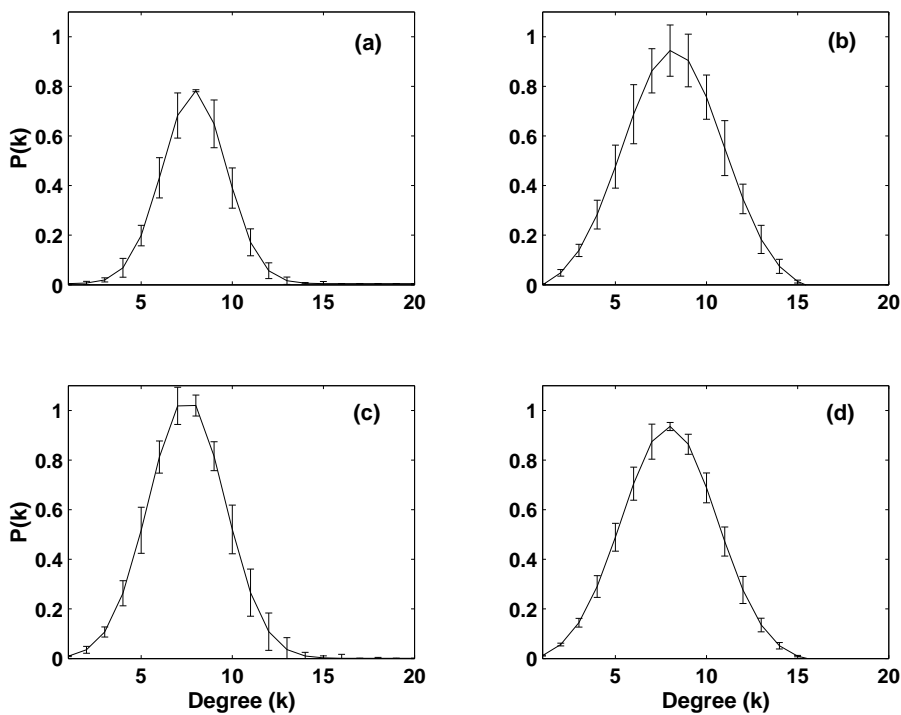} 
\end{tabular}
\end{center}
\caption[Degree distributions for (a) $\alpha$, (b) $\beta$, (c)
$\alpha+\beta$, and (d) $\alpha/\beta$ proteins] 
{Degree distributions for (a) $\alpha$, (b) $\beta$, (c)
  $\alpha+\beta$, and (d) $\alpha/\beta$ proteins, $20$ 
of each class. The error-bars show 1-$\sigma$ standard deviation
around the mean value.}
\label{chap02:degreedist02}
\end{figure*}

Figure~\ref{chap02:degreedist02} shows the 1-$\sigma$ standard
deviation of the data of $20$ PCNs for the normalised degree
distribution of respective classes. 
Solid line is a Gaussian fit to $\langle P(k) \rangle$, the mean of
$P(k)$ for each   
value of $k$.
The Gaussian fit was obtained with 
$$ y(x) = \frac{A}{w\sqrt{\pi/2}} \exp{\frac{-2(x-x_c)^2}{w^2}}.$$

Table~\ref{tab:fit:parameters} gives parameter values of the goodness
of fit. Here, $R^2$ is the coefficient of determination.

\begin{table}[!tbh]
\begin{center}
\begin{tabular}{|c|r|r|r|r|} \hline 
Class    & $x_c$   & $w$    & $A$     & $R^2$ \\\hline
$\alpha$ & $7.922$ & $3.524$& $3.443$ & $0.9126$ \\
$\beta$ & $8.175$ & $5.429$& $6.555$ & $0.9189$ \\
$\alpha+\beta$ & $7.506$ & $4.216$& $5.535$ & $0.9732$ \\
$\alpha/\beta$ & $7.961$ & $5.192$& $6.146$ & $0.9684$ \\\hline
\end{tabular}
\end{center}
\caption[Parameters: Degree Distribution Curve Fitting] 
{Degree Distribution Curve Fitting. Parameters and goodness of fit.} 
\label{tab:fit:parameters}
\end{table}

\begin{figure*}[!htb]
\begin{center}
\begin{tabular}{c}
\includegraphics[width=12cm]{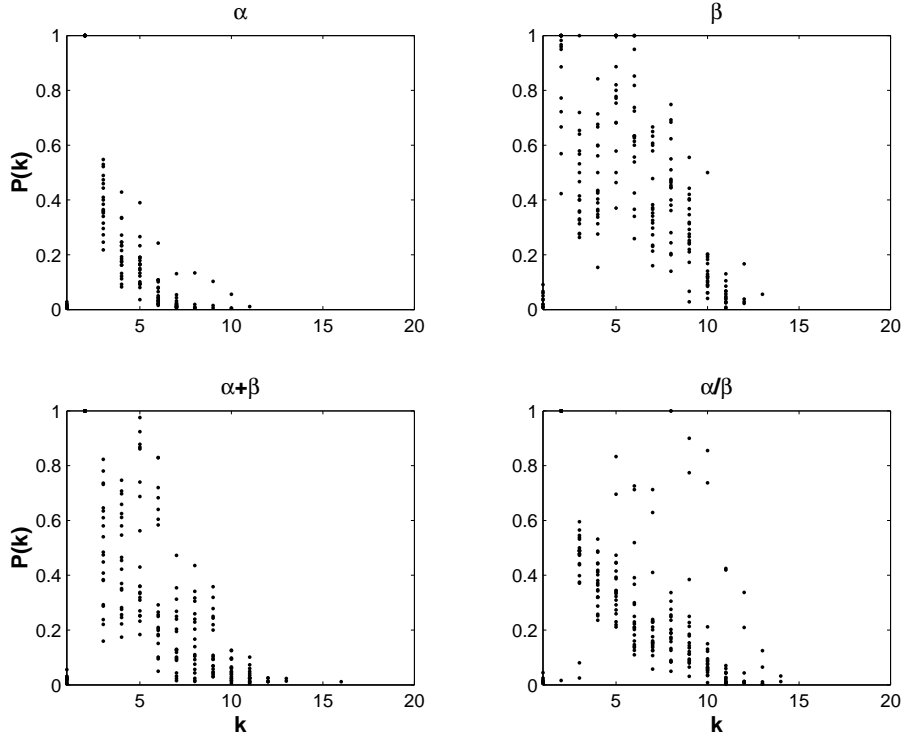}
\end{tabular}
\end{center}
\caption[Scatter plot of degree distributions for LINs of (a) $\alpha$, (b) $\beta$,
(c) $\alpha+\beta$, and (d) $\alpha/\beta$ proteins] 
{Scatter plot of degree distributions for LINs of (a) $\alpha$, (b)
  $\beta$, (c) $\alpha+\beta$, and (d) $\alpha/\beta$ proteins, $20$  
of each class. Data points in each plot indicate $P(k)$ values for all
the residues of $20$ proteins of the LINs of the PCNs of respective class.} 
\label{chap02:degreedist03}
\end{figure*}

Degree distribution of LINs is shown in
Fig.~\ref{chap02:degreedist03}. As seen the $P(k)$ of LINs show a 
single-scale decay with no typical node present in them.

\section{Diameter of PCNs and LINs}
\begin{figure*}[!htb]
\begin{center}
\begin{tabular}{c}
\includegraphics[width=10cm]{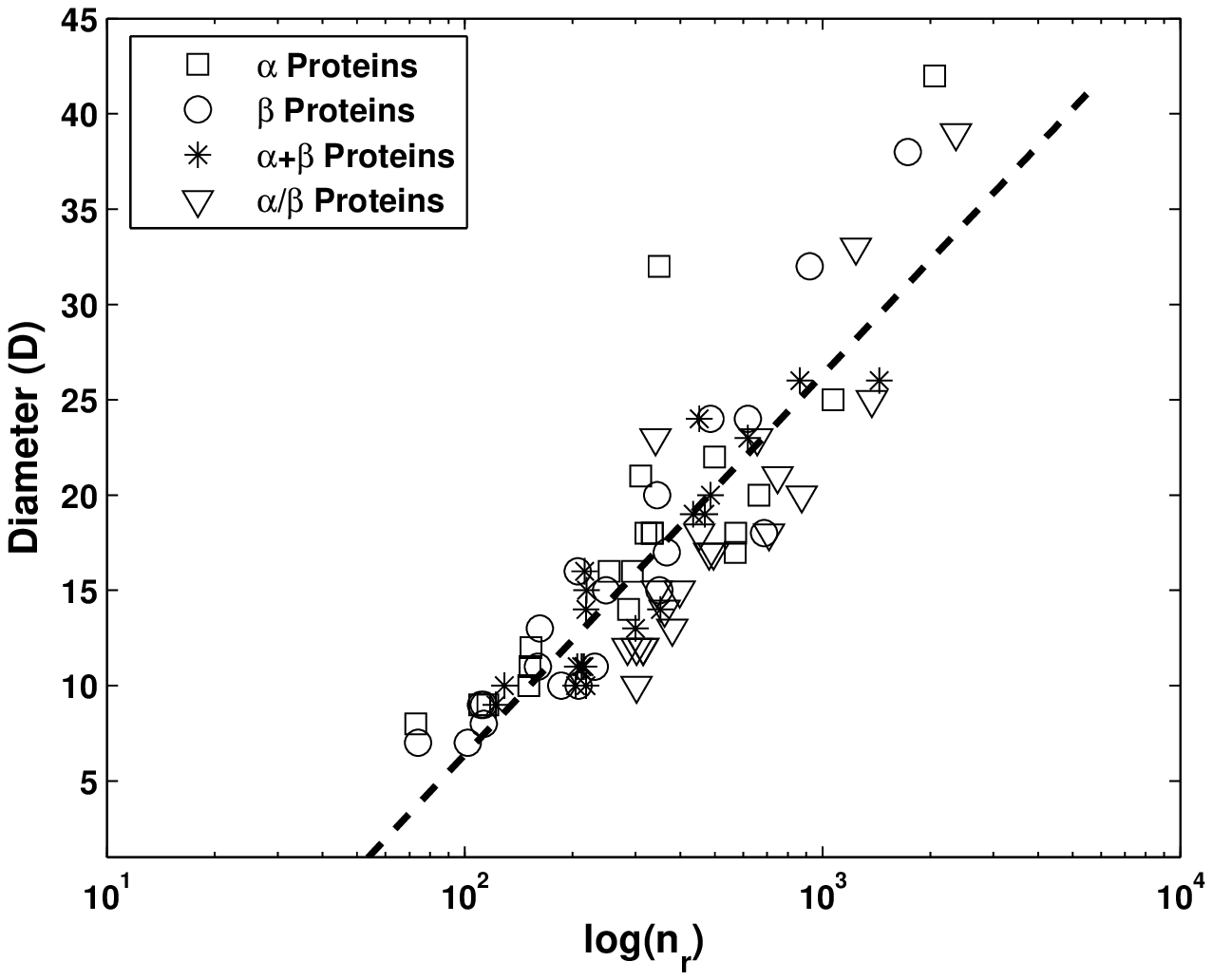}
\end{tabular}
\end{center}
\caption[Diameter of 2PDD network.]
{Diameter of the PCNs. (a) $\alpha$, (b) $\beta$, (c) $\alpha+\beta$, and (d) $\alpha/\beta$ proteins, $20$ of each class.}
\label{chap02:diamter}
\end{figure*}

The concept of diameter, strictly speaking, is applicable
only to single-component graphs. Owing to the presence of the
backbone connectivity, PCNs and its other versions are always
single-component. Diameter is expected to scale with the number of
nodes in the same way as the characteristic path length
($L$). Fig.~\ref{chap02:diamter} shows that $D$ does scale
logarithmically with $n_r$. Diameter, since it is maximal of the
distances between two nodes, the growth of $D$ with $n_r$ imposes 
upper limit on the rate of growth of $L$ with $n_r$.

\section{$C$--$n_r$ Plot}
\begin{figure}[!htb]
\begin{center}
\begin{tabular}{c}
\includegraphics[width=10.0cm]{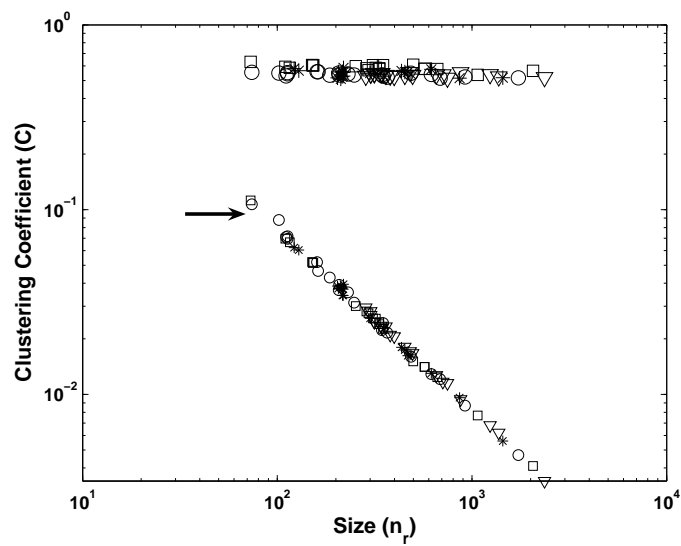}
\end{tabular}
\end{center}
\caption[$C$--$n_r$ plot for the proteins indicating the hierarchical
nature of the protein structures] 
{Change in $C$ of PCNs with increasing $n_r$, indicating the
  hierarchical nature of the protein structures. Random Control data
  is pointed with an arrow.}  
\label{chap02:cn}
\end{figure}

Clustering coefficient is essentially the probability of formation of 
triangles in the network. In a random network the probability that a
given node's two first-neighbours themselves are connected is equal to
that of any two randomly selected nodes are connected.
Therefore, clustering coefficient ($C_{rand}$) of a
random graph is given by
\begin{equation}
C_{rand} = p = \frac{\langle k \rangle}{n_r}.
\label{eq:cn}
\end{equation}
Therefore, according to Eq.~\ref{eq:cn} when $C_{rand}/\langle k
\rangle$ of random networks is plotted as a function of $n_r$ for
varying sizes of the network, the data will show a linear nature with
slope $-1$. The random controls of PCN show such behaviour as shown in
Fig.~\ref{chap02:cn} (The data pointed with an arrow).  

Figure~\ref{chap02:cn} also shows the change in $C$ with changing size of PCNs.
Here the $C$ of PCNs do not change with the size of the network
($n_r$) which indicates that the PCNs, far from being random, show an
indication of hierarchical structure~\cite{ravsaz:science} in them.  


\section{Discussion}
Our results show that protein networks have ``small-world'' property
regardless of their structural classification  
($\alpha$, $\beta$, $\alpha+\beta$, and $\alpha/\beta$) and tertiary 
structures (globular and fibrous proteins). 
Small world nature implies that PCNs have high degree of
clustering~\cite{protnet:PRE,protnet:JMB,protnet:Biophys,Bagler2005} 
(compared to their random counterparts). Clustering, for protein
structures, represents the extent/density of packing in the
network. Thus higher order compaction, observed in proteins, is in
agreement with what is expected from globular polymer chains in
contrast to `randomly folded control polymers'. 
%

Though small but definite differences exist between $\alpha$ and
$\beta$ classes, and fibrous and globular proteins.  
The size independence of the clustering coefficient in proteins
indicates a departure from the random nature and an inherent modular
organisation in the protein networks.  

It is interesting to note that unlike other networks, PCNs while being
small-world are not characterised by scale-free degree distribution.
The absence of hubs in PCNs is understandable as there is a physical
limit on the number of amino acids that can occupy the space within a
certain distance around another amino acid. Such system-specific
restrictions have been identified to be responsible for the emergence
of different classes of networks with characteristic
degree-distributions by Amaral et al.\/~\cite{amaral:PNAS}. They 
observed that preferential attachment to vertices in many real
scale-free networks~\cite{reka:thesis} can be hindered by factors like
ageing of the vertices (e.g. actors networks), cost of adding links to
the vertices, or, the limited capacity of a vertex (e.g. airports 
network).

\chapter{\label{chap:protnet03} Assortative Mixing in Protein Networks} 
\section{Introduction}
In recent years, there has been considerable
interest~\cite{reka:thesis,dorogovtsev:book} in structure and dynamics
of networks, with application to systems of diverse origins such as
society (actors' network, collaboration networks, etc.), technology
(world-wide web,  Internet, transportation infrastructure), biology
(metabolic networks, gene regulatory networks, protein--protein
interaction networks, food webs) etc. The aim of these studies has
been to identify correlation between general network parameters to the
structure, function, and evolution of the wide variety of systems. 

\subsubsection{Assortative Mixing}
While analysing and later modelling the evolution and structure of
real-world complex networks many features have been taken into
account: the path length, clustering, degree distribution, and degree
correlations. A lot of emphasis has been given to degree
distribution. 
The pattern of connectivity among the nodes of varying degrees also
affects the interaction dynamics of the network.
Degree correlations is a measure that computes the strength and
pattern of connectivity. 
Degree correlations were largely neglected until it was
emphasised, as shown in Table~\ref{tab:data:ass01}, that most
real-world (except social) networks are
disassortative~\cite{r:newman}. It is evident that all real-world 
networks of diverse origin are characterised by disassortative
mixing~\cite{r:newman}. 

\begin{table}[!tbh]
\begin{center}
\begin{tabular}{|l|r|r|} \hline 
  network &   $n$ &   $r$  \\ \hline
  physics coauthorship & $52,909$ & $0.363$ \\
  biology coauthorship & $1,520,251$ & $0.127$ \\
  mathematics coauthorship & $253,339$ & $0.120$ \\
  film actor collaborations & $449,913$ & $0.208$ \\ 
  company directors & $7,673$ & $0.276$ \\ \hline
  Internet & $10,697$  & $-0.189$ \\
  World-Wide Web & $269,504$ & $-0.065$ \\
  protein interactions & $2,115$ & $-0.156$ \\
  neural network & $307$ & $-0.163$ \\
  food web & $92$ & $-0.276$ \\ \hline
\end{tabular}
\end{center}
\caption[Size ($n$) and assortativity coefficient ($r$) of a number of
real-world networks] {Size ($n$) and assortativity coefficient ($r$) of a number of
real-world networks. Data adopted from~\cite{r:newman}. Except for
social networks, which have positive $r$s, all other networks are disassortative.}
\label{tab:data:ass01}
\end{table}

Social networks with their assortative nature, imply that they are  
fundamentally different from other networks and the property has been
claimed~\cite{newman:society01} to be originating from their  
unusually high clustering coefficients and community structure. 
Recently, assortative mixing has been demonstrated in brain functional
network~\cite{brainnw:eguiluz}, but no biological basis has  
been assigned to the property.
The disassortative degree mixing in most complex networks is an
unsolved riddle, and questions regarding the origin of this property  
and whether this is an universal property of complex networks has been
adjudged as ``one of the ten leading questions for network  
research''~\cite{EPJB:round_table}.

Biological networks are of special interest as they are the
products of long evolutionary history.  
The protein contact network is exclusive among other intra-cellular
networks (such as metabolic networks, gene regulatory networks,  
protein--protein interaction networks) for their unique method of
synthesis as a linear chain  of amino acids, and then folding into  
a stable three-dimensional structure through short- and long-range
contacts among the residues.  

\begin{table}[!tbh]
\begin{center}
\begin{tabular}{|c|c|c|c|c|c|c|c|} \hline \hline
PDB~ID	&Class	        &$n_r$	&$n_c$	&$k_{max}$	 &$\langle
k\rangle$	&$r_{PCN}$ &$r_{LIN}$	\\ \hline \hline
1HRC	&$\alpha$	&104	&488	&17	&9.3846	 &0.1821 & 0.3531	        \\ \hline
1IMQ	&$\alpha$	&86	&411	&17	&9.5581	 &0.2586 & 0.3675	        \\ \hline
1YCC	&$\alpha$	&108	&505	&17	&9.3519	 &0.2449 & 0.3379	        \\ \hline
2ABD	&$\alpha$	&86	&405	&17	&9.4186	 &0.2874 & 0.2736	        \\ \hline
2PDD	&$\alpha$	&43	&175	&14	&8.1395	 &0.1436 & 0.2616	        \\ \hline \hline

1AEY	&$\beta$ 	&58	&271	&15	&9.3448	 &0.1145 & 0.2829	        \\ \hline
1CSP	&$\beta$ 	&67	&308	&16	&9.194	 &0.2929 & 0.384	        \\ \hline
1MJC	&$\beta$ 	&69	&315	&16	&9.1304	 &0.3027 & 0.4115	        \\ \hline
1NYF	&$\beta$ 	&58	&262	&15	&9.0345	 &0.1752 & 0.4006	        \\ \hline
1PKS	&$\beta$ 	&76	&385	&17	&10.1316 &0.1872 & 0.3326	        \\ \hline
1SHF	&$\beta$ 	&59	&269	&16	&9.1186	 &0.1511 & 0.5789	        \\ \hline
1SHG	&$\beta$ 	&57	&265	&16	&9.2982	 &0.1503 & 0.4414	        \\ \hline
1SRL	&$\beta$ 	&56	&260	&16	&9.2857	 &0.2101 & 0.4433	        \\ \hline
1TEN	&$\beta$ 	&89	&415	&17	&9.3258	 &0.1645 & 0.5649	        \\ \hline
1TIT	&$\beta$ 	&89	&430	&17	&9.6629	 &0.2048 & 0.1212	        \\ \hline
1WIT	&$\beta$ 	&93	&489	&17	&10.5161 &0.0884 & 0.4072	        \\ \hline
2AIT	&$\beta$ 	&74	&374	&17	&10.1081 &0.1827 & 0.437	        \\ \hline
3MEF	&$\beta$ 	&69	&316	&15	&9.1594	 &0.3359 &0.3133	        \\ \hline \hline

1APS	&$\alpha\beta$ 	&98	&489	&16	&9.9796	 &0.193  & 0.4822              \\ \hline
1CIS	&$\alpha\beta$ 	&66	&304	&17	&9.2121	 &0.2935 & 0.4823	        \\ \hline
1COA	&$\alpha\beta$ 	&64	&274	&17	&8.5625	 &0.2805 & 0.3432	        \\ \hline
1FKB	&$\alpha\beta$ 	&107	&539	&15	&10.0748 &0.1704 & 0.4269	        \\ \hline
1HDN	&$\alpha\beta$ 	&85	&428	&16	&10.0706 &0.1678 & 0.4305	        \\ \hline
1PBA	&$\alpha\beta$ 	&81	&345	&14	&8.5185	 &0.3228 & 0.2856	        \\ \hline
1UBQ	&$\alpha\beta$ 	&76	&326	&13	&8.5789	 &0.1782 & 0.2977	        \\ \hline
1URN	&$\alpha\beta$ 	&96	&444	&18	&9.25	 &0.3568 & 0.1949	        \\ \hline
1VIK	&$\alpha\beta$ 	&99	&430	&15	&8.6869	 &0.5191 & 0.2061	        \\ \hline
2HQI	&$\alpha\beta$ 	&72	&407	&18	&11.3056 &0.145	 & 0.1623              \\ \hline
2PTL	&$\alpha\beta$ 	&78	&334	&14	&8.5641	 &0.5179 & 0.3125	        \\ \hline
2VIK	&$\alpha\beta$ 	&126	&616	&19	&9.7778	 &0.4144 & 0.464	        \\ \hline

\end{tabular}
\end{center}
\caption[Data table for 30 single-domain two-state folding
proteins. $\alpha$ and $\beta$ class proteins.] 
{Data table for 30 single-domain two-state folding proteins of $\alpha$,
$\beta$, and  $\alpha\beta$ class.}
\label{tab:data01a} 
\end{table}

Proteins are characterised by the covalent backbone
connectivity. Short- as well as long-range contacts are made 
in the process of folding. It is known that short-range contacts are
responsible for well-defined secondary structures such as
$\alpha$-helix and $\beta$-sheets. The structure into which the
protein chain folds and many of its properties hinge upon the
long-range contacts that are made on various `scales', as specified by
the separation distance between the contacting residues.
 
Thus, PCNs are a special class of network systems. 
Small-world property of proteins, as studied in 
Chapter~\ref{chap:protnet02}, is a reflection on the   
compact nature of the protein molecules. Other than that we
investigated various network features of protein contact networks at
different length scales (viz.\ PCN and LIN) in an attempt 
to get a better understanding of its structure, function, and
stability. In this chapter we analyse assortative mixing of PCNs and
LINs.  

\section{Data}
In the earlier study (Chapter~\ref{chap:protnet02}) we had considered
a set of $80$ proteins, $20$  proteins each from four SCOP structural
classifications. Here, we  considered $30$ separate proteins to
study. These $30$ proteins (Table No.~\ref{tab:data_30ptns_A} and
\ref{tab:data_30ptns_B}) were single-domain, two-state folding
proteins. Note: Henceforth, unless and otherwise mentioned, we 
use these $30$ proteins for our analyses, while supplementing it with
results from other data when required. Table~\ref{tab:data01a} gives
the following details:  no. of nodes, $n_r$, no. of contacts, $n_c$, maximum
degree, $k_{max}$ and the average degree $\langle k \rangle$ of the
PCNs, and the network parameters studied in this chapter--the
coefficient of assortativity of PCNs and their LINs, $r_{PCN}$ \&
$r_{LIN}$. 

\section{Degree Distributions}

\begin{figure*}
\begin{center}
\begin{tabular}{cc}
\includegraphics[width=6.5cm]{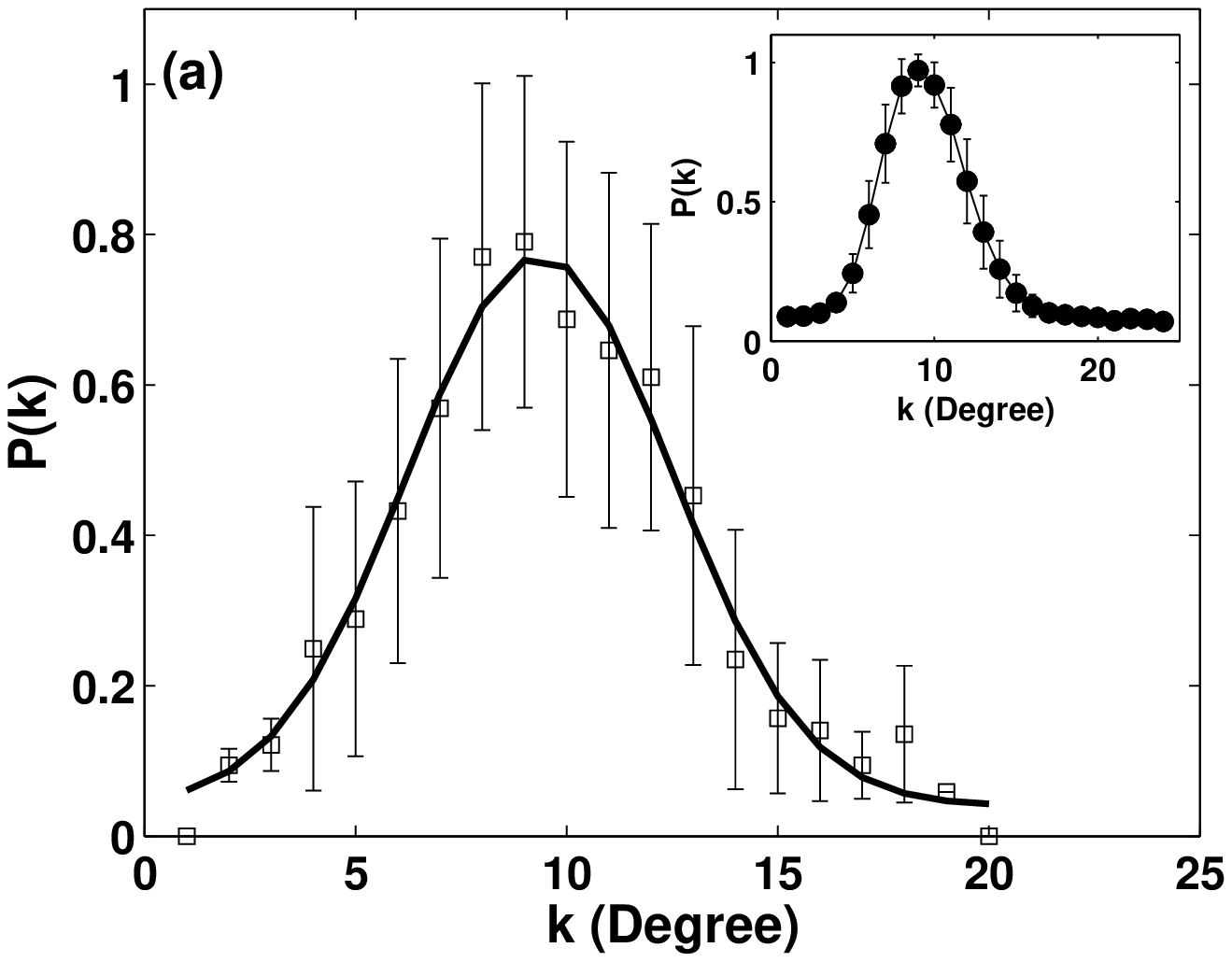} &
\includegraphics[width=6.5cm]{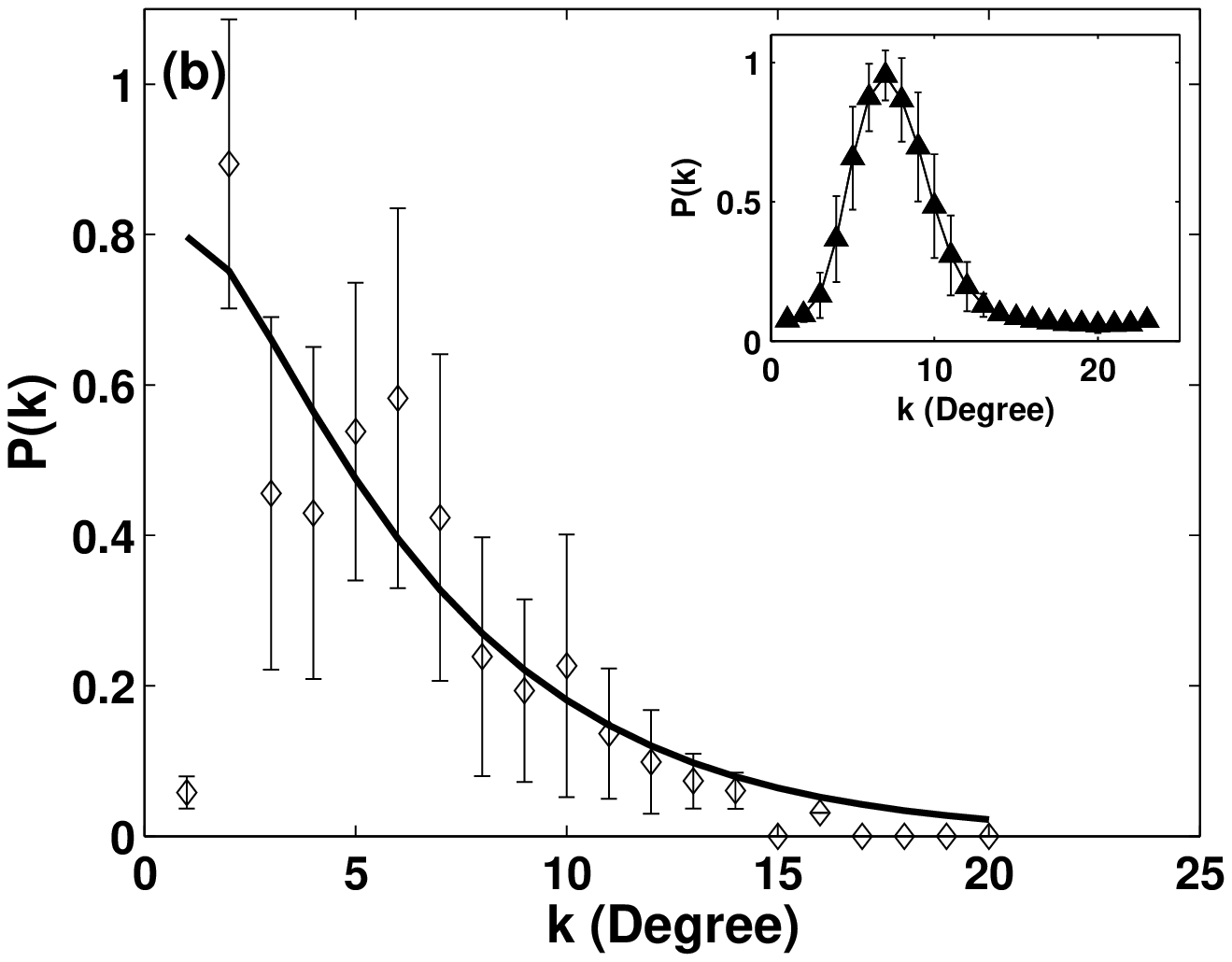}
\end{tabular}
\end{center}
\caption[Normalised degree distributions~$P(k)$ of PCNs,  LINs.]
{Normalised degree distributions~$P(k)$ of (a) PCNs and (b) 
  LINs. Shown in the insets are (a) Type I Random Controls of PCNs
  and (b) their LINs. Thick lines are the best-fit curves for the
  means of the data. Error-bars indicate standard deviation of the
  data for $P(k)$ of nodes with degree $k$ across the $30$ proteins
  analysed.} 
\label{fig:degree_dist}
\end{figure*}

As done with other proteins, we first studied the normalised degree
distributions of PCNs and LINs of these $30$ proteins. As seen in
Fig.~\ref{fig:degree_dist}(a), the PCNs have Gaussian degree
distribution. The parameters and expression with which the best fit
was obtained are: 
$$ y(x) = \frac{A}{w\sqrt{\pi/2}} \exp{\frac{-2(x-x_c)^2}{w^2}}$$ with
$A=5.538$, $w=6.265$, and $x_c=9.373$. 

On the other hand, Fig.~\ref{fig:degree_dist}(b) shows
that the degree distribution of LINs is very different than  
those of PCNs. In LINs, most nodes were populated in the low-degree
region and very few of them have high degrees. The best-fit for the
LINs represents a single-scale exponential function~\cite{protnet:JMB},
$$P(k) \sim k^{-\gamma} \exp{(-k/k_c)},$$ 
with $\gamma=0.24$ and $k_c=4.4$. 

The nodes of degree $1$ in the degree distributions of LINs are the N- and
C-terminal amino acids that are at the either end of the protein
backbone. As expected~\cite{bollobas1981}, the Type I random controls
of the PCNs (Fig.~\ref{fig:degree_dist}(a), inset) have a Poisson
degree distribution. LINs of Type I random controls
(Fig.~\ref{fig:degree_dist}(b), inset) too have a Poisson degree
distribution. The figure clearly shows that these properties are the
same for all the proteins~\cite{protnet:JMB,Bagler2005}.

\section{Assortative nature of PCNs and LINs}
We studied these $30$ single-domain, two-state folding proteins for
the existence of degree-degree correlations in PCNs and LINs.     
We first studied $\langle k_{nn}(k) \rangle$ versus $k$ profiles of
these proteins. As mentioned earlier a trend in degree correlation
profile is a signature of appropriate degree mixing in the network. 

\begin{figure}[!tbh]
\begin{center}
\begin{tabular}{cc}
\includegraphics[width=6.5cm]{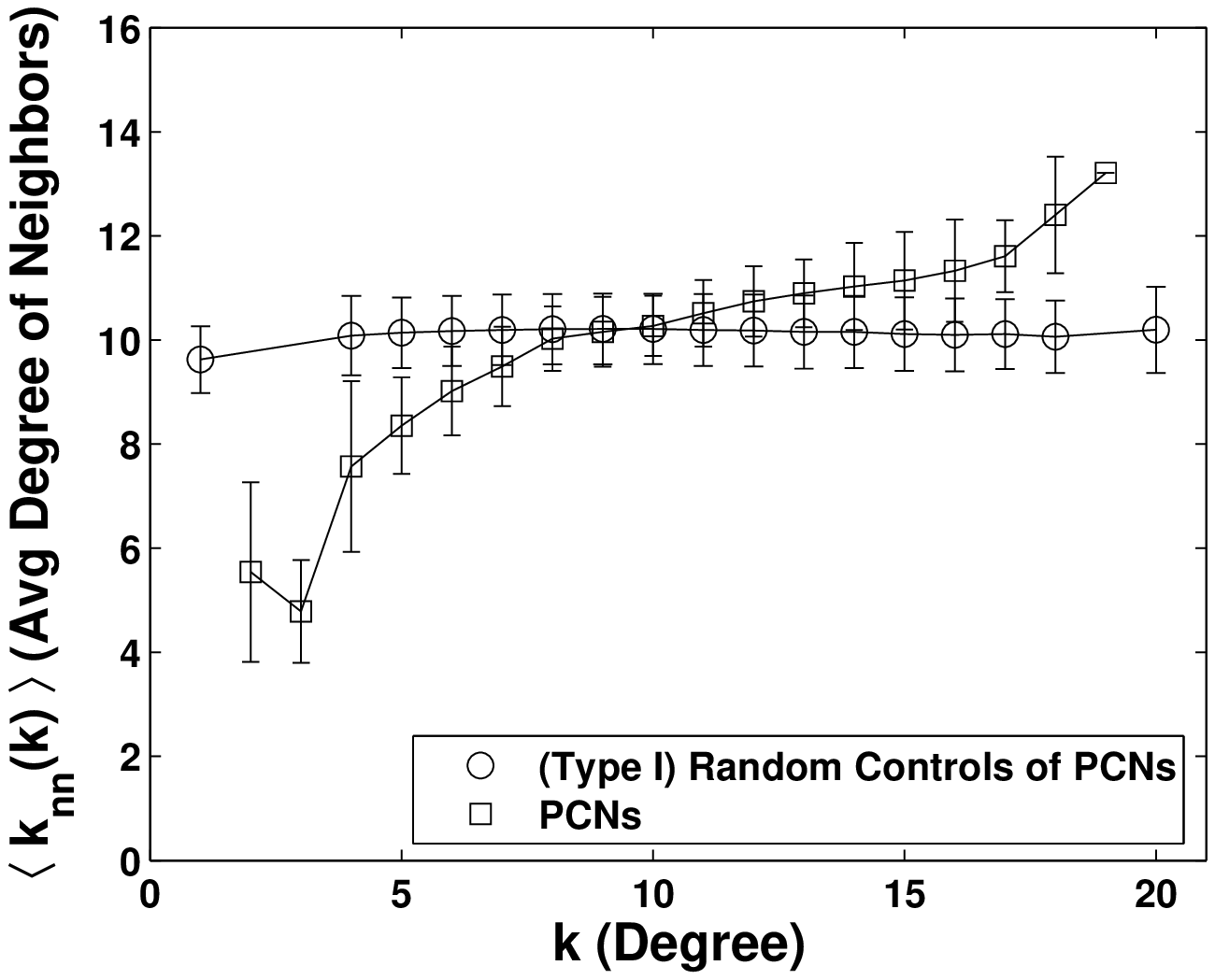} &
\includegraphics[width=6.5cm]{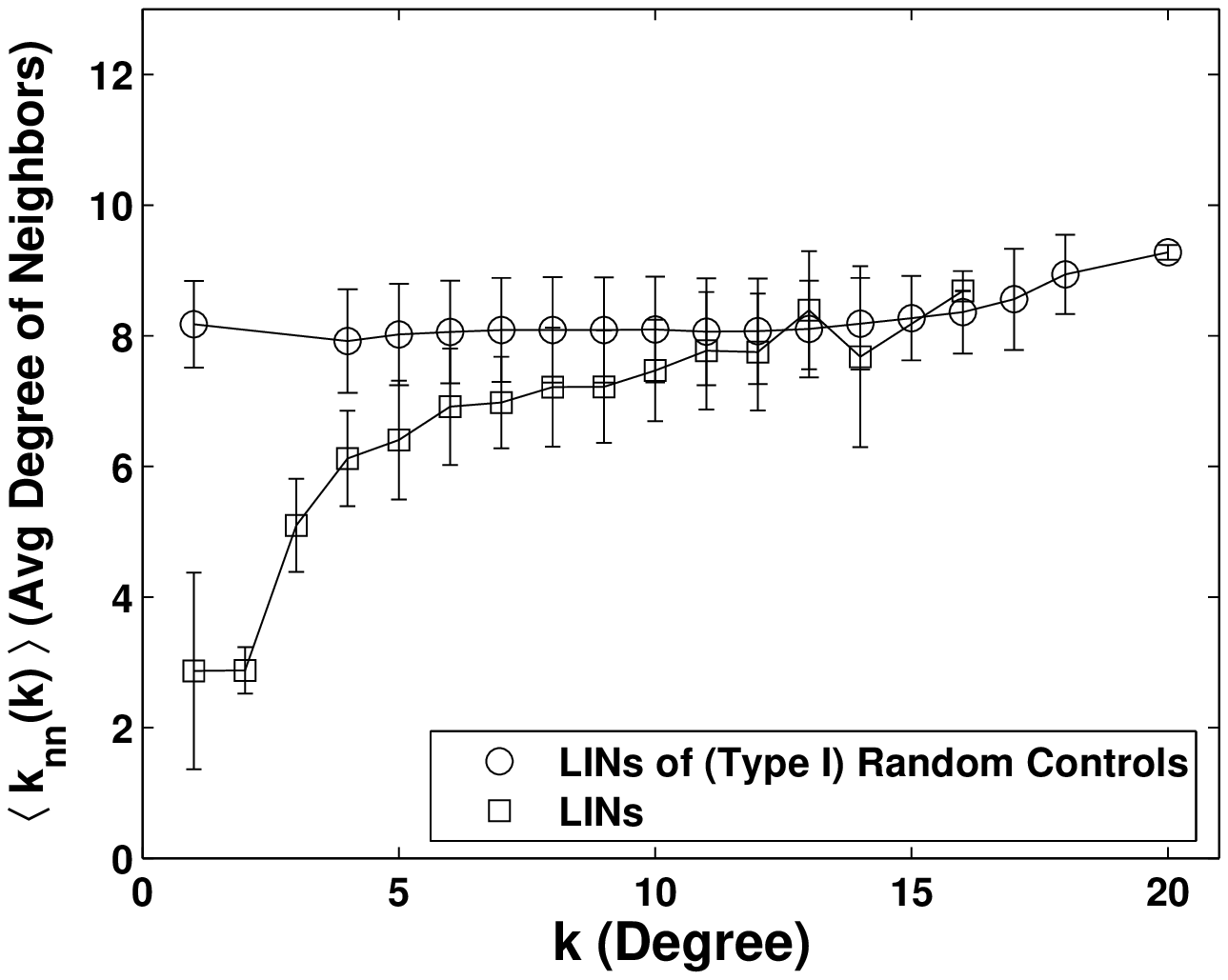} 
\end{tabular}
\end{center}
\caption[Degree correlation pattern comparison between PCN and Type-I Random
Control. Degree correlation pattern comparison between LINs and LINs
of Type-I Random Controls.]
{(a) Degree correlation pattern comparison between PCN and their Type-I Random
Controls. (b) Comparison of degree correlation pattern of LINs and LINs of
Type-I Random Controls.} 
\label{fig:degree_corr01a}
\end{figure}

\begin{figure}[!tbh]
\begin{center}
\begin{tabular}{cc}
\includegraphics[width=6.5cm]{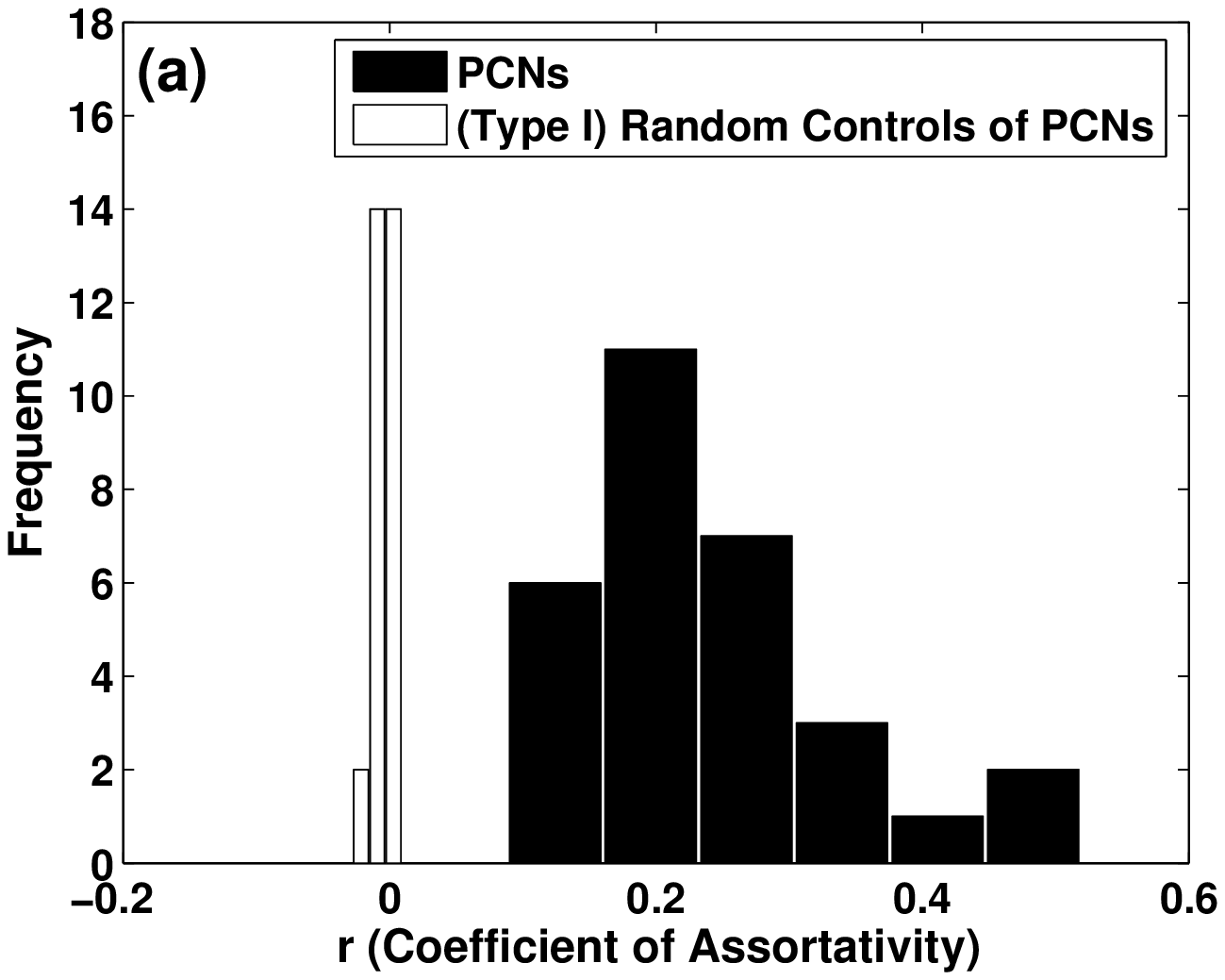} &
\includegraphics[width=6.5cm]{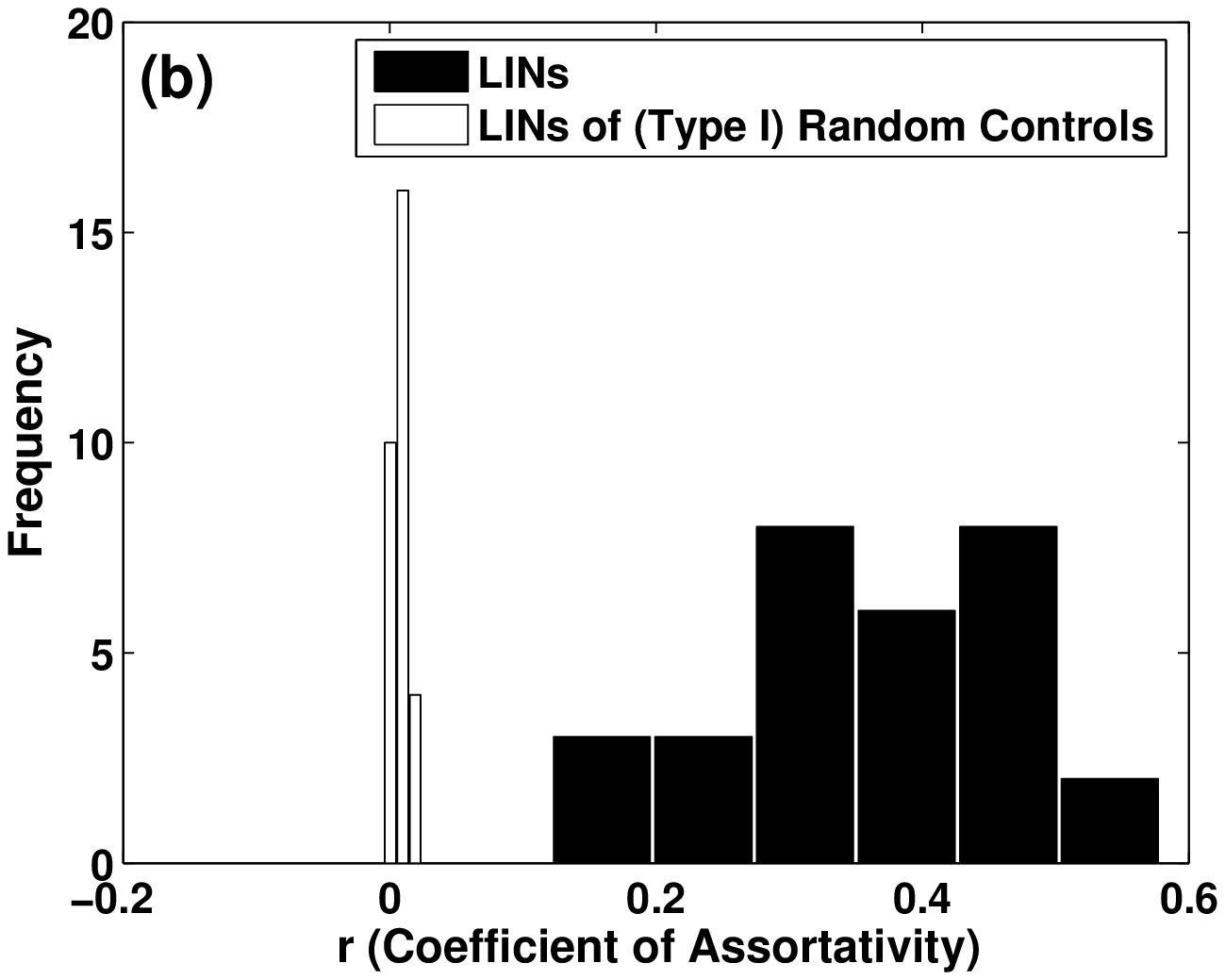} 
\end{tabular}
\end{center}
\caption[Histograms of `Coefficient of Assortativity ($r$)' (with
Type-I Random Controls)]
{Histograms of `Coefficient of Assortativity ($r$)' of 
(a) and (b) PCNs and LINs ($\blacksquare$) and their (Type I) Random
Controls ($\square$).} 
\label{fig:degree_corr02a}
\end{figure}

Fig.~\ref{fig:degree_corr01a} shows $\langle k_{nn}(k) \rangle$ versus
$k$ plots for the  PCNs ($\square$ in Fig.~\ref{fig:degree_corr01a}(a))
and LINs ($\square$ in Fig.~\ref{fig:degree_corr01a}(b)).  The nature
of these curves shows that both PCNs and LINs were characterised with
`assortative mixing', as the average degree of the neighbouring nodes
increased with $k$. In comparison, the $\langle k_{nn}(k) \rangle$
remained almost constant for the Type I random control of
PCNs~($\fullmoon$ in  Fig.~\ref{fig:degree_corr01a}(a)) and
LINs of PCNs~($\fullmoon$ in  Fig.~\ref{fig:degree_corr01a}(b)),
indicating lack of correlations among the nodes' connectivity in these
controls.  Fig.~\ref{fig:degree_corr01a}~(a) and (b) very clearly
brings forth the assortative nature of PCNs as well as their LINs. 

The normalised degree correlation function, $r$, is zero for no
correlations among nodes' connectivity, and positive or negative for
assortative or disassortative mixing, respectively.  
We computed $r_{PCN}$s and $r_{LIN}$s of the proteins (Table
~\ref{tab:data01a}).
The $r$ for both, PCNs and LINs of the $30$ proteins, were found
to be positive, indicating that the networks are assortative. 
Fig.~\ref{fig:degree_corr02a} shows the histograms of $r$ of (a) PCNs,
(b) LINs, and their Type I random controls.  
The $r$ values of both PCNs as well as LINs of all the proteins show
significantly high positive values (range: $0.09<r<0.52$ for  PCNs,
and $0.12<r<0.58$ for LINs).
Thus these naturally-occurring, biological networks, are clearly
characterised by high degree of assortative mixing.  
The Type I random controls in Fig.~\ref{fig:degree_corr02a}~(a \& b),
for both PCNs and their LINs, are distributed around zero,  
confirming the observation of lack of degree correlations of the
controls, made in Fig.~\ref{fig:degree_corr01a}.  

\begin{figure}[tbh]
\includegraphics[scale=0.8]{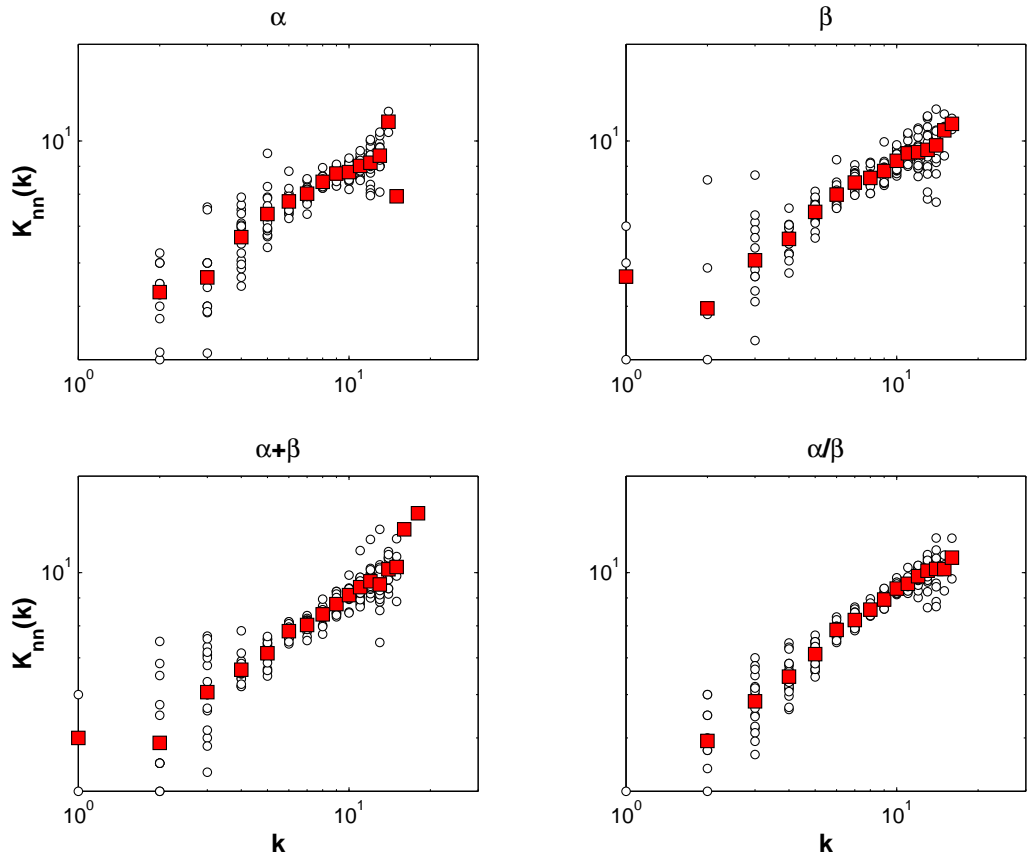}
\caption[Degree correlation pattern for PCNs, showing their
assortative nature for structurally and functionally diverse class of proteins.] 
{Degree correlation pattern of PCNs. Assortative mixing of PCNs. 
The circles ($\circ$) represent $\langle k_{nn}(k) \rangle$ for a give
value of $k$ across all proteins; Filled squares averages of these
values showing the trend of degree correlations.} 
\label{fig:degree_corr03}
\end{figure}

These properties of positive $r$ and assortative degree correlations
were also observed~(See Figure~\ref{fig:degree_corr03}) for a large
number of protein structures used in studies in earlier chapter 
belonging to diverse structural categories.

\section{Degree Distribution partially accounts for assortativity}
\begin{figure}[!tbh]
\begin{center}
\begin{tabular}{cc}
\textbf{(a)}\includegraphics[width=6.5cm]{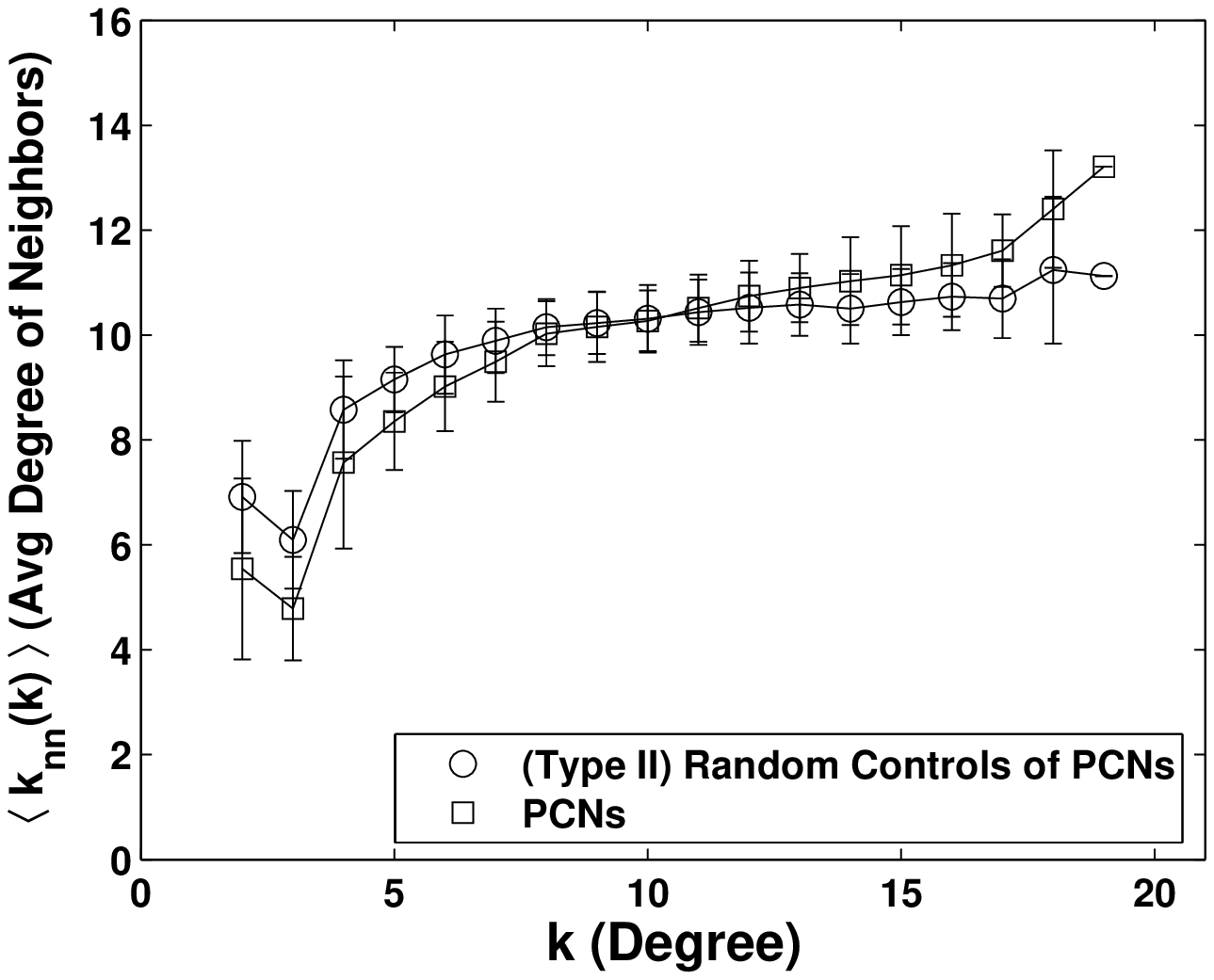} &
\textbf{(b)}\includegraphics[width=6.5cm]{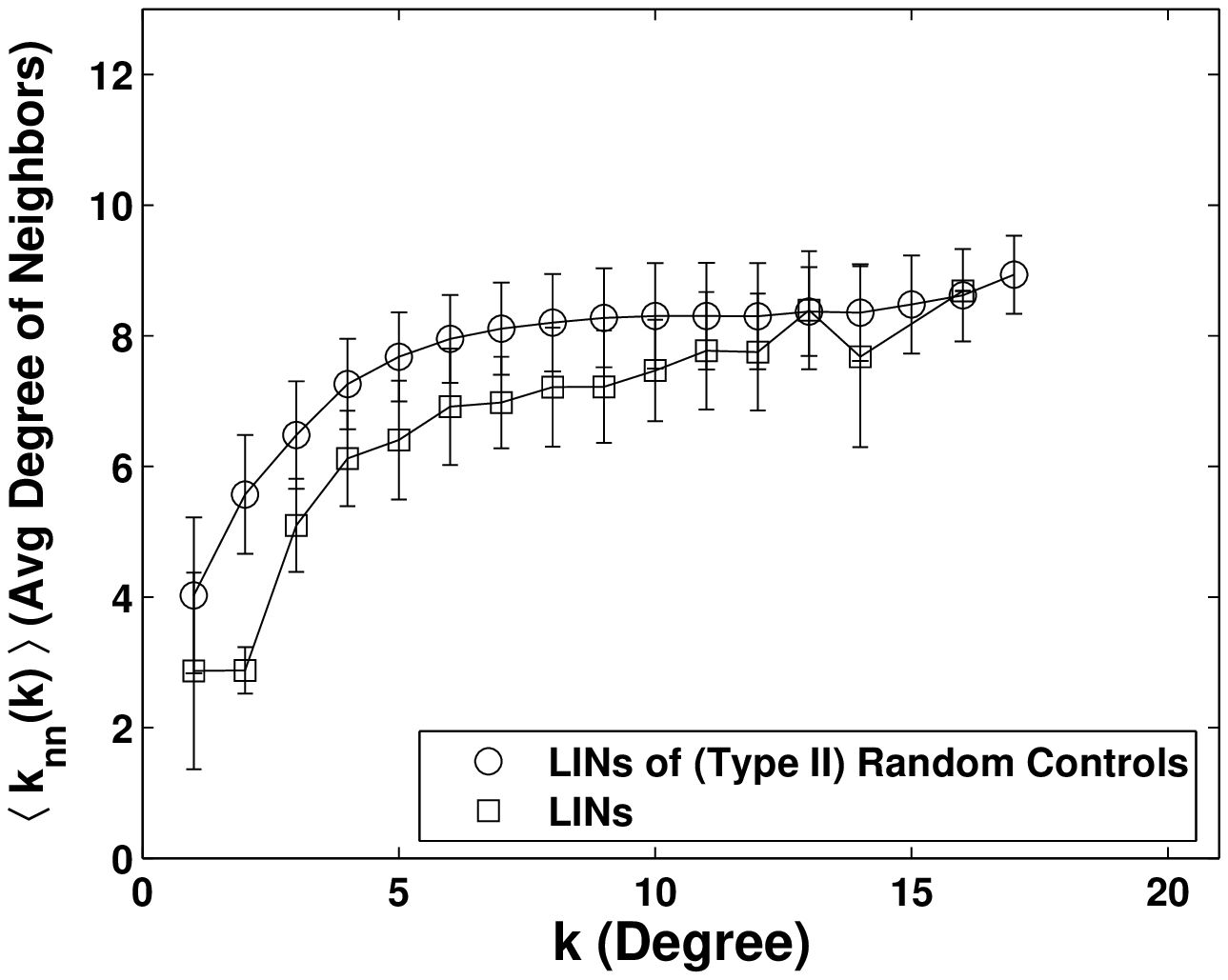} 
\end{tabular}
\end{center}
\caption[Recovery of Degree correlation pattern by Type-II Random
Controls of PCNs as well as LINs of Type-II Random Controls]
{(a) Recovery of Degree correlation pattern by Type-II Random
Controls of PCNs. (b) Recovery of Degree correlation pattern by
LINs of Type-II Random Controls.} 
\label{fig:degree_corr01b}
\end{figure}

\begin{figure}[!tbh]
\begin{center}
\begin{tabular}{cc}
\includegraphics[width=6.5cm]{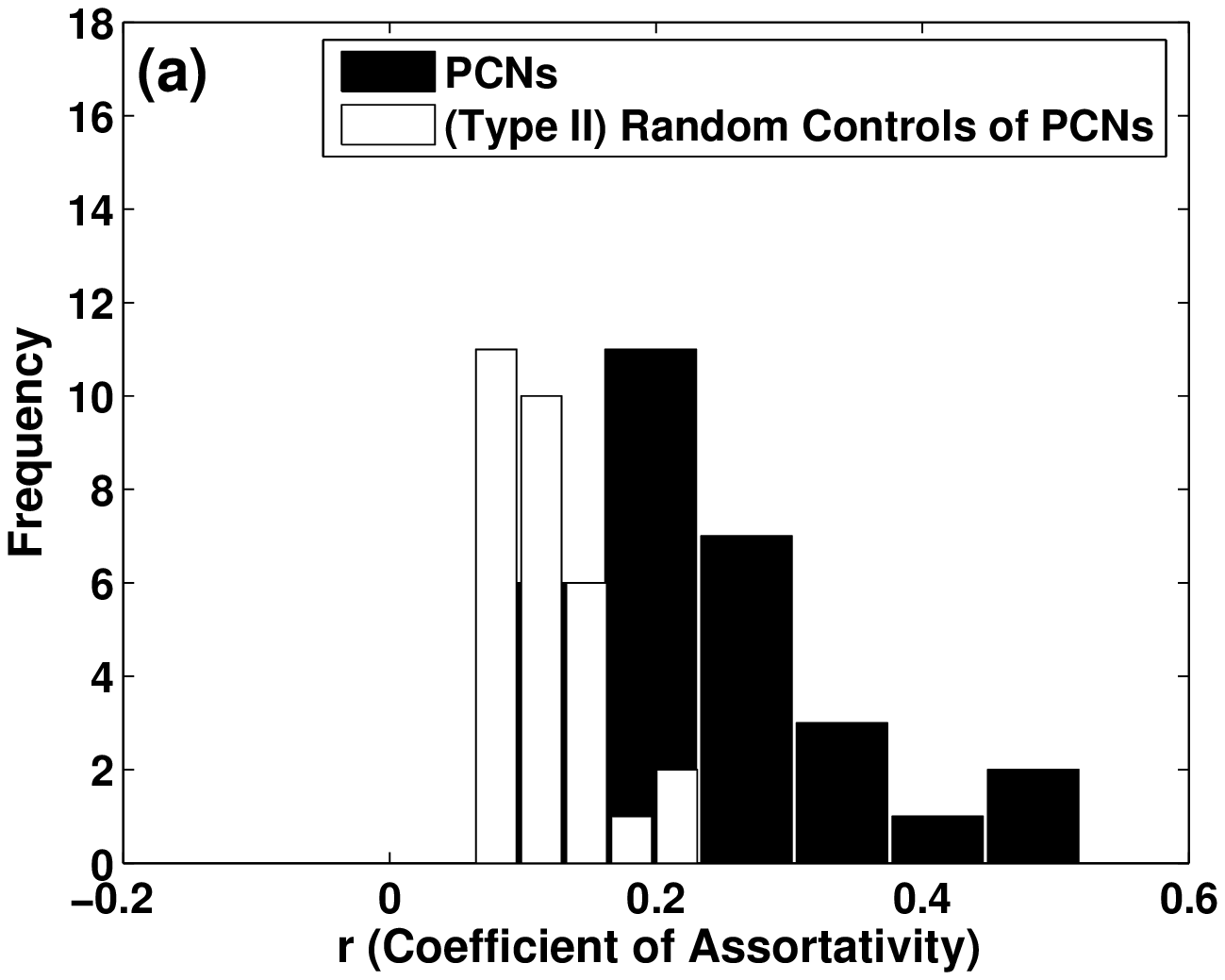} &
\includegraphics[width=6.5cm]{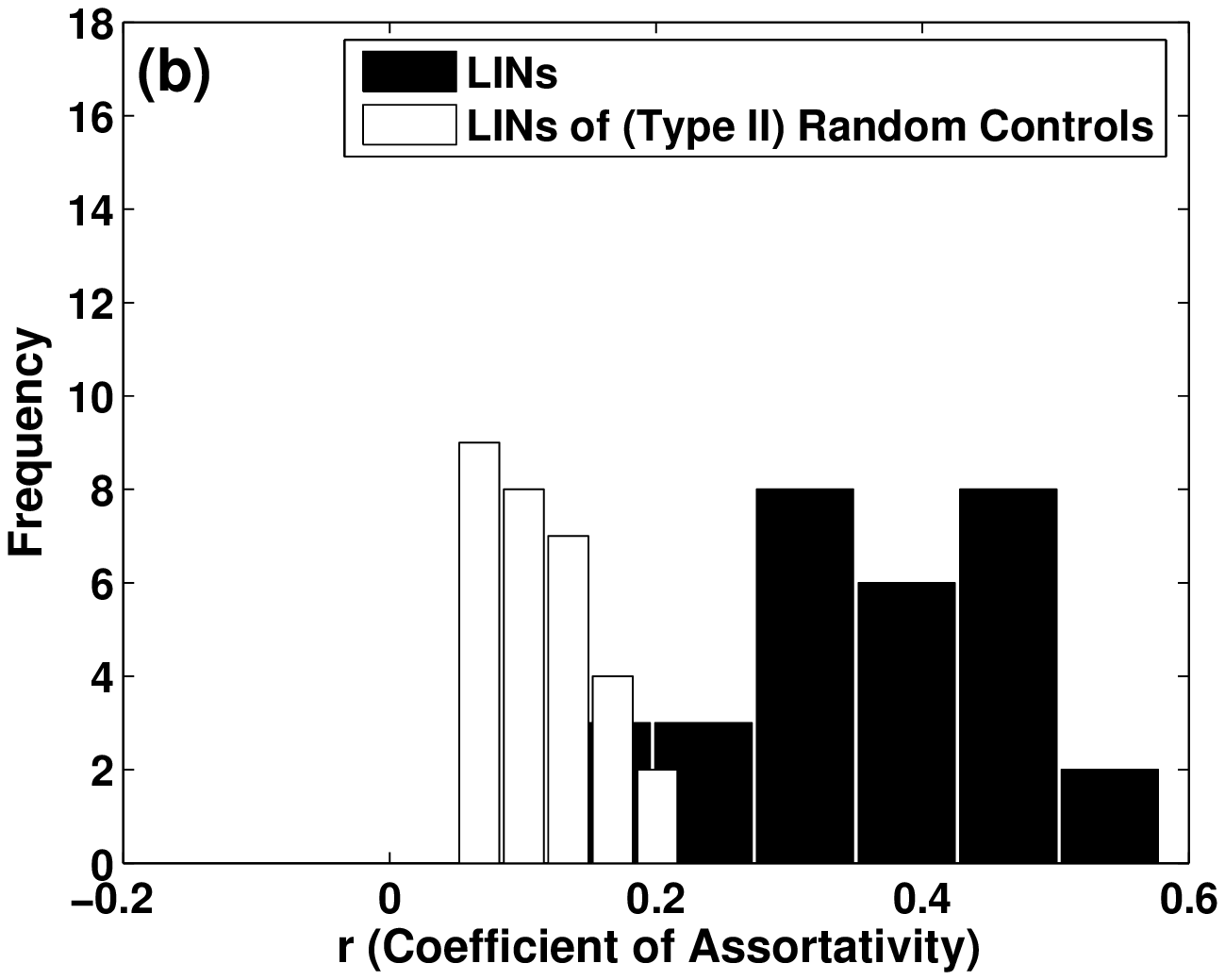} 
\end{tabular}
\end{center}
\caption[Histograms of `Coefficient of Assortativity ($r$)' (with
Type-II Random Controls)]
{Histograms of `Coefficient of Assortativity ($r$)' of 
(a) PCNs ($\blacksquare$) and (b) LINs ($\blacksquare$) and their (Type II) Random
Controls ($\square$).} 
\label{fig:degree_corr02b}
\end{figure}

To investigate the possible role of different network features we
built appropriate controls as discussed in
Subsec.~\ref{chap01:subsec:Random_Controls}.  
Specifically we investigated whether the distribution of degrees has
any effect on observed assortativity in PCNs and LINs. We studied  
the `coefficient of assortativity' of Type II random controls of the
PCNs, in which, apart from having same number of nodes and contacts, 
we also preserved their degree distribution while randomising the
pair-connectivities.  
In Fig.~\ref{fig:degree_corr01b} and Figs.~\ref{fig:degree_corr02b} 
we show that the assortativity is partially recovered in  
the Type II random controls for both PCNs and their LINs. 
Thus degree distribution partially explains the observed assortative
mixing. 
This implies that preserving the degree distribution of PCN, even while
randomising the pair-connectivities, is important in order to partially  
restore the assortative mixing in the random controls of PCNs as well
as their LINs. 
The recovery of assortative mixing in the LINs by Type II random
controls of PCNs is even more surprising, as the degree distribution  
of LINs (Fig.~\ref{fig:degree_dist}(b)) is very different compared to
the PCNs (Fig.~\ref{fig:degree_dist}(a)).  
This is especially significant in the light of the
observation~\cite{brunet_2004pre,Xulvi-Brunet2005} that one can
rewire the links  
in a (scale-free) network to obtain assortativity or disassortativity,
to any degree, without any change in the degree distribution. 

\section{Discussion}

Our coarse-grained complex network model of protein structures
uncovers, for the first time in a naturally evolved biological 
system, the interesting, and exceptional topological feature of
assortativity. The assortative nature is found to be a generic
feature of protein structures. 

Our discovery of assortativity in the amino acid networks in protein
structures questions the invoked generality of disassortativity
property in natural networks. By constructing appropriate random
controls, we show~(Figures.~\ref{fig:degree_corr01b}~(a, b) and
Figures.~\ref{fig:degree_corr02b}(a, b)) that degree distribution can 
partially explain the observed assortative nature of PCNs as  
well their LINs. 
Thus, this novel feature could be a reflection on the mechanisms of
contact formations in proteins while folding that have evolved through 
natural selection. 
An obvious question would be, ``What are the processes by which a
typical protein acquires a Gaussian-like degree distribution?''  

A large number of networks are shown~\cite{reka:thesis} to have
scale-free degree distributions. 
The scale-free distribution, characterised by a power law, \mbox{$P(k)
  \sim k^{-\gamma}$}, with a scaling exponent $\gamma$, is explained
with the help of a growing network model with `preferential attachment
of the nodes,' which are being added to the network~\cite{SFNW:barabasi}. 
In addition to others'~\cite{protnet:JMB,dokholyan_pnas2002}, our
results on the degree distributions (Fig.~\ref{fig:degree_dist})  
also show that the process underlying the formation of the PCNs does
not follow the `preferential attachment' mode. 
This is understandable as the PCNs differ from other networks in many
aspects. PCNs are characterised by covalent backbone connectivity
which constrains the connectivity pattern. As opposed to other
networks, PCNs evolve by changing the connectivity  pattern through
noncovalent contacts, while keeping the number of nodes
constant~\cite{Brinda2005}.  
Also in PCNs, steric hindrance limits the number of contacts an amino
acid can have.  
All this could lead to the observed degree distribution in PCNs.    

From computational studies, it has been
observed~\cite{r:newman,brunet_2004pre} that assortative networks
percolate easily, i.e.,  information gets easily transferred through
the network as compared to that in disassortative networks.  
Protein folding is a cooperative phenomenon, and hence, communication
amongst nodes is essential, so that appropriate noncovalent  
interactions can take place to form the stable native state
structure~\cite{foldon_englander_PNAS2005}. 
Thus percolation of information is very much essential and could lead
to the observed cooperativity and fast folding of the proteins.  
Hence assortative mixing observed in proteins could be an essential
prerequisite for facilitating folding of proteins. 

Disassortative mixing is observed in certain networks of biological
origin such as metabolic signalling pathways network, and  
gene regulatory network~\cite{maslov:science2002}. 
This disassortativity is conjectured to be responsible for decreasing
the likelihood of crosstalk between different functional modules of  
the cell, and increasing the overall robustness of a network by
localising effects of deleterious perturbations.  
In contrast to these two networks, for the PCN one may put forward the
possibility of the backbone chain connectivity as a means of  
conferring greater robustness against perturbations. 
It would also be interesting to study the role of ``community
structure'' in conferring assortativity in these molecular  
networks~\cite{newman:society01,community_vicsek_nature}.

Here we have shown that the assortative mixing in PCNs and LINs
is a generic feature of protein structures. Also the $r$ values
observed are quite high compared to other real-world networks~(See
Tables~\ref{tab:data:ass01} and \ref{tab:data01a}).
It may be pointed out that this is the first instance of the presence
of assortative mixing in a naturally occurring biological network, as
all other networks studied~\cite{r:newman} (except for social
networks) have been shown to exhibit disassortative mixing. The role,
if any, the assortative nature of the protein contact networks may
play in their kinetics of folding process is discussed in the next
chapter. 
\chapter[Correlation of topological parameters to the rate of folding]
{\label{chap:protnet04} Correlation of Topological Parameters to the 
  Rate of Folding of Two-State, Single-Domain Proteins} 

\section{Introduction} 
In the previous chapter (Chapter~\ref{chap:protnet03}), we showed that
proteins, in general, are characterised by assortative mixing. 
Given that, in general, networks are known to be
characterised by disassortative mixing~\cite{r:newman}, it brings
forth an exceptional feature of proteins. It also calls for an
explanation as to the purpose, if any, served by this special
property. In this chapter we seek answer to this question.

In Chapter~\ref{chap:protnet02} we also showed that clustering
coefficients($C$), which enumerate local compactness of PCNs, can not be
used to distinguish proteins from each other (see the $L$--$C$ plot in 
Fig.~\ref{chap02:lcln}). In reality, proteins are unique,   
function-specific and have biophysical properties which distinguish
them despite structural similarities. Clearly the observed trend in
characteristics path length ($L$)and clustering coefficient ($C$) are
generic features that capture their compact nature and 
ease of communication within the structures. This indicates that either
the complex network studies are limited by the coarse-grained approach
to draw conclusions about the specific functionalities of protein or,
that of their residues; 
or we need to look at different parameter(s) that capture relevant
features even at this level of coarse-graining. There is
evidence~\cite{baker99a} to suggest that coarse-graining indeed is a
useful way of simplifying protein structure data while not losing the
relevant biophysical details. Keeping this in view, in this chapter,
we proceed to investigate both the PCNs and their long-range
counterparts (LINs). 

\subsubsection{Rate of folding and Native-state topology}
Though the proteins are comprised of thousands of atoms and hence
potentially millions of inter-atomic interactions are possible,
folding rates and mechanisms appear to be largely determined by the
topology of the native (folded) state~\cite{baker:nature}. 
For network analyses and biophysical comparison we choose proteins
that are structurally and kinetically simple. Hence we use
single-domain, two-state folding proteins for our studies. 
Many geometrical parameters viz.\ Contact Order (CO)~\cite{co},
Long-range Order (LRO)~\cite{lro}, Total Contact Distance
(TCD)~\cite{tcd}, that have been defined based on the native-state
structure of the protein have been shown to have negatively correlated
with the rate of folding ($ln(k_F)$). Contact Order, as well as LRO
and TCD which are its variants, essentially measure the average
sequence separation between residues that make contacts in the 3-D
structure. The correlation is remarkable given that it holds over a
million-fold range of folding rates and for diverse structures. 
The observed negative correlation has been explained in terms of
increased time needed to span the conformational space with increase
in the value of these parameters. It is a reasonable explanation given
that all these parameters enumerate the average normalised separation
between residues those are in `spatial contact' in the protein's
native-state structure.

This evidence was one of the two reasons we conjectured that our
topology-based parameters may have bearing on rate of folding. 
The other was that our novel property of assortative mixing is
independent of short-range contacts as shown in
Chapter~\ref{chap:protnet03}. Zhou and Zhou~\cite{tcd} reported that
``the accuracy of total contact distance in predicting folding rates
is essentially unchanged if `short'-ranged contacts ($|i-j| \leq 14$)
are not included in the calculations''. Given their observation we
proceeded to check if the assortativity coefficient ($r$) could have a
bearing on rate of folding.  

\section{Data}
The analyses was conducted with $30$ single-domain, two-state folding,
globular proteins listed in Table Nos.~\ref{tab:data_30ptns_A} and
\ref{tab:data_30ptns_B}. Table~\ref{tab:rate_of_folding01}
provides the data for Figure~\ref{fig:lc_LRI}. 
Table~\ref{tab:rate_of_folding02} provides the data for
Figures~\ref{fig:lnkr_r01a}, \ref{fig:lnkr_r01}, \ref{fig:lnkr_c01a},
and \ref{fig:lnkr_c01}.  In Table~\ref{tab:rate_of_folding01} are
listed the PDB IDs of the $30$ single-domain two-state folding
proteins that have been used in this study and the $L$ and $C$ of
their corresponding PCNs and LINs. Table~\ref{tab:rate_of_folding02}
lists the coefficient of assortativity and clustering coefficients of
PCNs and LINs as well as the rate of folding of these proteins.

\begin{table}[!htp]
\begin{center}
\begin{tabular}{|c|c|c|c|c|c|} \hline \hline
PDB~ID&    Class&       $L_{PCN}$&$C_{PCN}$& $L_{LIN}$& $C_{LIN}$\\ \hline \hline
1HRC&	$\alpha$&	3.427&	0.5816&	5.5174&	0.2713\\
1IMQ&	$\alpha$&	3.0711&	0.6027&	5.2774&	0.1902\\
1YCC&	$\alpha$&	3.4574&	0.5693&	5.4283&	0.2766\\
2ABD&	$\alpha$&	3.22&	0.5882&	4.5447&	0.2289\\
2PDD&	$\alpha$&	2.4352&	0.6171&	4.9767&	0.1612\\ \hline

1AEY&	$\beta$&	2.5529&	0.6147&	3.6122&	0.3503\\
1CSP&	$\beta$&	2.7291&	0.5954&	3.8471&	0.3271\\
1MJC&	$\beta$&	2.7715&	0.5933&	4.0273&	0.3333\\
1NYF&	$\beta$&	2.5983&	0.5987&	3.9734&	0.3457\\
1PKS&	$\beta$&	2.7421&	0.5855&	3.9537&	0.3646\\
1SHF&	$\beta$&	2.6236&	0.6059&	4.2303&	0.3628\\
1SHG&	$\beta$&	2.5175&	0.5949&	3.6028&	0.3824\\
1SRL&	$\beta$&	2.5156&	0.5887&	3.6338&	0.382\\
1TEN&	$\beta$&	3.2824&	0.5738&	4.3011&	0.4309\\
1TIT&	$\beta$&	3.0904&	0.552&	4.2515&	0.4519\\
1WIT&	$\beta$&	3.1346&	0.5753&	3.9589&	0.4488\\
2AIT&	$\beta$&	2.8297&	0.5922&	3.726&	0.4396\\
3MEF&	$\beta$&	2.7626&	0.5952&	3.9757&	0.3136 \\\hline

1APS&	$\alpha\beta$&	3.1273&	0.5676&	3.9405&	0.4158\\
1CIS&	$\alpha\beta$&	2.8154&	0.5892&	3.8424&	0.349\\
1COA&	$\alpha\beta$&	2.8358&	0.5768&	3.9559&	0.3901\\
1FKB&	$\alpha\beta$&	3.319&	0.5821&	4.5636&	0.3626\\
1HDN&	$\alpha\beta$&	2.8538&	0.5534&	3.779&	0.3243\\
1PBA&	$\alpha\beta$&	3.2256&	0.5859&	4.7802&	0.2801\\
1UBQ&	$\alpha\beta$&	3.0996&	0.6074&	4.9846&	0.3243\\
1URN&	$\alpha\beta$&	3.2529&	0.5864&	4.5465&	0.2971\\
1VIK&	$\alpha\beta$&	3.6199&	0.5849&	5.2247&	0.3138\\
2HQI&	$\alpha\beta$&	2.5767&	0.5928&	3.3521&	0.3935\\
2PTL&	$\alpha\beta$&	3.972&	0.599&	6.982&	0.2541\\
2VIK&	$\alpha\beta$&	3.4179&	0.565&	4.575&	0.3095\\ \hline
\end{tabular}
\end{center}
\caption[The PDB IDs, $L$ and $C$ values for PCN and LIN of $30$
single-domain two-state folding proteins of $\alpha$,
  $\beta$, and $\alpha\beta$ class of proteins.] 
{The PDB IDs, $L$ and $C$ values for PCN and LIN of $30$ single-domain
  two-state folding proteins of $\alpha$, $\beta$, and $\alpha\beta$ class of proteins. }
\label{tab:rate_of_folding01}
\end{table}

\begin{table}[htp]
\begin{center}
\begin{tabular}{|c|c|c|c|c|c|c|c|} \hline \hline
PDB~ID&    Class&     $r_{PCN}$&   $C_{PCN}$&   $r_{LIN}$& $C_{LIN}$&    $ln(k_F)$& Ref.\\ \hline \hline

1HRC&	$\alpha$&	0.1821&	0.5816&	0.3531&	0.2713&	8.76& \cite{Chan1997} \\
1IMQ&	$\alpha$&	0.2586&	0.6027&	0.3675&	0.1902&	7.31& \cite{Ferguson1999}\\
1YCC&	$\alpha$&	0.2449&	0.5693&	0.3379&	0.2766&	9.62& \cite{Mines1996}\\
2ABD&	$\alpha$&	0.2874&	0.5882&	0.2736&	0.2289&	6.55& \cite{Kragelund1995}\\
2PDD&	$\alpha$&	0.1436&	0.6171&	0.2616&	0.1612&	9.8& \cite{Spector1999}\\ \hline

1AEY&	$\beta$&	0.1145&	0.6147&	0.3133&	0.3503&	2.09& \cite{Viguera1994,Viguera1996}\\
1CSP&	$\beta$&	0.2929&	0.5954&	0.4822&	0.3271&	6.98& \cite{Perl1998}\\
1MJC&	$\beta$&	0.3027&	0.5933&	0.4823&	0.3333&	5.24& \cite{lro}\\
1NYF&	$\beta$&	0.1752&	0.5987&	0.3432&	0.3457&	4.54& \cite{Plaxco1998a}\\
1PKS&	$\beta$&	0.1872&	0.5855&	0.4269&	0.3646&	-1.05&\cite{Guijarro1998}\\
1SHF&	$\beta$&	0.1511&	0.6059&	0.4305&	0.3628&	4.55& \cite{Plaxco1998}\\
1SHG&	$\beta$&	0.1503&	0.5949&	0.2856&	0.3824&	1.41& \cite{Viguera1996}\\
1SRL&	$\beta$&	0.2101&	0.5887&	0.2977&	0.382&	4.04& \cite{Grantcharova1997}\\
1TEN&	$\beta$&	0.1645&	0.5738&	0.1949&	0.4309&	1.06& \cite{Clarke1997}\\
1TIT&	$\beta$&	0.2048&	0.552&	0.2061&	0.4519&	3.47& \cite{Fowler2001}\\
1WIT&	$\beta$&	0.0884&	0.5753&	0.1623&	0.4488&	0.41& \cite{Clarke1999}\\
2AIT&	$\beta$&	0.1827&	0.5922&	0.3125&	0.4396&	4.2& \cite{Schonbrunner1997}\\
3MEF&	$\beta$&	0.3359&	0.5952&	0.464&	0.3136&	5.3& \cite{lro}\\\hline

1APS&	$\alpha\beta$&	0.193&	0.5676&	0.2829&	0.4158&	-1.48& \cite{Nuland1998}\\
1CIS&	$\alpha\beta$&	0.2935&	0.5892&	0.384&	0.349&	3.87& \cite{lro}\\
1COA&	$\alpha\beta$&	0.2805&	0.5768&	0.4115&	0.3901&	3.87& \cite{Jackson1991}\\
1FKB&	$\alpha\beta$&	0.1704&	0.5821&	0.4006&	0.3626&	1.46& \cite{Main1999}\\
1HDN&	$\alpha\beta$&	0.1678&	0.5534&	0.3326&	0.3243&	2.7& \cite{Nuland1998a}\\
1PBA&	$\alpha\beta$&	0.3228&	0.5859&	0.5789&	0.2801&	6.8& \cite{Villegas1995}\\
1UBQ&	$\alpha\beta$&	0.1782&	0.6074&	0.4414&	0.3243&	7.33& \cite{Khorasanizadeh1996}\\
1URN&	$\alpha\beta$&	0.3568&	0.5864&	0.4433&	0.2971&	5.76& \cite{Silow1997}\\
1VIK&	$\alpha\beta$&	0.5191&	0.5849&	0.5649&	0.3138&	6.8& \cite{Choe1998}\\
2HQI&	$\alpha\beta$&	0.145&	0.5928&	0.1212&	0.3935&	0.18& \cite{tcd}\\
2PTL&	$\alpha\beta$&	0.5179&	0.599&	0.4072&	0.2541&	4.1& \cite{Scalley1997}\\
2VIK&	$\alpha\beta$&	0.4144&	0.565&	0.437&	0.3095&	6.8& \cite{Choe1998}\\\hline
\end{tabular}
\end{center}
\caption[The $r$, $C$, of PCNs and LINs, and the corresponding rate of folding
  $ln(k_F)$ for 30 single-domain two-state folding proteins. $\alpha$,
  $\beta$, and  $\alpha\beta$ class of proteins.]
{The $r$, $C$, of PCNs and LINs, and the corresponding rate of folding
  $ln(k_F)$ for 30 single-domain two-state folding proteins. $\alpha$,
  $\beta$, and  $\alpha\beta$ class of proteins.} 
\label{tab:rate_of_folding02}
\end{table}

\section{Clustering Coefficients of PCNs and LINs}
\begin{figure}[!tbh]
\begin{center}
\includegraphics[scale=0.6]{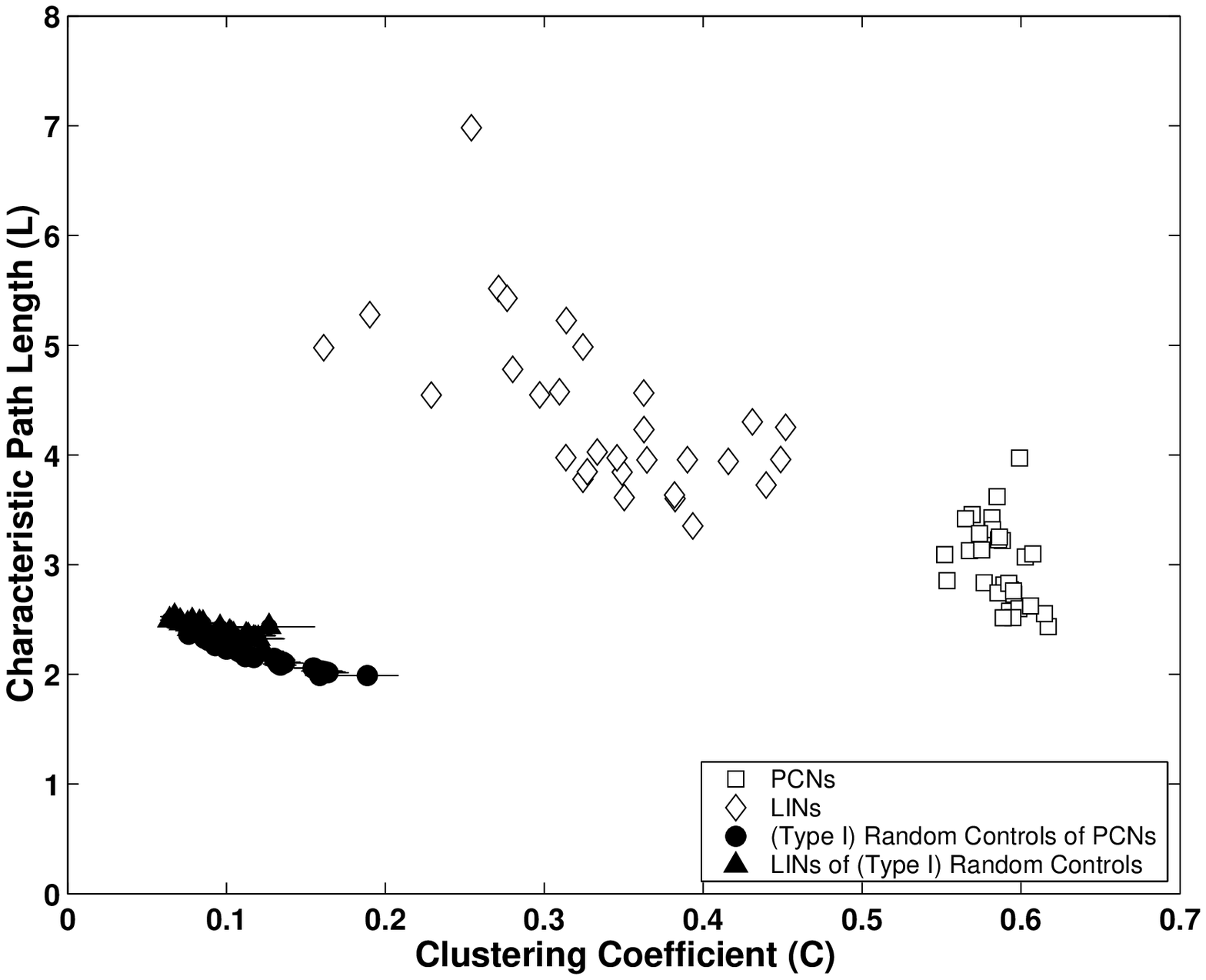}
\end{center}
\caption[L-C plot for 30 single domain, two-state proteins: PCNs,
LINs, Type I Random Controls of PCNs their and LINs.] 
{L-C plot for 30 single domain, two-state proteins: PCNs ($\square$),
  LINs ($\Diamond$), Type I Random  
Controls of PCNs ($\newmoon$) and LINs($\blacktriangle$). 
Error-bars in the random controls data indicate standard deviations in
$L$ and $C$ for each protein computed over $100$ instances.} 
\label{fig:lc_LRI}
\end{figure}

As shown in Chapter~\ref{chap:protnet02}, here also we studied the
$L$--$C$ properties of the $30$ proteins under consideration.  
To study if the PCNs of the $30$ proteins and their corresponding LINs
have similar topological properties such as, characteristic  
path length ($L$) and clustering coefficient ($C$), we plotted the
data of $L$ and $C$ from Table~\ref{tab:rate_of_folding01} in
Fig.~\ref{fig:lc_LRI}.   
The plot shows their corresponding Type I random controls. 
The Type II random controls were found to be indistinguishable from
the Type I controls and not shown in Fig.~\ref{fig:lc_LRI}. 

The results indicate two major differences between the topological
properties of the PCNs and their corresponding LINs. 
The PCNs of these proteins have high clustering coefficients
($C>0.55$) compared to their random controls, whereas the LINs show  
distribution in $C$ over a range ($0.16$ to $0.45$) even though their
random counterparts were almost indistinguishable from those of
PCNs. $L$ and $C$ of random controls of PCNs were $2.168\pm0.11$ \&
$0.1224\pm0.0284$ and that of their LINs were $2.395\pm0.0699$ \&
$0.0942\pm0.0178$. 
The LINs also have marginally higher characteristic path lengths
($4.379\pm0.7677$) than PCNs ($3\pm0.371$) owing to their reduced
number of contacts as compared to those in PCNs. 

Notice that these differences in $C_{LIN}$s compared to that in
$C_{PCN}$s assign specificity to the network models of proteins which
is other is otherwise missing in $C_{PCN}$s.

\section{Correlation of protein network parameters to protein folding rates}
We have shown that even though the PCNs and their LINs differ in their
clustering coefficients ($C$) (Fig.~\ref{fig:lc_LRI}), both show  
high coefficient of assortativity
($r$)~(Table~\ref{tab:rate_of_folding02}) with $r_{PCN}$ being
marginally lower ($0.2412\pm0.1082$) as compared to $r_{LIN}$ ($0.36\pm0.1102$). 
We now study the correlation of the network parameters to the rate of
folding $ln(k_F)$ of the corresponding proteins
(Table~\ref{tab:rate_of_folding02}). 

\subsection{Coefficient of Assortativity and Rate of Folding}
\begin{figure}
\begin{center}
\includegraphics[scale=0.53]{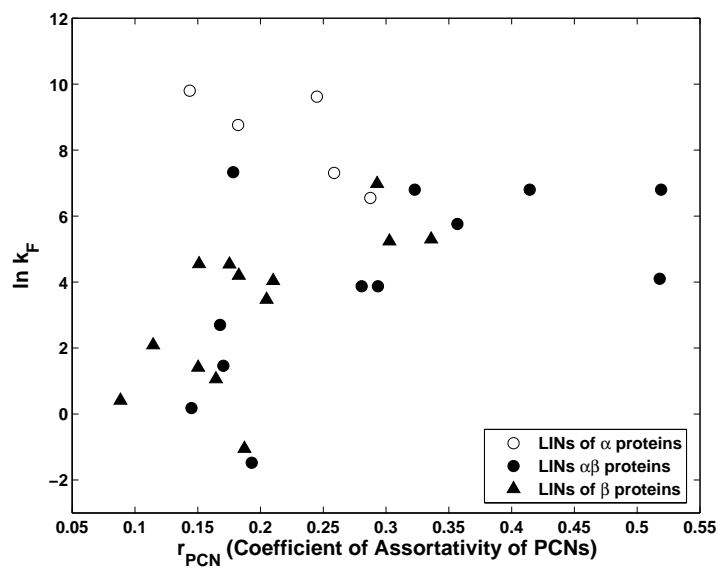}
\end{center}
\caption[Plot of the rate of folding ($ln(k_F)$) versus the assortativity
  coefficient of PCNs ($r_{PCN}$) of $30$ proteins.] 
{Plot of the rate of folding ($ln(k_F)$) versus the assortativity
  coefficient of PCNs ($r_{PCN}s$) of $30$ proteins.}
\label{fig:lnkr_r01a}
\end{figure}

Figure~\ref{fig:lnkr_r01a} shows the plot of $r_{PCN}$ with $ln(k_F)$. As
seen in the figure, though there is a positive trend in the data, the
correlation is poor. We find that the correlation coefficient for PCNs
to be $0.3776$ ($p<0.04$). The correlation becomes better ($0.5943$;
$p<0.005$) after the five $\alpha$ proteins are not considered.

\begin{figure}
\begin{center}
\includegraphics[scale=0.68]{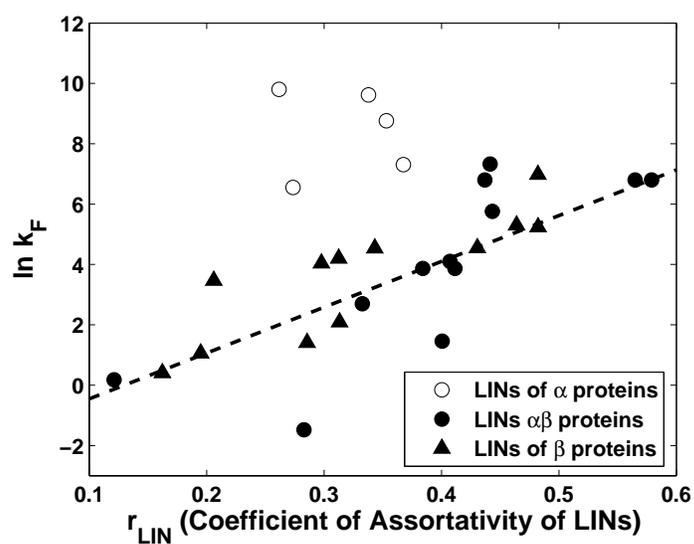}
\end{center}
\caption[Plot of rate of folding, $ln(k_F)$,
and the Coefficient of Assortativity of LINs~($r_{LIN}$) of the PCNs
of the $30$ proteins.] 
{Plot of rate of folding, $ln(k_F)$,
and the Coefficient of Assortativity of LINs~($r_{LIN}$) of the PCNs
of the $30$ proteins. The trend-line is shown as a dashed line.}
\label{fig:lnkr_r01}
\end{figure}

In Fig.~\ref{fig:lnkr_r01}, the rate of folding of the 30 proteins are
plotted as a function of the coefficient of assortativity of  
their LINs. 
There is an increasing trend of $\ln(k_F)$ with increase in $r$
(correlation coefficient=$0.4221$; $p<0.02016$). 
Though $\beta$ and $\alpha\beta$ proteins show an increasing trend,
the $5$ $\alpha$ proteins have high $ln(k_F)$ values.
The correlation coefficient between the rate of folding ($ln(k_F)$) and
$r$ of their LINs, excluding the five $\alpha$ proteins,  
is $0.6981$ ($p<0.0005$). 
This implies that, along with showing assortative mixing, the
PCNs and particularly their LINs show significant positive correlations  
with the rate of folding. 
Thus the generic property of assortative mixing in proteins tend to
contribute positively towards their kinetics of folding and is  
fairly independent of the short and long range of interactions.

\begin{figure}
\begin{center}
\includegraphics[scale=0.53]{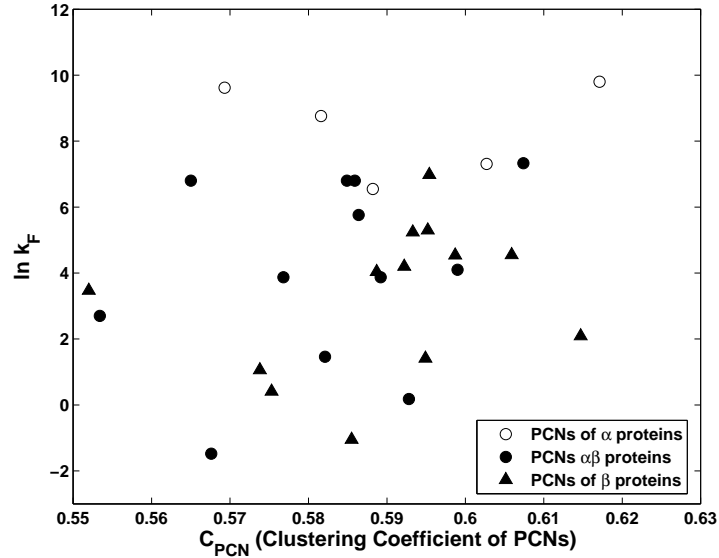}
\end{center}
\caption[Plot of the rate of folding, $ln(k_F)$  with Clustering
Coefficient of PCNs~($C_{PCN}$).] 
{Plot of the rate of folding, $ln(k_F)$  with Clustering Coefficient
  of PCNs~($C_{PCN}$).} 
\label{fig:lnkr_c01a}
\end{figure}

\begin{figure}
\begin{center}
\includegraphics[scale=0.68]{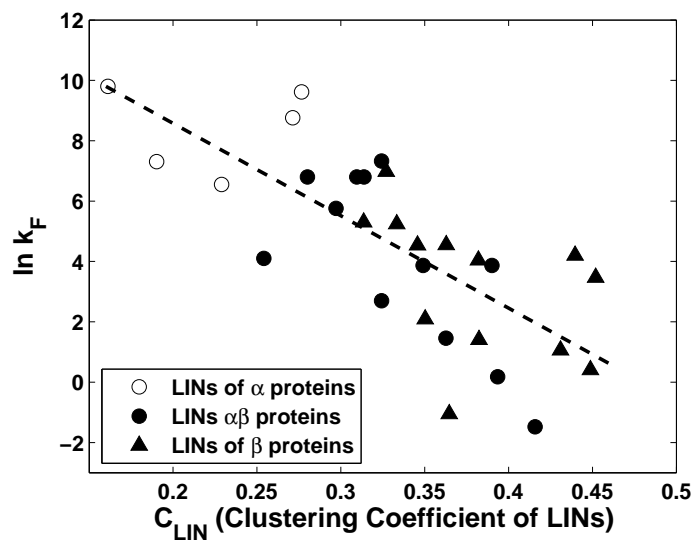}
\end{center}
\caption[Plot of rate of folding, $ln(k_F)$, with Clustering
Coefficient of LINs~($C_{LIN}$). ]
{Plot of rate of folding, $ln(k_F)$, with Clustering Coefficient of
  LINs~($C_{LIN}$). The trendline is indicated by a dashed line.} 
\label{fig:lnkr_c01}
\end{figure}

\begin{figure}[!htb]
\begin{center}
\includegraphics[scale=0.71]{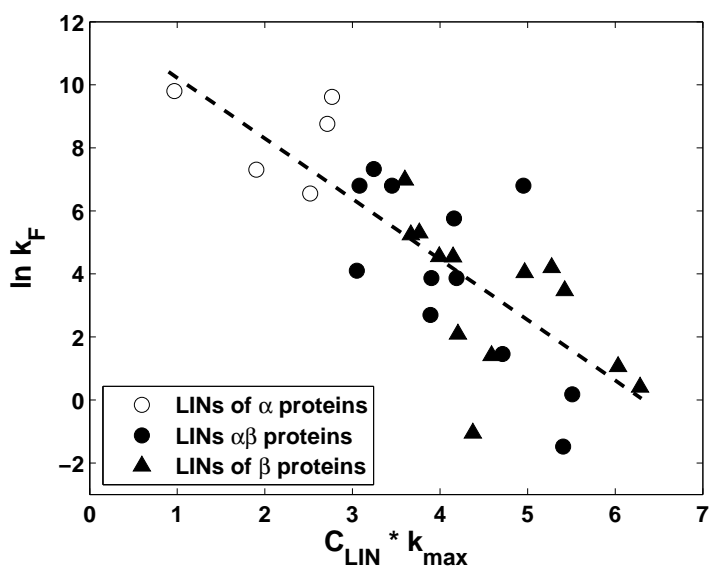}
\end{center}
\caption[The plot of $ln(k_F)$, with $C_{LIN}*k_{max}$.]
{The plot of $ln(k_F)$, with $C_{LIN}*k_{max}$.} 
\label{fig:lnkr_c01b}
\end{figure}

\subsection{Average Clustering Coefficient and Rate of Folding}
Figure~\ref{fig:lnkr_c01a} shows the plot of $ln(k_F)$ with the
clustering coefficient of the PCNs ($C_{PCN}$) of the $30$
proteins. As is obvious from the plot there is hardly any correlation
between the two parameters. This is borne out by the correlation
coefficient that we compute as $-0.2437$ ($p<0.2$)

Figure~\ref{fig:lnkr_c01} shows the plot of $ln(k_F)$ with the
clustering coefficient of the LINs ($C_{LIN}$). 
The $ln(k_F)$ show high negative correlation (corr. coeff.
$=-0.7337$; $p<0.0001$) with the $C_{LIN}$ for all the  
proteins. 
 
$C_{LIN}$ enumerates number of triads made among the nodes of the
Long-range Interaction Network.  
Thus $C_{LIN}$ essentially correlates to the number of `distant' amino
acids (nodes), separated by a minimum of $12$ or more  
other amino acids along the backbone, brought in mutual `contact' with
each other in the native state structure of the protein.  
Understandably, more the number of such long-range mutual contacts 
required to be made in order to achieve the native state,  
more is the time taken to fold, and hence slower is the rate of folding.

The calculation of clustering coefficient is dependent on the degree
of the node. In Figure~\ref{fig:lnkr_c01b} we scale $C_{LIN}$s with $k_{max}$ as
$C_{LIN}*k_{max}$ and plot it with $ln(k_F)$. We find that, after
scaling, the correlation coefficient improves to $-0.7712$ ($p<0.0001$).
Thus $C_{LIN}$s show significantly negative correlation with
single-domain, two-state folding proteins--a property completely
neutralised in PCNs.

\section{Discussion}
Here we have studied two important network parameters ($C$ and $r$) of
$30$ single-domain two-state folding proteins at two
length-scales--PCNs and the LINs. The results show that even though
the PCNs show ``small-world'' property in their $L$--$C$ plot, the
$C_{LIN}$s have comparatively low values and are distributed over a
range between the random controls and PCNs. 
We have studied the correlation of two widely-used topological parameters
(assortativity and clustering) to the kinetics of folding at different
length scales.  

A `positive' correlation of $ln(k_F)$ with $r$ (Coefficient of
Assortativity) is an important feature of this network as the property  
is maintained for both PCNs and LINs.
Thus, it is a generic feature of proteins that needs fast network
transmission of information for functional versatility in the cell.  
Apart from helping in fast folding, assortative mixing, with its
role in percolation of information, could also be important for
allostery and signalling in proteins that require transfer of binding
information among different parts of the protein for further
function. It may be noted that `positive' correlation of $ln(k_F)$
with $r$ is an exceptional feature of coefficient of assortativity as all
other measures described so far~\cite{co,lro,tcd} have been known to
have a negative correlation.
Given the genetic basis and mode of formation of protein chains, the
signature of assortativity as an indicator to the rate of folding  
is clear. 
It will be interesting to see which physico-chemical factors could be
responsible for a positive correlation with $r$,  
thus speeding up the rate of folding with increasing assortative
mixing of the proteins.


Most networks having high degree of clustering consist of nodes such
that any two neighbours of these nodes have a high probability of  
themselves being linked to each other. 
The PCNs have been shown to have high degree of clustering, which
contributes to their small-world nature helping in efficient  
and effective dissipation of energy needed in their
function~\cite{protnet:Biophys,Bagler2005}. 
Though the LINs have significantly lower clustering coefficients than
their PCNs~(Fig.~\ref{fig:lc_LRI}) they  
show~(Fig.~\ref{fig:lnkr_c01}) a negative correlation with the rate of
folding of the proteins.  
This indicates that clustering of amino acids that participate in the
long-range interactions, into `cliques' slows down the  
folding process. 
However, the clustering coefficient of PCNs \emph{does not} have any
significant correlation to the rate of folding, indicating that the
short-range interactions may be playing a constructive and active role
in the determination of  the rate of the folding process by reducing
the negative contribution of the LINs.  
Our results clearly show that the separation of the types of contacts
in the PCNs and LINs clearly delineate the length scale of  
contacts that play crucial role in protein folding.

\chapter{\label{chap:protnet05}Conclusions}

After the synthesis in the cell, folding of the amino acid chain is
important for attaining the structure required to reach a functional
state as soon as possible.  
This happens through the formation of short- as well as long-range
interactions.  
While the former are largely responsible for formation of secondary
structure units, The latter bring spatially distant (along the chain) residues
closer.  
Secondary and tertiary structures are formed primarily by noncovalent
interactions. Our graph theoretical representations of proteins structure,
Proteins Contact Network (PCN) and Long-range Interaction Network
(LIN), model various aspects of the three-dimensional structure of a
protein in an attempt to understand it's function and kinetics.

\subsubsection{The Small World Nature}
\begin{figure}[!tbh]
\begin{center}
\begin{tabular}{c}
\includegraphics[scale=0.8]{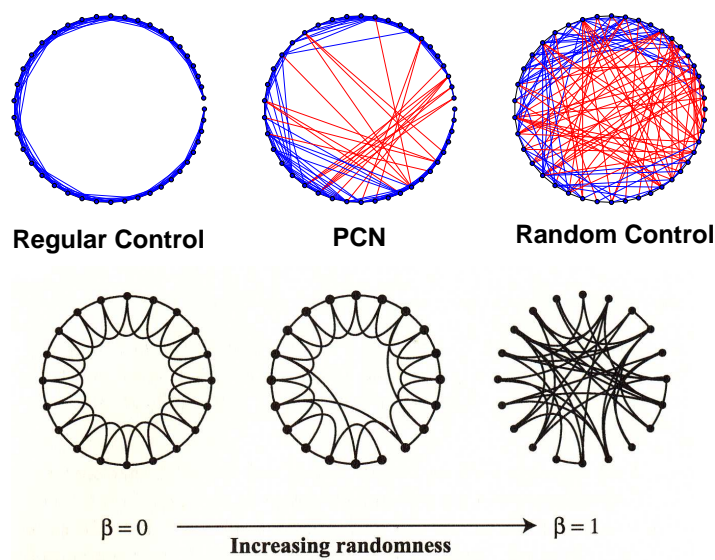}\\
\includegraphics[scale=0.8]{ws_scan.eps}
\end{tabular}
\end{center}
\caption[Small World Nature of PCNs.]
{PCN and Random \& Regular control of 2PDD. Comparison with
  Watts-Strogatz model.}  
\label{fig:Small_World_PCN}
\end{figure}

We found that proteins of diverse structural and functional
classification have small-world nature with low characteristic path
length ($L$) and high level of clustering ($C$) as shown in
Figure~\ref{fig:Small_World_PCN}. In this regard PCNs 
are similar to most other real-world
networks. Interestingly, we find that LINs depart from the small-world
nature. The LINs have medium range of $C$ in the proteins studied. 
The implication of small-world nature of PCNs is attributed to the
case of dissipation of energy upon complexation. Such a property may
have important role in efficient allosteric regulation of protein
functions.

\subsubsection{Hierarchy, Modularity, and Community structure}
We find that PCNs are characterised by hierarchical nature, as shown
by the independence of their clustering coefficient with size
(Figure~\ref{chap02:cn}).  
This observation is in accord with earlier findings by other
researchers~\cite{Rose1979,Crippen1978,cabios}, and more
investigation needs to be done on this line. We would like to point
out that it has been found in many real networks~\cite{ravsaz:science}
the hierarchy and modular architecture go hand in hand. Our
preliminary results suggest modular architecture in PCNs. It would be
interesting to see what significance, if any, the modules thus found
in protein structures would have. Such modules could be identified by 
the network community structure algorithms. 

\subsubsection{Assortative Nature of PCNs and LINs}
In contrast to all other naturally evolved intracellular networks
studied so far, we found that contact networks of proteins show
assortative mixing at both short and long length scales 
i.e.\ rich nodes tend to connect to other rich nodes. This is an
exceptional property as all other real-world networks known (except
for social networks) are disassortative. Interestingly, we find that
LINs too are assortative, which implies that assortativity is
independent of short-range interactions. We built appropriate random
controls to identify the appropriate network feature that possibly
contributes towards assortativity. We found that degree distribution
contributes significantly towards assortative mixing in PCNs as well
as their LINs. 

The predominance of disassortativity in real-world networks have been
alluded to confer the property of robustness (reduced spread of
perturbations) in the network. Then why are the contact networks of
protein structures assortative?
Communication among the residues of the protein is important.
It is known that ``network of residues'' mediate allosteric
communication in proteins~\cite{Suel2002,Dima2006}. It is also
proposed that allostery is an intrinsic property of \emph{all} dynamic
proteins~\cite{Gunasekaran2004}. We propose that assortativity is an
indicator of `allosteric communication network' established within the
protein structure and is important enough to be found in all proteins.

The role of specific residues in protein folding and their
evolutionary conservation is highly
debated~\cite{mirny_shakh_PNAS_1998,ScalleyKim_Baker_JMB_2004,Larson_etal_JMB_2002}.
Mirny \emph{et al.}~\cite{mirny_shakh_PNAS_1998} found that conserved
residues that are part of folding nucleus, across proteins, were found
to be in contact with each other. Based on this finding, we propose that
folding nucleus of a protein could be a subset of the set of residues
that form assortative group.  

Our observation of assortative mixing and hence a set of residues that
are part of a assortative network opens up new directions of work.

\subsubsection{Biophysical implication of topological parameters}
One would expect to have biophysical implications of the exceptional
network properties that we observe. We found that for both PCNs and
LINs, coefficient of assortativity, a measure of the assortativity,
has positive correlation with the rate of folding of single-domain,
two-state folding proteins. Similarly, we find that clustering
coefficient of LINs has a high negative correlation to the rate of
folding of these proteins, though that of PCNs show no significant
correlation.
Other workers have developed parameters specific for proteins ($CO$,
$LRO$, $TCD$) and correlated with rate of folding. Our aim was to show
the relevance of general network parameters to a kinetic property of
the proteins.  
Indices such as closeness, betweenness offer more local and hence
residue-specific information. By combining our general, global
parameters with such local ones one could address broader questions
related to protein structure and function. 

\subsubsection{Advantages and Limitations of PCN Model}
PCN is, by virtue of coarse-graining, a simple model. It doesn't
involve the time evolution of the protein structure. Rather it models
the static native state structure. We don't explicitly consider
information about the chemical nature of the side-chains in this
model. Although, since the final native-state structure is an outcome
of the chemical interactions happening among various amino acids
the model implicitly does consider the chemical interactions involved.  

Each of the twenty amino acids has different numbers of atoms and
hence has different size. Hence the nature of noncovalent contacts 
depends on the specific amino acids involved. We haven't included
that information in our model so far. 
 
The time evolution of the protein structure can be considered by
building the weighted network using the Transition State Ensemble 
(TSE) structures. Depending on the question being asked and its
sensitivity to the above-mentioned details one may consider adding  
further details to the PCN, thereby enhancing it.

Thus complex network analyses offers to be an important tool in
studying the structure-function of proteins--the fascinating molecule 
of life. 

\appendix
\chapter{Pseudocode of the Algorithms Implemented}
\label{app:pseudocode}

In this Appendix we list pseudocodes of some of the important
algorithms we implemented for the complex network analyses of the
proteins structures. Algorithm~\ref{main} lists various experiments to
be performed on a single protein structure. Following is a list of
frequently used variables across all the algorithms.
\begin{center}
\begin{tabular}{cl}
$n_r$ & Number of nodes in the network\\
$n_c$ & Number of links in the network\\
$Adj$ & The $n_r$x$n_r$ Adjacency Matrix\\
\end{tabular}
\end{center}

\begin{pseudocode}[ruled]{ComplexNetworksAnalysis}{PDB File}
\label{main}
\COMMENT{The `main' function for Complex Network analyses.}\\
  \BEGIN
    \CALL{Analyse-Pcn}{PDB File}\\
    \CALL{Analyse-Pcn-Lin}{Adj}\\
    \CALL{Analyse-RandomControlTypeI}{n_r,n_c}\\
    \CALL{Analyse-RandomControlTypeI-LIN}{Adj}\\
    \CALL{Analyse-RandomControlTypeII}{Adj}\\
    \CALL{Analyse-RandomControlTypeII-LIN}{Adj}
  \END
\end{pseudocode}

\begin{pseudocode}[ruled]{AnalysePcn}{PDB File}
\COMMENT{Function for analyses of Protein Contact Network.}\\
  \BEGIN
    \CALL{GetCoordinates}{PDB File}\\
     \quad \RETURN{XYZ, n_r}\\
    \CALL{GetAdjacencyMatrix}{XYZ,n_r,R_c}\\ 
     \quad \RETURN{Distances, Adj,n_c}\\
    \CALL{GetDegree}{Adj, n_r}\\ 
     \quad \RETURN{Degree,k_{max}}\\
    \CALL{GetDegreeDistribution}{Degree,n_r}\\ 
     \quad \RETURN{DegreeDist}\\
    \CALL{GetDegreeCorrelations}{Degree,n_r}\\
     \quad \RETURN{DegreeCorrelations}\\
  \END
\end{pseudocode}

\begin{pseudocode}[ruled]{GetAdjacencyMatrix}{XYZ,n_r,R_c}
\COMMENT{Function for construction of adjacency matrix.}\\

\PROCEDURE{ComputeDist}{XYZ,i,j}
  Dist \GETS  \sqrt(\sum_{k=1}^{3} (XYZ(i,k)-XYZ(j,k))^2 )\\
  \RETURN{Dist}
\ENDPROCEDURE

Adj \GETS 0\\
Dist \GETS 0\\
\FOR i \GETS 1 \TO n_r-1 \DO
\FOR j \GETS i+1 \TO n_r \DO
\BEGIN
Dist=\CALL{ComputeDist}{XYZ,i,j};\\
\IF (Dist<= R_c\AA) \THEN
     \BEGIN  Adj(i,j) \GETS 1;\\  Adj(j,i) \GETS 1\\   
             Distance(i,j) \GETS  Dist;\\  Distance(j,i) \GETS Dist\\  \END
\END\\
\\
\RETURN{Adj, Dist}
\end{pseudocode}

\begin{pseudocode}[ruled]{GetDegree}{Adj,n_r}
Degree(n_r,2) \GETS 0\\
\FOR i \GETS 1 \TO n_r \DO
\BEGIN Degree(i,1) \GETS 1\\ \END\\ \\

\FOR i \GETS 1 \TO n_r \DO
\FOR j \GETS 1 \TO n_r \DO
\BEGIN Degree(i,2) \GETS Degree(i,2)+1\\ \END\\
\\
DegreeMax \GETS \CALL{GetDegreeMaximum}{Degree};\\
\\
\RETURN{Degree,DegreeMax}
\end{pseudocode}

\begin{pseudocode}[ruled]{GetDegreeDistribution}{Degree,n_r}
DegreeDistribution(1:n_r,1:2) \GETS 0\\
DegreeDistributionNorm(1:n_r,1:2) \GETS 0\\

\FOR i \GETS 1 \TO n_r \DO
\BEGIN DegreeDistribution(i,1) \GETS i\\ 
       DegreeDistributionNorm(i,1) \GETS i\\
\END\\ \\

\FOR i \GETS 1 \TO n_r \DO
\BEGIN DegreeDistribution(Degree(i,2),2) \GETS \\
DegreeDistribution(Degree(i,2),2) + 1\\ \END\\
\\
MaxDegreeDist \GETS \CALL{GetMax}(DegreeDistribution(:,2))\\
DegreeDistributionNorm(:,2)\GETS \\ DegreeDistribution(:,2)/ MaxDegreeDist\\
\\
\RETURN{DegreeDistribution, DegreeDistributionNorm}
\end{pseudocode}

\begin{pseudocode}[ruled]{GetDegreeCorrAvg}{Degree,Adj,n_r}
DegreeCorr(1:n_r,1:2) \GETS 0\\
DegreeCorrAvg(1:n_r,1:2) \GETS 0\\

\FOR i \GETS 1 \TO n_r \DO \BEGIN
DegreeCorr(i,1) =  Degree(i,2)\\

\FOR j \GETS 1 \TO n_r \DO \BEGIN
\IF (Adj(i,j) == 1) \THEN
     \BEGIN DegreeCorr(i,2) = DegreeCorr(i,2) + Degree(j,2) \END\\
\END \END
\\ \\

\FOR i \GETS 1 \TO n_r \DO
\BEGIN DegreeCorrAvg(i,1) \GETS i\\ 
       DegreeCorrAvg(DegreeCorr(i,1),2) \GETS \\
       DegreeCorrAvg(DegreeCorr(i,1),2) + DegreeCorr(i,2)\\
\END\\ \\

\FOR i \GETS 1 \TO n_r \DO \BEGIN
\IF (DegreeDist(i,2) /= 0) \THEN
     \BEGIN DegreeCorrAvg(i,2) = DegreeCorrAvg(i,2) / DegreeDist(j,2) \END\\
\END\\
\\
\RETURN{DegreeCorr, DegreeCorrAvg}
\end{pseudocode}

\begin{pseudocode}[ruled]{GetCoeffAssortativity}{Degree,Adj,n_r}

Term01 \GETS 0 \\
Term02 \GETS 0 \\
Term03 \GETS 0 \\

\FOR i \GETS 1 \TO n_r \DO \\ \BEGIN
\FOR j \GETS 1 \TO n_r \DO \\ \BEGIN

\IF (Adj(i,j) == 1) \THEN
     \BEGIN Term01 \GETS Term01 + (Degree(i,2) * Degree(j,2))\\
            Term02 \GETS Term02 + (Degree(i,2) * Degree(j,2))\\ 
            Term03 \GETS Term03 + (Degree(i,2)^2 + Degree(j,2)^2)\\
     \END\\
\END \END\\
\\
TotalEdges=\CALL{SUM}(Adj)\\
Temp01 = \frac{Term01}{TotalEdges}\\
Temp02 = {\left(\frac{Term02}{2*TotalEdges}\right)}^2\\
Temp02 = \frac{Term03}{2*TotalEdges}\\
CoeffAssortativity = \frac{Temp01-Temp02}{Temp03-Temp02}\\
\\
\RETURN{CoeffAssortativity}
\end{pseudocode}

\begin{pseudocode}[ruled]{GetClusteringCoeff}{Degree,Adj,n_r}

ClusteringCoeff(1:n_r,1:2) \GETS 0\\
AvgCC \GETS 0\\
\FOR i \GETS 1 \TO n_r \DO \\ \BEGIN
ClusteringCoeff(i,1) \GETS i\\

\FOR i \GETS 1 \TO n_r \DO \\ \BEGIN
\FOR j \GETS 1 \TO n_r \DO \\ \BEGIN

\IF (Degree(i,2) >= 2) \THEN
\BEGIN 
 ClusteringCoeff(i,2) \GETS ClusteringCoeff(i,2) + \\ (Adj(i,j)* Adj(i,k) * Adj(j,k))
     \END\\
\END \END\\
\\
\IF (Degree(i,2) >= 2) \THEN
\BEGIN 
 ClusteringCoeff(i,2) \GETS \frac{ClusteringCoeff(i,2)}{Degree(i,2) * (Degree(i,2)-1)}
\END
\END\\
\\
\FOR i \GETS 1 \TO n_r \DO \\ \BEGIN
  AvgCC = AvgCC + ClusteringCoeff(i,2)
\END\\
AvgCC=AvgCC/n_r
\\ \\
\RETURN{ClusteringCoeff,AvgCC}
\end{pseudocode}

\begin{pseudocode}[ruled]{GetClusteringCoeffDist}{ClusteringCoeff}
ClusteringCoeffDist(1:10,1:2) \GETS 0\\
\FOR i \GETS 1 \TO 10 \DO \BEGIN
ClusteringCoeffDist(i,1)=i\\
\END\\ \\

\FOR i \GETS 1 \TO n_r \DO \\ \BEGIN
\IF ClusteringCoeff(i,2)>=0~\&~ClusteringCoeff(i,2)<0.1\\
 ClusteringCoeffDist(1,2) = ClusteringCoeffDist(1,2) + 1\\

\ELSEIF ClusteringCoeff(i,2)>=0.1~\&~ClusteringCoeff(i,2)<0.2\\
 ClusteringCoeffDist(2,2) = ClusteringCoeffDist(2,2) + 1\\

\ELSEIF ClusteringCoeff(i,2)>=0.2~\&~ClusteringCoeff(i,2)<0.3\\
 ClusteringCoeffDist(3,2) = ClusteringCoeffDist(3,2) + 1\\

\ELSEIF ClusteringCoeff(i,2)>=0.3~\&~ClusteringCoeff(i,2)<0.4\\
 ClusteringCoeffDist(4,2) = ClusteringCoeffDist(4,2) + 1\\

\ELSEIF ClusteringCoeff(i,2)>=0.4~\&~ClusteringCoeff(i,2)<0.5\\
 ClusteringCoeffDist(5,2) = ClusteringCoeffDist(5,2) + 1\\

\ELSEIF ClusteringCoeff(i,2)>=0.5~\&~ClusteringCoeff(i,2)<0.6\\
 ClusteringCoeffDist(6,2) = ClusteringCoeffDist(6,2) + 1\\

\ELSEIF ClusteringCoeff(i,2)>=0.6~\&~ClusteringCoeff(i,2)<0.7\\
 ClusteringCoeffDist(7,2) = ClusteringCoeffDist(7,2) + 1\\

\ELSEIF ClusteringCoeff(i,2)>=0.7~\&~ClusteringCoeff(i,2)<0.8\\
 ClusteringCoeffDist(8,2) = ClusteringCoeffDist(8,2) + 1\\

\ELSEIF ClusteringCoeff(i,2)>=0.8~\&~ClusteringCoeff(i,2)<0.9\\
 ClusteringCoeffDist(9,2) = ClusteringCoeffDist(9,2) + 1\\

\ELSEIF ClusteringCoeff(i,2)>=0.9~\&~ClusteringCoeff(i,2)<1.0\\
 ClusteringCoeffDist(10,2) = ClusteringCoeffDist(10,2) + 1\\

\END\\
\\
\RETURN{ClusteringCoeffDist}
\end{pseudocode}

\begin{pseudocode}[ruled]{GetTypeIRandomControl}{n_r,n_c}
adjTypeI(1:n_r,1:n_r) \GETS 0\\
RandomEdges = n_c - (n_r-1)\\

\FOR i \GETS 1 \TO n_r-1 \DO \\ \BEGIN
   adjTypeI(i,i+1) \GETS 1\\
   adjTypeI(i+1,i) \GETS 1\\
\END\\ 
\\
RandomEdgesCounter=0\\
\WHILE RandomEdgesCounter \leq RandomEdges \DO
\BEGIN
iRan=\CALL{RandomNumber}{iSeed};\\
jRan=\CALL{RandomNumber}{jSeed};\\
i \GETS iRan*(n_r-1) + 1\\
j \GETS jRan*(n_r-1) + 1\\

\IF (i \neq j~\&\&~|i-j| \neq 1~\&\&~adjTypeI(i,j) \neq 1) \THEN
\BEGIN 
  adjTypeI(i,j) \GETS 1\\
  adjTypeI(j,i) \GETS 1\\
\END
\END
\\
\RETURN{adjTypeI}
\end{pseudocode}

\begin{pseudocode}[ruled]{GetLin}{adj,n_r}

\FOR i \GETS 1 \TO n_r-1 \DO
\FOR j \GETS i+1 \TO n_r \DO
\BEGIN 
\IF (adj(i,j)==1 ~\&\&~ j \neq i+1 ~\&\&~ j <= i+LRIThreshold) \THEN
\BEGIN
   adj(i,j) \GETS 0\\
   adj(j,i) \GETS 0\\
\END
\END\\
\\
\RETURN{adj}
\end{pseudocode}

\begin{pseudocode}[ruled]{GetTypeIIRandomControl}{adj,n_r,n_c,EdgeRewirings}
EdgeRewiringsCounter=0\\
\WHILE EdgeRewiringsCounter < EdgeRewirings \DO
\BEGIN
iRan=\CALL{RandomNumber}{iSeed};\\
jRan=\CALL{RandomNumber}{jSeed};\\
pRan=\CALL{RandomNumber}{pSeed};\\
qRan=\CALL{RandomNumber}{qSeed};\\
i \GETS iRan*(n_r-1) + 1\\
j \GETS jRan*(n_r-1) + 1\\
p \GETS pRan*(n_r-1) + 1\\
q \GETS qRan*(n_r-1) + 1\\ \\
   \WHILE adj(i,j)== 0 ~||~ adj(p,q)==0  ~||~ adj(i,q)==1 ~||~\\
          adj(p,j)==1 ~||~ |i-j|<2 ~||~ |p-q|<2 ~||~ |i-q|<2 ~||~\\ |p-j|<2 \DO
   \BEGIN
   iRan=\CALL{RandomNumber}{iSeed};\\
   jRan=\CALL{RandomNumber}{jSeed};\\
   pRan=\CALL{RandomNumber}{pSeed};\\
   qRan=\CALL{RandomNumber}{qSeed};\\
   i \GETS iRan*(n_r-1) + 1\\
   j \GETS jRan*(n_r-1) + 1\\
   p \GETS pRan*(n_r-1) + 1\\
   q \GETS qRan*(n_r-1) + 1\\
   \END\\ \\
    adj(i,j) \GETS 0; adj(j,i) \GETS 0;\\
    adj(p,q) \GETS 0; adj(q,p) \GETS 0;\\
    adj(i,q) \GETS 1; adj(q,i) \GETS 1;\\
    adj(p,j) \GETS 1; adj(j,p) \GETS 1;\\
    \\
    EdgeRewiringsCounter \GETS EdgeRewiringsCounter + 1
\END
\\ \\
\RETURN{adj}
\end{pseudocode}


\nocite{*}
\bibliographystyle{ieeetr}
\addcontentsline{toc}{chapter}{Bibliography}


\printindex
\end{document}